\documentclass[fleqn,usenatbib]{mnras}

\usepackage{newtxtext,newtxmath}

\usepackage[T1]{fontenc}
\usepackage{ae,aecompl}

\usepackage{graphicx}	
\usepackage{amsmath}	
\usepackage{amsfonts}
\usepackage{caption}
\usepackage{pdflscape} 
\usepackage{graphicx}
\usepackage{comment}
\usepackage[dvipsnames]{xcolor}
\usepackage[normalem]{ulem}

\newcommand{\kms}{km~s$^{-1}$}

\newcommand{\unitlum}{erg~s$^{-1}~$}

\newcommand{\ha}{H$\upalpha$}
\newcommand{\hb}{H$\upbeta$}
\newcommand{\hg}{H$\upgamma$}
\newcommand{\hd}{H$\updelta$}
\newcommand{\nii}{[\ion{N}{ii}]}
\newcommand{\niii}{\ion{N}{iii}}
\newcommand{\oiii}{[\ion{O}{iii}]}
\newcommand{\heii}{\ion{He}{ii}}
\newcommand{\chisq}{$\rm \chi^2$}
\newcommand{\dsg}{AT\,2019dsg}

\newcommand{\ppxf}{\textsc{ppxf}}

\title[AT2019dsg]{Accretion disc cooling and narrow absorption lines in the tidal disruption event AT\,2019dsg}

\author[G. Cannizzaro et al.]{
G. Cannizzaro$^{1,2}$ \thanks{E-mail: g.cannizzaro@sron.nl},
T. Wevers$^{3,4}$,
P.~G. Jonker$^{2,1}$,
M.~A. P\'{e}rez-Torres$^{5}$,
J. Moldon$^{5,6}$,
\newauthor
D. Mata-S\'anchez$^{6}$,
G. Leloudas$^{7}$,
D.~R. Pasham$^{8}$,
S. Mattila$^{9}$,
I. Arcavi$^{10,11}$,
\newauthor
K. Decker French$^{12}$
F. Onori$^{13}$,
C. Inserra$^{14}$,
M. Nicholl$^{15,16}$
M. Gromadzki$^{17}$,
T.-W. Chen$^{18}$,
\newauthor
T.~E. M\"uller-Bravo$^{19}$,
P. Short$^{16}$,
J.~P. Anderson$^{4}$,
D.~R. Young $^{20}$,
K.~C. Gendreau$^{21}$,
\newauthor
Z. Arzoumanian$^{21}$,
M. L\"{o}wenstein$^{21,22}$,
R. Remillard$^{8}$,
R. Roy$^{23}$,
D. Hiramatsu$^{24,25}$
\\
$^{1}$SRON, Netherlands Institute for Space Research, Sorbonnelaan, 2, NL-3584CA Utrecht, the Netherlands\\
$^{2}$Department of Astrophysics/IMAPP, Radboud University, P.O. Box 9010, 6500 GL Nijmegen, the Netherlands\\
$^{3}$Institute of Astronomy, University of Cambridge, Madingley Road, Cambridge CB3 0HA, UK\\
$^{4}$European Southern Observatory, Alonso de C\'ordova 3107, Casilla 19, Santiago, Chile\\
$^{5}$Instituto de Astrof\'{i}sica de Andaluc\'{i}a (CSIC), Glorieta de la Astronom\'{i}a s/n, E-18080 Granada, Spain\\
$^{6}$Jodrell Bank Centre for Astrophysics, School of Physics and Astronomy, The University of Manchester, Manchester, M13 9PL, UK\\
$^{7}$DTU Space, National Space Institute, Technical University of Denmark, Elektrovej 327, 2800 Kgs. Lyngby, Denmark\\
$^{8}$MIT Kavli Institute for Astrophysics and Space Research, Cambridge, MA 02139, USA\\
$^{9}$Tuorla Observatory, Department of Physics and Astronomy, FI-20014 University of Turku, Finland\\
$^{10}$The School of Physics and Astronomy, Tel Aviv University, Tel Aviv 69978, Israel\\
$^{11}$CIFAR Azrieli Global Scholars program, CIFAR, Toronto, Canada\\
$^{12}$Department of Astronomy, University of Illinois, Urbana IL, 61801, USA\\
$^{13}$INAF - Osservatorio Astronomico d'Abruzzo via M. Maggini snc, I-64100 Teramo, Italy\\
$^{14}$School of Physics \& Astronomy, Cardiff University, Queens Buildings, The Parade, Cardiff, CF24 3AA, UK\\
$^{15}$Birmingham Institute for Gravitational Wave Astronomy and School of Physics and Astronomy, University of Birmingham, Birmingham B15 2TT, UK \\
$^{16}$Institute for Astronomy, University of Edinburgh, Royal Observatory, Blackford Hill, EH9 3HJ, UK \\
$^{17}$Astronomical Observatory, University of Warsaw, Al. Ujazdowskie 4, 00-478 Warszawa, Poland\\
$^{18}$The Oskar Klein Centre, Department of Astronomy, Stockholm University, AlbaNova, SE-10691 Stockholm, Sweden\\
$^{19}$School of Physics and Astronomy, University of Southampton, Southampton, Hampshire, SO17 1BJ, UK\\
$^{20}$Astrophysics Research Centre, School of Mathematics and Physics, Queens University Belfast, Belfast BT7 1NN, UK\\
$^{21}$Astrophysics Science Division, NASA Goddard Space Flight Center, Greenbelt, MD 20771, USA\\
$^{22}$Department of Astronomy, University of Maryland, College Park, MD 20742, USA\\
$^{23}$The Inter-University Centre for Astronomy and Astrophysics, Ganeshkhind, Pune - 411007, India\\
$^{24}$Las Cumbres Observatory, 6740 Cortona Drive, Suite 102, Goleta, CA 93117-5575, USA\\
$^{25}$Department of Physics, University of California, Santa Barbara, CA 93106-9530, USA\\
}
\date{Accepted XXX. Received YYY; in original form ZZZ}

\pubyear{2020}

\begin{document}
\label{firstpage}
\pagerange{\pageref{firstpage}--\pageref{lastpage}}
\maketitle

\begin{abstract}
We present the results of a large multi-wavelength follow-up campaign of the Tidal Disruption Event (TDE) \dsg, focusing on low to high resolution optical spectroscopy, X-ray, and radio observations. 
The galaxy hosts a super massive black hole of mass $\rm (5.4\pm3.2)\times10^6\,M_\odot$ and careful analysis finds no evidence for the presence of an Active Galactic Nucleus, instead the TDE host galaxy shows narrow optical emission lines that likely arise from star formation activity. The transient is luminous in the X-rays, radio, UV and optical. The X-ray emission becomes undetected after $\sim$125 days, and  the radio luminosity density starts to decay at frequencies above 5.4 GHz by $\sim$180 days. Optical emission line signatures of the TDE are present up to $\sim$250 days after the discovery of the transient. The medium to high resolution spectra show traces of absorption lines that we propose originate in the self-gravitating debris streams. At late times, after $\sim$200 days, narrow Fe lines appear in the spectra. The TDE was previously classified as N-strong, but after careful subtraction of the host galaxy's stellar contribution, we find no evidence for these N lines in the TDE spectrum, even though O Bowen lines are detected. The observed properties of the X-ray emission are fully consistent with the detection of the inner regions of a cooling accretion disc. The optical and radio properties are consistent with this central engine seen at a low inclination (i.e., seen from the poles). 
\end{abstract}

\begin{keywords}
Accretion:accretion discs -- transients:tidal disruption events -- galaxies:nuclei
\end{keywords}

\section{Introduction}

In the nuclear regions of a galaxy, a star whose orbital pericenter passes too close to the central supermassive black hole (SMBH) will be torn apart by the tidal forces \citep{hills75,rees88, evans89}. During this so-called Tidal Disruption Event (TDE), roughly half of the stellar matter will spiral towards the SMBH, confined into self-gravitating streams \citep{Guillochon2014} that can self-intersect and dissipate energy. 
This phenomenon gives rise to a luminous flare, typically peaking in the UV or soft X-rays, that can exceed the Eddington luminosity of the SMBH, over timescales of months to a year.
While TDEs were originally discovered in the X-ray band \citep[see][for a review]{Komossa2002}, recently significant numbers of TDE candidates are being discovered through wide field time domain optical surveys \citep[e.g.][]{holoien18,Leloudas2016,Blagorodnova2017,wyrzykowski17,onori19,blagorodnova19,vanvelzen20}. The optical emission of TDEs has shown a wide variety of properties, but they are usually characterised by broad H and He emission lines \citep{Arcavi2014}, a strong blue continuum, black body (BB) temperatures of order 10$^4$ K and luminosities of order $10^{44}$ \unitlum. A subset of TDEs have shown metal lines: either N and O lines \citep{blagorodnova19,Leloudas2019,onori19} excited through the Bowen fluorescence mechanism \citep{bowen34,bowen35} and/or low ionisation Fe lines \citep{wevers2019b}. The presence/absence of H, He and N lines led to the phenomenological classification of \citet{vanvelzen20} into three classes: TDE-H, TDE-He and TDE-Bowen.
There is no consensus on the origin of the optical emission of TDEs, be it either reprocessing of X-ray light through an atmosphere \citep{Guillochon2014,Stone2016,Dai2018} or originating from shocks due to the self-intersections of the debris stream \citep{Piran2015,Shiokawa2015,Bonnerot2019}. 

While the presence of an accretion disc has been inferred also in optically selected TDEs, through the emergence of metal lines, double-peaked line profiles \citep{short20, hung2020} and their X-ray properties \citep{Jonker2020,Wevers2020}, the TDE system has yet to be fully understood. This is hindered by the low numbers of TDE candidates (a few dozens), lack of high cadence, detailed spectroscopic monitoring, possible presence of reprocessing dust \citep{mattila18}, the presence of (non)relativistic radio outflows \citep[see][for a review]{Alexander20} and/or the dependency on the viewing angle of the observed properties \citep{Dai2018,nicholl19}.

We present the analysis of an intensive follow-up campaign of \dsg, a nuclear transient discovered on 2019 April 09 by the Zwicky Transfient Facility (ZTF)\footnote{https://wis-tns.weizmann.ac.il/object/2019dsg/discovery-cert}, with the name ZTF19aapreis and subsequently classified as a TDE, at a magnitude r=18.9. The transient was discovered within a galaxy at a redshift z=0.0512 \citep{atel_class_dsg}, which translates into a distance of D = 224 Mpc (we do not consider uncertainties in the luminosity distance), assuming a cosmology with $H_{\rm 0}$=67.7 km s$^{-1}$ Mpc$^{-1}$, $\Omega_{\rm M}$=0.309, $\Omega_{\rm \Lambda}$=0.691 \citep{Planck2014a}.
In \citet{vanvelzen20} \dsg\ was studied as part of a larger sample of TDEs discovered by the ZTF. \dsg\ is categorised as a "TDE-Bowen" due to the perceived presence of H Balmer, \heii\ as well as Bowen (N/O) fluorescence lines \citep{bowen34,bowen35}. The host galaxy of \dsg\ is in the "Green Valley" \citep{schawinski14}, a transition area of the colour-mass diagram between star-forming and quiescent galaxies. Galaxies in this region can show strong Balmer lines and the post-starburst, E+A galaxies, which are known to be preferential hosts for TDEs \citep{Arcavi2014,French2016}, are also in this region. 
\dsg\ was also listed as a candidate counterpart for a neutrino event \citep{atel_neutrino}. A study of the possible neutrino emission from \dsg\ and a more detailed analysis of the Spectral Energy Distribution (SED) is presented in \cite{stein20}, where they propose a "multi-zone" model to explain the various emission components of the transient: a central engine for the X-ray emission, an UV/Optical photosphere and an outflow that powers the radio and  neutrino emission.
In this work, we focus on the X-ray emission and the optical spectra, on which we perform subtraction of the stellar component of the galaxy. Through this, we find that the N emission lines previously identified with the TDE are instead due to the host galaxy. Finally, in our medium and high resolution spectra, we find evidence for absorption lines, potentially due to the streams of disrupted stellar material.
Throughout the paper, we use as a reference the discovery date of the transient: 2019 Apr 09 (MJD 58\,582.46). All uncertainties are reported as 1$\sigma$, unless stated otherwise.

\section{Observations}

\subsection{Swift UVOT/XRT}
\dsg\ was observed with the UVOT and XRT instruments on board the Neil Gehrels \textit{Swift} satellite \citep{swift:gehrels} with a cadence of around 3 days starting on 2019 May 21 (42 days after discovery) until 2019 Oct 15.
We reduce and extract Swift/UVOT measurements using the {\sc uvotsource} task in {\sc heasoft} version 6.24. We use a standard 5 arcsec aperture to extract flux measurements, and a 50 arcsec aperture centred on an empty nearby region to estimate sky background levels. We correct for Galactic extinction assuming E(B$-$V) = 0.087 \citep{schlafly11}.

We use the online Swift/XRT pipeline tool\footnote{https://www.swift.ac.uk/user\_objects/} to reduce the X-ray data and determine source count rates in the 0.3--10 keV band, using the source optical coordinates. Due to the low number of counts, we create a single stacked spectrum with a total exposure time of 20 ks for spectral analysis.

\subsection{NICER}
Following the {\it Swift} detection of X-rays from \dsg, the Neutron star Interior Composition ExploreR ({\it NICER}) \citep{Gendreau16} made several observations between 21 May 2019 and 6 June 2019. A second set of observations were obtained several months later (3-5 October 2019) in response to the IceCube alert of a neutrino detection from a sky region containing \dsg\ \citep{icecube}. All observations of \dsg\ (OBSIDs: 200680101-2200680112) are reprocessed using the gain file {\it nixtiflightpi20170601v005.fits} by applying the \textsc{nicerl2 ftool} with the default filtering criteria. 

To build up statistics, data from observation IDs 2200680103-2200680104 and 2200680107-2200680108 are combined due to observational proximity in time and/or short exposure times. Observation IDs 2200680110-2200680112 are also combined. 
The total spectra for these seven groups are extracted and the background spectra are estimated using empirical background spectral libraries, constructed from observations of source-free areas of sky.

\subsection{Optical spectroscopic observations}
\subsubsection{WHT/ISIS}
Observations of \dsg\ were obtained with the Intermediate dispersion Spectrograph and Imaging System (ISIS) mounted at the Cassegrain focus of the William Herschel Telescope (WHT) in Roque de los Muchachos observatory (La Palma, Spain), Spain on 2019 August 22 under program SW19b01. Using a 1 arcsec slit in combination with the R600B grating provides wavelength coverage between 3600 and 5100 \AA\ in the host rest frame, while the R600R grating covers the 6300 - 7800 \AA\ range. The seeing was variable between 0.5 and 0.9 arcsec during the 2$\times$2700s observations, leading to a seeing limited full width half maximum (FWHM) spectral resolution of $\sim$ 94 km s$^{-1}$ (i.e., $\sigma_{instr}$ = 40 km s$^{-1}$), measured at $\sim$4000 \AA\ from the arc frame. The observations were carried out with the slit at parallactic angle \citep{filippenko82} and with binning 1x1.

After performing the standard reduction tasks, such as a bias level subtraction and flat field correction, in \textsc{iraf} \citep{tody86}, we extract spectra using an extraction box with width of 1 arcsec in the spatial dimension. Wavelength solutions are applied using CuAr+CuNe arc frames obtained prior and after the science exposures. We combined the two exposures into a single, averaged spectrum using weights set to the average signal to noise ration (SNR) of the individual spectra, and subsequently we fit cubic splines to normalise the averaged spectrum to the continuum (the spectra were not flux calibrated due to the absence of a standard star observation). 
A journal of the spectroscopic observations is presented in Table \ref{tab:spec_obs}.

\subsubsection{WHT/ACAM}
\label{sec:acam}
\dsg\ was observed several times with the Auxiliary-port CAMera (ACAM) low-resolution spectrograph mounted at the Cassegrain focus of the WHT under program W19AN003. The V400 grating in combination with the GG395A order-sorting filter provides a wavelength coverage of 3950--9400 \AA\ and the resolution is R$\sim$430 and $\sim$580 for a 1.0 and 0.75 arcsec slit, respectively (from the technical manual of the instrument, measured at 5650 \AA). The data were reduced using a pipeline based on standard \textsc{iraf} data reductions procedures: flat field and bias correction, cosmic-ray cleaning, wavelength and flux calibration with arc lamps and standard stars. All observations were carried out with the slit at parallactic angle and with 1x1 binning.
\subsubsection{TNG/DOLORES}
We observed AT2019dsg twice with the Device Optimized for the LOw RESolution (DOLORES), installed at the Nasmyth B focus of the Telescopio Nazionale Galileo (TNG) in Roque de los Muchachos observatory (La Palma, Spain). The wavelength coverage of the LR-B grating is $\sim$3000-8430 \AA, with a resolution R$\sim$580 for a 1.0 arcsec slit (nominal value, measured at 5850 \AA). The slit was oriented at parallactic angle for all observations and the binning was 1x1. The spectra were reduced using standard \texttt{iraf} procedures (bias and flat-field correction, wavelength and flux calibration with arc lamps and standard stars).

\subsubsection{NTT/EFOSC2}
\label{sec:efosc}
Observations of \dsg\ were carried out in the framework of the advanced Public ESO Spectroscopic Survey for Transient Objects (ePESSTO+; \citealt{Smartt2015}), starting with the classification spectrum taken on 2019 May 13 \citep{atel_class_dsg}, with the ESO Faint Object Spectrograph and Camera v.2 (EFOSC2), an instrument for low resolution spectroscopy mounted at the Nasmyth B focus of the New Technology Telescope (NTT) at the La Silla observatory, Chile. All observations were performed using grism Gr\#13, that provides a nominal wavelength coverage of 3685-9315 \AA\ and a resolution with R$\sim$355 for a 1 arcsec slit, measured at 5600 \AA\ (nominal value), with the slit at parallactic angle and with 1x1 binning. Data were reduced using pipelines based on standard \textsc{iraf} tasks, such as bias and flat-field correction and wavelength and flux calibration using arc lamps and standard stars. When necessary, multiple spectra taken on the same night have been averaged with weights set to the mean value of the SNR of the individual exposures. 

\subsubsection{VLT/X-SHOOTER}
\label{sec:xsh}
We obtained one spectrum with the medium resolution spectrograph X-Shooter \citep{Vernet11}, mounted at the Cassegrain focus of the second Unit Telescope (UT2) at the Very Large Telescope (VLT). X-shooter covers the wavelength range from 3000 to 25000 \AA. The observation was carried out on 2019 August 29 with slit widths of 1.0, 0.9 and 0.9 arcsec for the UVB, VIS and NIR arms, respectively, with the slit oriented at parallactic angle. The set up used yields resolution of $\rm R \simeq 5400$ (UVB), $\rm R \simeq 8900$ (VIS) and $\rm R \simeq 5600$ (NIR).s The data were reduced using the \textsc{reflex} X-shooter pipeline version 2.9.3 \citep{freudling13} and are not flux calibrated. The spectra were also corrected for atmospheric absorption features using synthetic transmission spectra with the \textsc{molecfit} software \citep{smette15,kausch15}. The NIR part of the spectrum shows no significant spectroscopic TDE signatures and is therefore not considered in this paper.

\subsubsection{VLT/UVES}
We observed \dsg\ with the Ultraviolet and Visual Echelle Spectrograph (UVES), mounted at the Nasmyth B focus of UT2 at the VLT. The observation was carried out on 2019 Jul 04 with the standard central wavelengths of 346+580 nm (dichroic 1) and 437+860 nm (dichroic 2) that provide an almost full coverage of the 300 -- 1060 nm wavelength range. With a 1 arcsec slit, the resolving power is $\sim$40000. The data were reduced using the UVES pipeline \citep{ballester00}, adjusting the pipeline parameters where necessary.\\

\subsubsection{Las Cumbres/FLOYDS}
Las Cumbres Observatory (LCO) optical spectra were taken with the Folded Low Order whYte-pupil Double-dispersed Spectrograph (FLOYDS) mounted on the 2m Faulkes Telescope North (FTN) and South (FTS) at Haleakala (USA) and Siding Spring (Australia), respectively. A 2$"$ slit was placed on the target at the parallactic angle. One-dimensional spectra were extracted, reduced, and calibrated following standard procedures using the FLOYDS pipeline\footnote{https://github.com/svalenti/FLOYDS\_pipeline} \citep{valenti14}. 
The FLOYDS spectra are not used in the analysis, due to a combination of low SNR, wavelength range covered and proximity in time with higher quality spectra. Nonetheless, the spectra are reported in appendix (\ref{app:floyds}).

\subsubsection{du Pont/WFCCD}
\begin{figure*}
        \includegraphics[scale=0.55, angle = 0]{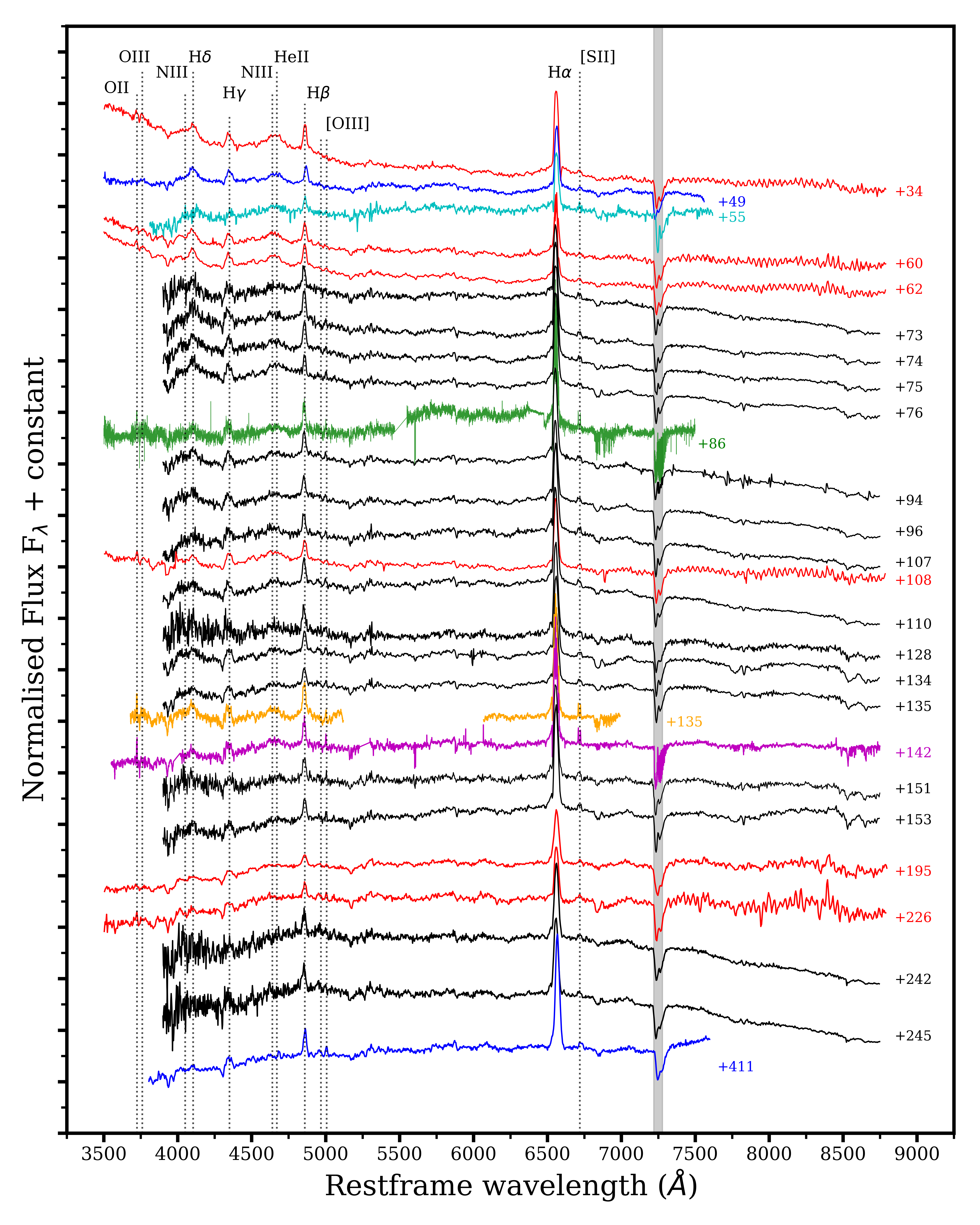} 
        \caption{Sequence of spectra taken with EFOSC2 (red), DOLORES (blue), Du Pont (cyan), UVES (green), X-shooter (magenta), ISIS (orange) and with ACAM (black). For each spectrum the phase with respect to the transient discovery date (2019 Apr 09, MJD 58582) is reported on the right. The dotted lines indicate the main emission lines and the grey band represents the area affected by telluric absorption. The \ha\ line is affected by telluric absorption, but this is not shown in the plot for clarity. The UVES spectrum has been smoothed for clarity. All spectra have been corrected for foreground extinction. For plotting purposes, all spectra have been divided by their median value. The X-shooter, ISIS and UVES spectra are not flux calibrated.}
        \label{fig:spectra_dsg}   
    \end{figure*}

Observations were taken using the Wide Field reimaging CCD Camera (WFCCD) on the du Pont 100-inch telescope at Las Campanas Observatory using a 1 arcsec slit. The data were calibrated using HeNeAr arcs and bias subtraction, flat fielding and standard star observations for flux calibration. Reduction was performed with a modified version of {\sc pydis} \citep{davenport16} to extract and sky-subtract the spectrum.
\\
\\
For a journal of all the spectroscopic observations, see Table~\ref{tab:spec_obs} and the complete sequence of spectra is shown in Fig.~\ref{fig:spectra_dsg}.

    \begin{figure*}
        \includegraphics[scale=0.55, angle = 0]{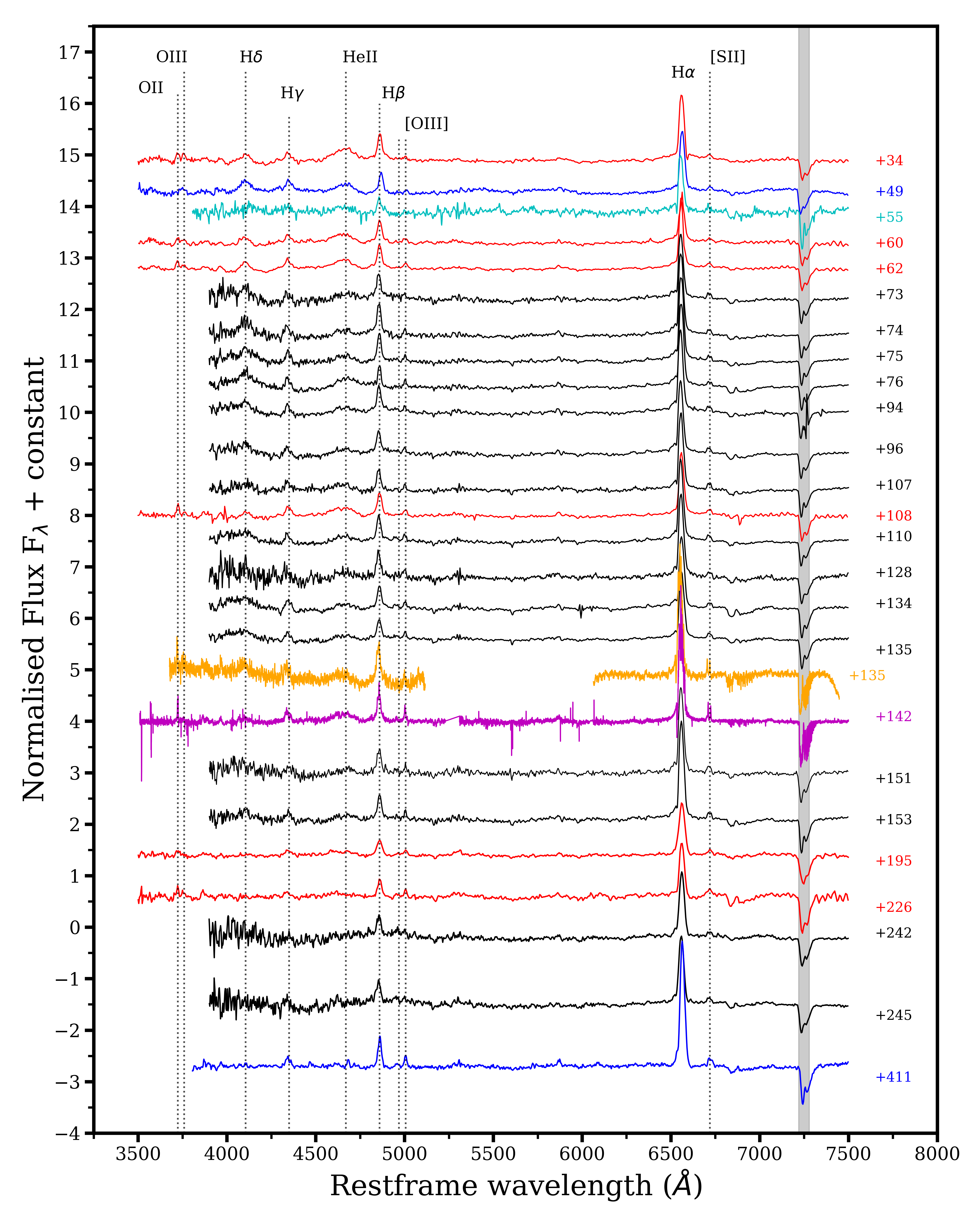} 
        \caption{Sequence of host-subtracted spectra taken with EFOSC2 (red), DOLORES (blue), Du Pont (cyan), X-shooter (magenta), ISIS (orange) and with ACAM (black). For each spectrum we annotate to the right the number of days passed since the transient discovery (2019 Apr 09, MJD 58\,582). The dotted lines indicate the main emission lines and the grey band represents the area affected by telluric absorption. The \ha\ line is affected by telluric absorption, but this is not shown in the plot for clarity. All spectra have been corrected for foreground extinction. For plotting purposes, all spectra have been normalised. The X-shooter and ISIS spectra are not flux calibrated.}
        \label{fig:spectra_dsg_sub}   
    \end{figure*}

\begin{table}
	\centering
	\small
 	\caption{Spectroscopic observations}

 	\label{tab:spec_obs}
 	\tabcolsep=0.11cm
 	\begin{center}
	\begin{tabular}{lccccl}
		\hline
		MJD$^{(1)}$ & phase$^{(2)}$ &UTC Date  & Instrument & exposure time   & slit   \\
        	{[days]} & {[days]}	 &       &   & [s]             &[$''$]     \\
        \hline
        58\,616.36 & +34  & 2019 May 13 & EFOSC2       & 900    & 1.0  \\
        58\,631.13 & +49  & 2019 May 27 & DOLORES      & 1800   & 1.5  \\
        58\,638.28 & +55  & 2019 Jun 03 & WFCCD        & 4x900  & 1.0 \\
        58\,642.35 & +60  & 2019 Jun 08 & EFOSC2       & 1800   & 1.0 \\
        58\,644.32 & +62  & 2019 Jun 10 & EFOSC2       & 2700   & 1.0  \\
		58\,655.19 & +73  & 2019 Jun 20 & ACAM         & 1800   & 0.75  \\ 
        58\,656.12 & +74  & 2019 Jun 21 & ACAM         & 1800   & 0.75 \\ 
        58\,657.09 & +75  & 2019 Jun 22 & ACAM         & 1800   & 0.75 \\ 
        58\,658.14 & +76  & 2019 Jun 23 & ACAM         & 1800   & 0.75  \\ 
        58\,668.22 & +86  & 2019 Jul 04 & UVES         & 1800  & 1.0 \\
        58\,676.19 & +94  & 2019 Jul 12 & ACAM         & 1800   & 1.0  \\
        58\,678.10 & +96  & 2019 Jul 13 & ACAM         & 2x1800 & 1.0  \\
        58\,689.25 & +107  & 2019 Jul 25 & EFOSC2       & 2700   & 1.0 \\
        58\,690.11 & +108  & 2019 Jul 25 & ACAM         & 1800   & 0.75 \\
        58\,692.12 & +110  & 2019 Jul 27 & ACAM         & 1800   & 1.0 \\
        58\,710.94 & +128  & 2019 Aug 15 & ACAM         & 1800   & 1.0 \\
        58\,715.88 & +134  & 2019 Aug 20 & ACAM         & 2x1800 & 1.0 \\
        58\,717.00 & +135  & 2019 Aug 21 & ACAM         & 5x1800 & 1.0 \\
        58\,717.89 & +135  & 2019 Aug 22 & ISIS         & 2x2700 & 1.0 \\
        58\,724.14 & +142  & 2019 Aug 29 & XSH/UVB      & 4x920  & 1.0 \\
        58\,724.14 & +142  & 2019 Aug 29 & XSH/VIS      & 4x920  & 0.9 \\
        58\,724.14 & +142  & 2019 Aug 29 & XSH/NIR      & 8x480  & 0.9 \\
        58\,733.93 & +151  & 2019 Sep 07 & ACAM         & 2x1800 & 1.0  \\
        58\,735.96 & +153  & 2019 Sep 09 & ACAM         & 2x1800 & 0.75 \\
        58\,777.02 & +195  & 2019 Oct 10 & EFOSC2       & 2x1800 & 1.5 \\
        58\,808.04 & +226  & 2019 Nov 21 & EFOSC2       & 2100   & 1.0 \\
        58\,824.83 & +242  & 2019 Dec 07 & ACAM         & 2200   & 1.5 \\
    	58\,827.82 & +245  & 2019 Dec 10 & ACAM         & 2700   & 1.5 \\
    	58\,994.12 & +411  & 2020 May 24 & DOLORES      & 2x2200 & 1.0 \\
		\hline
\end{tabular}
\end{center}
\textit{Note.}(1) Modified Julian Day of observations; (2) calculated with respect to the discovery date MJD 58\,582. 
\end{table}

\subsection{Optical photometric observations}
\subsubsection{Las Cumbres Sinistro}
We observed AT2019dsg with the Las Cumbres Observatory Global Telescope Network \citep{brown13}. Observations were performed with the Sinistro cameras, mounted at the focus of 1-meter telescopes, at the following sites: South African Astronomical Observatory (CPT), Siding Spring Observatory (COJ), Cerro Tololo Interamerican Observatory (LSC) and McDonald Observatory (ELP). The target was observed with the Johnson B and V filters and the Sloan filters \textit{g, r, i}. Data was reduced with the \textsc{banzai} pipeline \citep{mccully18} which performs standard data reduction routines such as bad-pixel masking, bias frames subtraction, dark subtraction, flat field correction and cosmic rays correction. The zero points of the images were calibrated using the American Association of Variable Star Observers (AAVSO) Photometric All-Sky Survey (APASS) DR10 \citep{henden19}. The magnitudes were estimated with aperture photometry, using the \textsc{iraf} task \textsc{apphot}, with aperture sizes depending on the seeing conditions.

\subsubsection{LT IO:O}
We observed \dsg\, using the Optical Wide Field camera (IO:O) at the Liverpool Telescope (LT) using the Sloan \textit{g, r, i, z} and Johnson B and V filters. The images were reduced with the LT pipeline and the zeropoints calculated using stars in the APASS catalogue. Magnitudes were estimated through aperture photometry (\textsc{iraf} task \textsc{apphot}), using apertures of variable size, depending on the seeing.

\subsection{Radio observations: e-MERLIN}

We observed our target source, \dsg, with the e-MERLIN\footnote{http://www.e-merlin.ac.uk/} radio
interferometer in the UK between 2019 June 3 and September 12 (project code
DD8006: PI P\'erez-Torres). We observed \dsg\ a total of 13 times: 10 at C-band 
(4.82--5.33~GHz) and 3 at L-band (1.25--1.77~GHz).
We summarize in Table~\ref{tab:emerlin-results} the start time of each observing run and its duration at each frequency band.
All observations had a total bandwidth of 512 MHz divided in eight spectral windows of 64 MHz with 512 channels per spectral window, except runs 07 through 10, which had four spectral windows of 128 MHz each. We used 3C286 and OQ208 as amplitude and bandpass calibrators, respectively. We correlated the phase calibrator, J2052+1619, 
at position $\alpha_{\rm J2000.0}=20^{\rm h} 52^{\rm m} 43\fs6199$ and $\delta_{\rm J2000.0}=16\degr 19\arcmin 48\farcs828$, which is separated  2.37 deg from our target, and we detected it clearly in all runs with a flux density of 0.32~Jy and 0.42~Jy at C and L-band, respectively.

We carried out all reduction steps using the e-MERLIN \textsc{casa}
pipeline\footnote{https://github.com/e-merlin/eMERLINCASApipeline} version v1.1.16
running on \textsc{casa} \citep{mcmullin07} version 5.6.2. We followed the default procedures of the pipeline using the default parameters, but adding manual flag commands to remove bad data that the pipeline could not remove. We used a common model for the phase reference calibrator to calibrate and image each run. We then used the \textsc{casa} task \textsc{tclean} to image the target source, using the e-MERLIN \textsc{casa} pipeline imaging procedure with Briggs weighting and a robust parameter of 0.5 for all runs except numbers 03 and 07, which required robust 0.0 and -1.0, respectively, due to the lack of the longest baselines. The cell size was 8~mas for the C-band data and and 20~mas for the L-band data.
We show in Table~\ref{tab:emerlin-results} the synthesized beam size, local rms of the residual image, flux density and luminosity of the target source measured as the peak of emission of a Gaussian fit to the deconvolved image.

\begin{table*}
	\centering
	\caption{Schedule of the e-MERLIN observations, with the run number, beam and total
    flux density at each frequency band. Values of the non-detections are 3-$\sigma$ level
    uncertainties with respect to the rms of the image. The last line is the 4.1$\sigma$ detection from the combined L-band data.}
	\label{tab:emerlin-results}
\begin{tabular}{lcccccccccr}
\hline
Run & MJD & Start date & Duration & Freq.  & Beam$_{\rm maj}$ & Beam$_{\rm minor}$ & Beam$_{\rm pa}$ &  rms & Flux density & Luminosity \\
    & {[days]} & & [h] & [GHz] & [mas]         & [mas]           & [deg]           &  [$\mu$Jy~b$^{-1}$] & [$\mu$Jy] & [$10^{37}$\unitlum]\\
\hline
01 & 58\,637.92 & 2019-06-03 22:00 &  12  & 5.1  & 67  & 32  & 28 & 32  & 160  $\pm$ 30  & 4.9  $\pm$ 0.9 \\
02 & 58\,654.88 & 2019-06-20 21:02 &  12  & 5.1  & 71  & 32  & 28 & 21  & 340  $\pm$ 20  & 10.4  $\pm$ 0.6 \\
03 & 58\,668.94 & 2019-07-04 22:32 &  10  & 5.1  & 764 & 580 & 76 & 86  & 410  $\pm$ 90  & 12.5 $\pm$ 2.7 \\
04 & 58\,669.94 & 2019-07-05 22:32 &  10  & 5.1  & 96  & 29  & 28 & 24  & 460  $\pm$ 20  & 14.1 $\pm$ 0.6 \\
05 & 58\,670.94 & 2019-07-06 22:32 &  10  & 5.1  & 98  & 30  & 27 & 23  & 490  $\pm$ 20  & 15.0 $\pm$ 0.6 \\
06 & 58\,671.94 & 2019-07-07 22:32 &  10  & 5.1  & 99  & 29  & 27 & 23  & 480  $\pm$ 20  & 14.7 $\pm$ 0.6 \\
07 & 58\,687.88 & 2019-07-23 21:17 &  10  & 5.1  & 177 & 54  & 14 & 128 & 680  $\pm$ 130 & 20.8 $\pm$ 4.0 \\
08 & 58\,691.89 & 2019-07-27 21:22 &  9   & 5.1  & 92  & 30  & 20 & 42  & 650  $\pm$ 40  & 19.9 $\pm$ 1.2 \\
09 & 58\,701.75 & 2019-08-06 18:00 &  11  & 5.1  & 69  & 33  & 25 & 27  & 770  $\pm$ 30  & 23.6 $\pm$ 0.9 \\
10 & 58\,738.68 & 2019-09-12 16:13 &  11  & 5.1  & 78  & 29  & 28 & 31  & 1210 $\pm$ 30  & 37.0 $\pm$ 0.9 \\
\hline
11 & 58\,688.88 & 2019-07-24 21:17 &  10  & 1.5  & 293 & 119 & 22 & 40  &      < 120     &      < 1.1     \\
12 & 58\,689.91 & 2019-07-25 21:50 &  9   & 1.5  & 322 & 115 & 21 & 44  &      < 130     &      < 1.2     \\
13 & 58\,690.89 & 2019-07-26 21:20 &  9   & 1.5  & 285 & 108 & 23 & 38  &      < 115     &      < 1.0     \\
\hline
   &            &                  &      & 1.5  &     &     &    &     & 145 $\pm$ 35   & 1.3$\pm$0.3   \\
\hline
\end{tabular}
\end{table*}

    \begin{figure}
    \includegraphics[width=1\columnwidth]{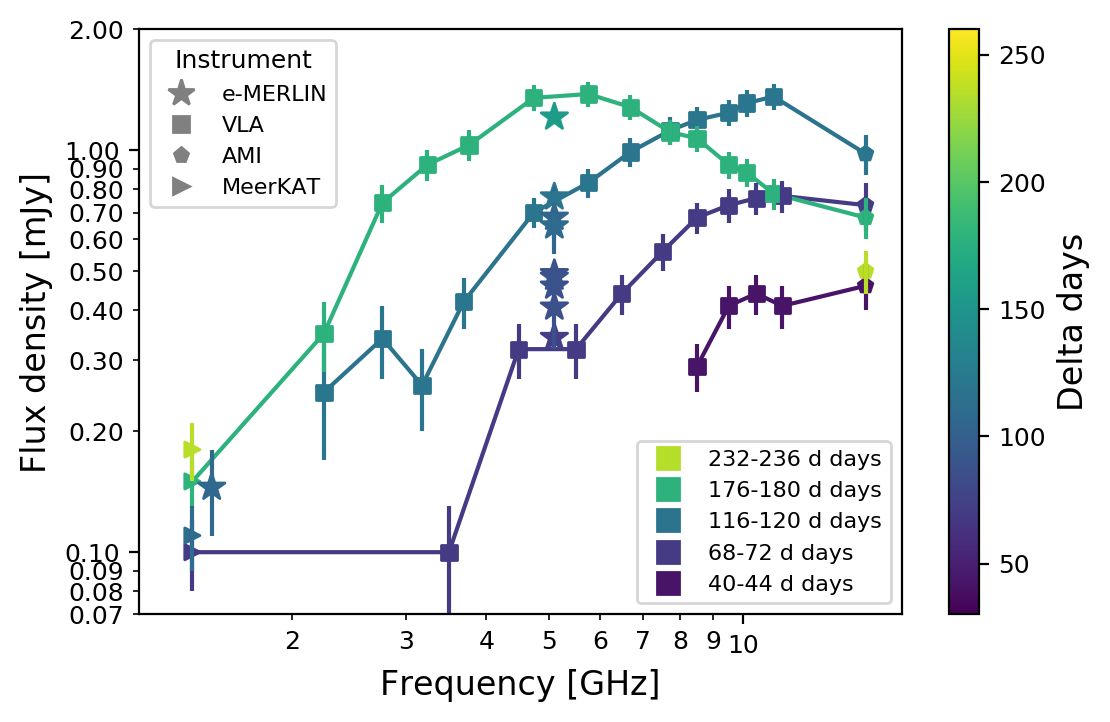}\\
     \includegraphics[width=1\columnwidth]{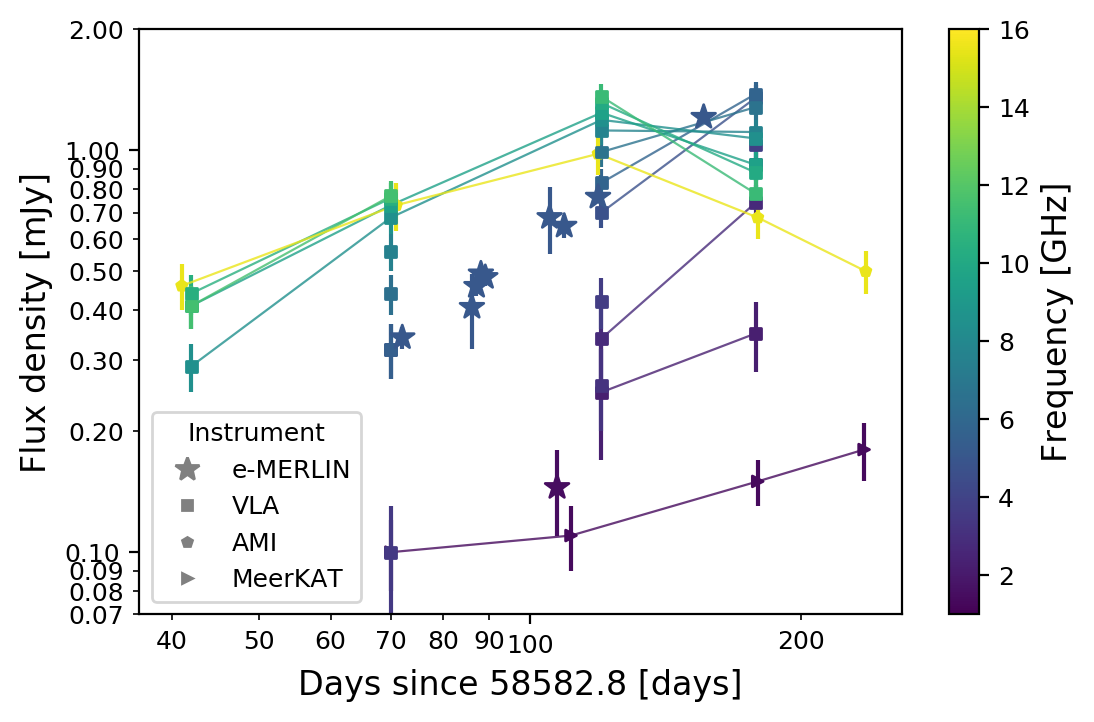}
        \caption{Radio evolution of \dsg. VLA, AMI, and MeerKAT data are from \citet{stein20}, while e-MERLIN data (stars) are our own data. The e-MERLIN data are not connected by solid lines for readability. Note the characteristic synchrotron shift of the peak frequency with time, so by day  $\sim$180 the radio emission at frequencies above $\sim$6.0 GHz is already in its optically thin, decaying phase, while at smaller frequencies is still in its optically thick, increasing phase.
        }
        \label{fig:radio}   
    \end{figure}

\section{Analysis and results}
\subsection{e-MERLIN observations}
The source is an unresolved point-like source in all detected epochs, with no significant hints of extended emission.  We computed the statistical astrometric uncertainty as the standard deviation of the positions of the centroid fitted to the images, resulting in a relative astrometric accuracy of about 6~mas (6.5 pc). We measured no significant astrometric displacement from the C-band detections.  The average position of \dsg\ is $\alpha_{\rm J2000.0}=20^{\rm h} 57^{\rm m} 02\fs9647$ and $\delta_{\rm J2000.0}=14\degr 12\arcmin 16\farcs305$, as measured with respect to the phase reference correlation position quoted above.
In L-band, we obtained 3$\upsigma$ upper limits on the flux density. We therefore combined the data from the three consecutive days of observations in the L-band and found a point-like source that is spatially coincident with the source in the C-band images. We obtain a 4 $\upsigma$ detection of the source in the L-band with a flux density of $\rm 145 \pm 35\,\mu Jy$, corresponding to a luminosity of $\rm 1.3 \pm 0.3 \times 10^{37}$ \unitlum. 

We show in Fig.~\ref{fig:radio} the radio lightcurve of \dsg\ for the first $\sim240$ days at multiple frequencies, including our e-MERLIN data.  
The e-MERLIN interferometer provides a better angular resolution than any of the other instruments. Therefore, the observed flux is not contaminated by background emission, so our observations provide bona fide flux densities against which to test the baseline of radio emission assumed in \citet{stein20}. Overall, it appears that the agreement is good.

\subsection{Host velocity dispersion measurement}

We use the X-shooter spectrum to constrain the width of the host galaxy absorption lines.
In particular, we use the penalized pixel fitting routine \ppxf \citep{cappellari17} to perform full spectrum template fitting with the Elodie stellar template library \citep{prugniel01}. We resample the X-shooter spectrum within the errors and measure velocity dispersions for 1000 realisations of the data (see \citealt{Wevers2017} for a detailed explanation of the method). Fitting the resulting velocity dispersion distribution with a Gaussian model, we find a mean FWHM of 94 km s$^{-1}$ with a standard deviation of 1 km s$^{-1}$, which we adopt as the velocity dispersion and its measurement uncertainty. Using the M--$\sigma$ relation of \citet{Gultekin09} and adding the measurement errors linearly with the scatter in the relation, this corresponds to a black hole mass of log(M$_{BH}$) = 6.73 $\pm$ 0.40 M$_{\odot}$. We note that by using both the M-sigma relations of \citet{Ferrarese2005} and of \citet{mcconnel13}, we obtain a value for the black hole mass which is compatible with the previous estimate within uncertainties. This corresponds to an Eddington luminosity $\rm L_{Edd}=(6.8\pm4.1)\times10^{44}$ \unitlum and a Schwarzschild radius of $R_S = (1.6\pm0.9)\times10^{13}$ cm, for a non-rotating BH.

\subsection{UV/optical lightcurve}
The host galaxy is marginally detected in the GALEX NUV band ($\lambda_{cen}$ = 2328 \AA; NUV = 21.1$\pm$0.4), indicating that even in the latest epochs (190 days after discovery) when the source was detected at a magnitude of UVM2 ($\lambda_{cen}$ = 2260 \AA) = 18.95, host contamination is not an issue in the UV bands. To check the host galaxy contamination in the optical, we use the Kron magnitudes from Pan-STARSS (PS) Data Release 2, using the filter transformations from \citet{jordi06}, when necessary.

\begin{figure*}\includegraphics[width=2\columnwidth]{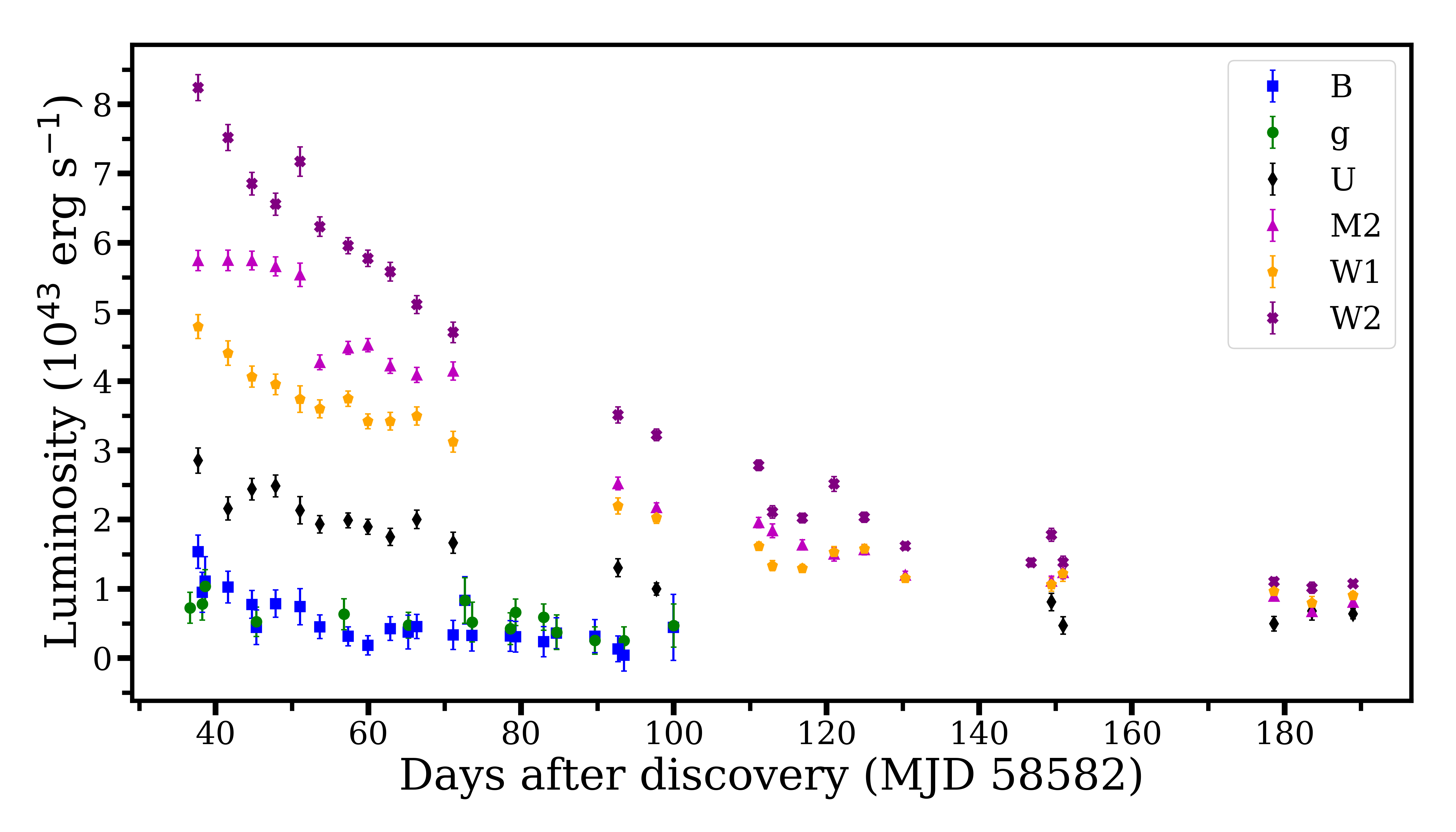} 
        \caption{Host-subtracted light curve of \dsg\ : B (blue squares), {\it g} (green circles), U (black diamonds), UVM2 (magenta triangles), UVW1 (orange pentagons) and UVW2 (purple crosses).}
        \label{fig:lc_19dsg}   
    \end{figure*}

In the optical, the transient is detected above the host galaxy light in the U band, and in B and {\it g} only for the first $\sim$80 days of observation. We then remove the host galaxy contribution by subtracting the flux derived from the Kron magnitudes from PS. We plot the host-subtracted UV/optical light curve in Fig.~\ref{fig:lc_19dsg} (only magnitudes above the galaxy light are plotted) and the observed magnitudes are reported in Table \ref{tab:mag}.

\subsubsection{UV/optical blackbody fitting}
Since in most of our additional photometric observations the transient light is not detected above the host galaxy level, we do not create an SED for the source, but we use the results from \citet{vanvelzen20} and \citet{stein20}. They find that the UV/optical lightcurve is well fit by a blackbody with roughly constant temperature $\rm T=(3.9\pm0.2)\times10^{4}$ K and radius $\rm R=(3.9\pm0.3)\times10^{14}$ cm. The maximum bolometric luminosity is $\rm L_{bb}=(2.9\pm0.3)\times10^{44}$ \unitlum $\rm \sim0.4L_{Edd}$.

\subsection{X-ray spectral analysis}
\subsubsection{Swift/XRT}
To fit the Swift X-ray spectrum we use \textsc{xspec} 12.10.0 \citep{arnaud} in {\sc HEASOFT} v6.24, and assume a Galactic column density of n$_H$ = 6.45 $\times$ 10$^{20}$ cm$^{-2}$ \citep{hi4pi}.
Due to the low number of counts in each bin, we use Cash statistics (the results are identical when rebinning the spectrum to 20 counts/bin and using Gaussian statistics). Confidence intervals are quoted at the 90 per cent level ($\Delta$C-stat=2.71). 
A 65$\pm$6 eV blackbody model (TBabs $\times$ zashift $\times$ bbodyrad) provides a good fit to the data (cstat = 137 for 682 dof or $\chi^2$=0.81 with 9 dof), with a  radius $\rm R=3.4^{+1.5}_{-0.9}\times10^{11}$ cm. Using instead a multi-temperature blackbody model (diskbb) yields kT = 80$\pm$7 eV (cstat = 135 for 682 dof). These results are in agreement with those presented in \citet{stein20}. To convert the count rates to unabsorbed 0.3--10 keV fluxes, we assume the simple blackbody model with Galactic extinction, which gives a conversion factor of 9.1$\times$10$^{-11}$ cts/erg/cm$^2$/s.

\subsubsection{NICER}
Background-subtracted NICER spectra are fit to the same absorbed, redshifted blackbody model used for the Swift/XRT spectrum. The source is not detected over the NICER background above 1 keV, and fits are conducted in the 0.3--1 keV bandpass, with luminosities and count rates referring to this energy range. The latest effective area ({\it nixtiaveonaxis20170601v004.arf}) and response matrix ({\it nixtiref20170601v002.rmf}) files are utilized. The blackbody temperature for the late-time spectrum is fixed at 30 eV. The results are shown in Table \ref{tab:nicertab}, and are consistent with the analysis of the Swift/XRT spectrum. 
    \begin{figure}\includegraphics[width=1\columnwidth]{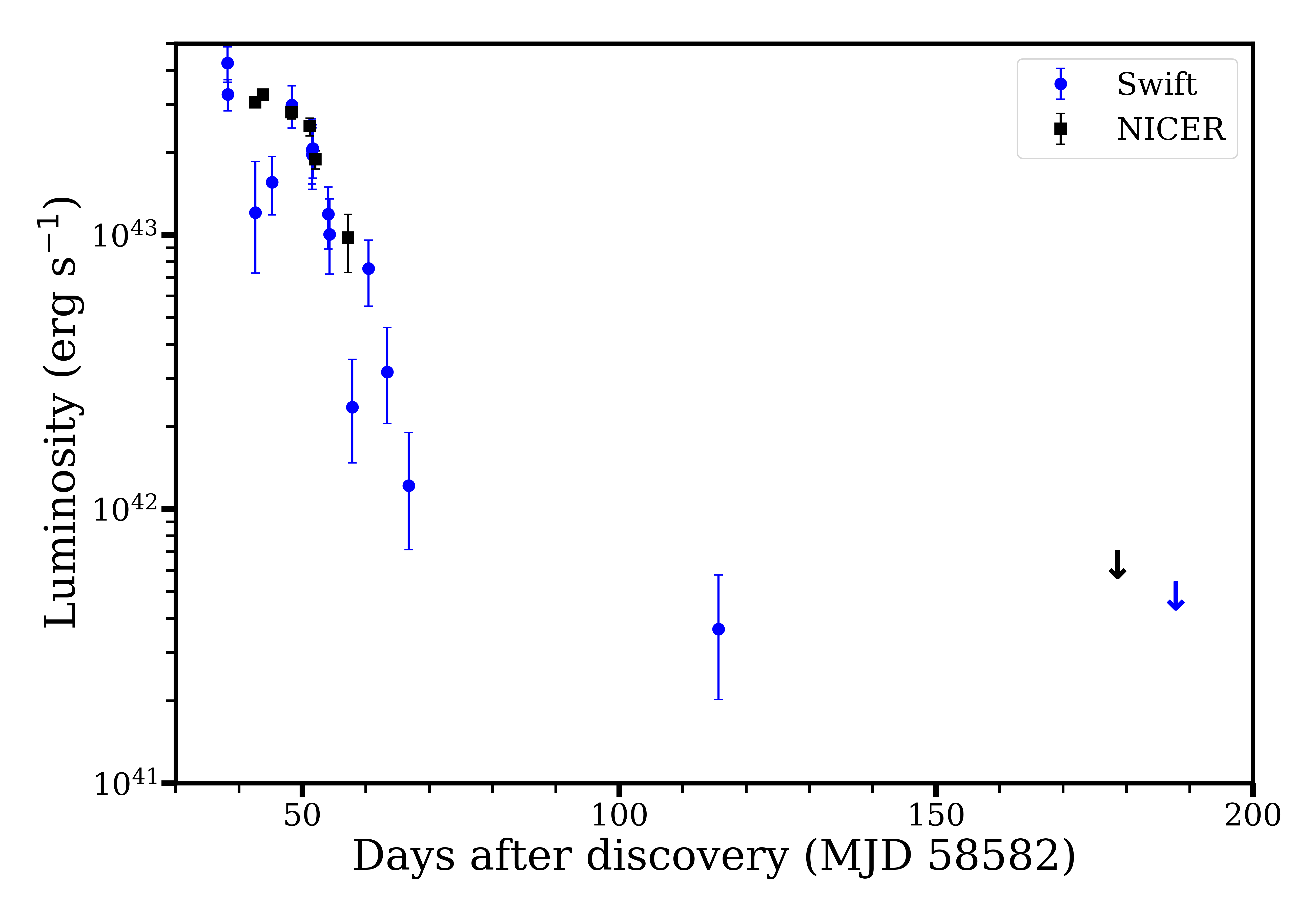} 
        \caption{The 0.3-10 keV X-ray luminosity evolution with time of \dsg. In blue (circles for detections), the Swift data, in black (squares for detections), the NICER data. The downward arrows represent 3$\upsigma$ upper limits.}
        \label{fig:xray_lc}
    \end{figure}

\begin{figure}
    \includegraphics[width=\linewidth]{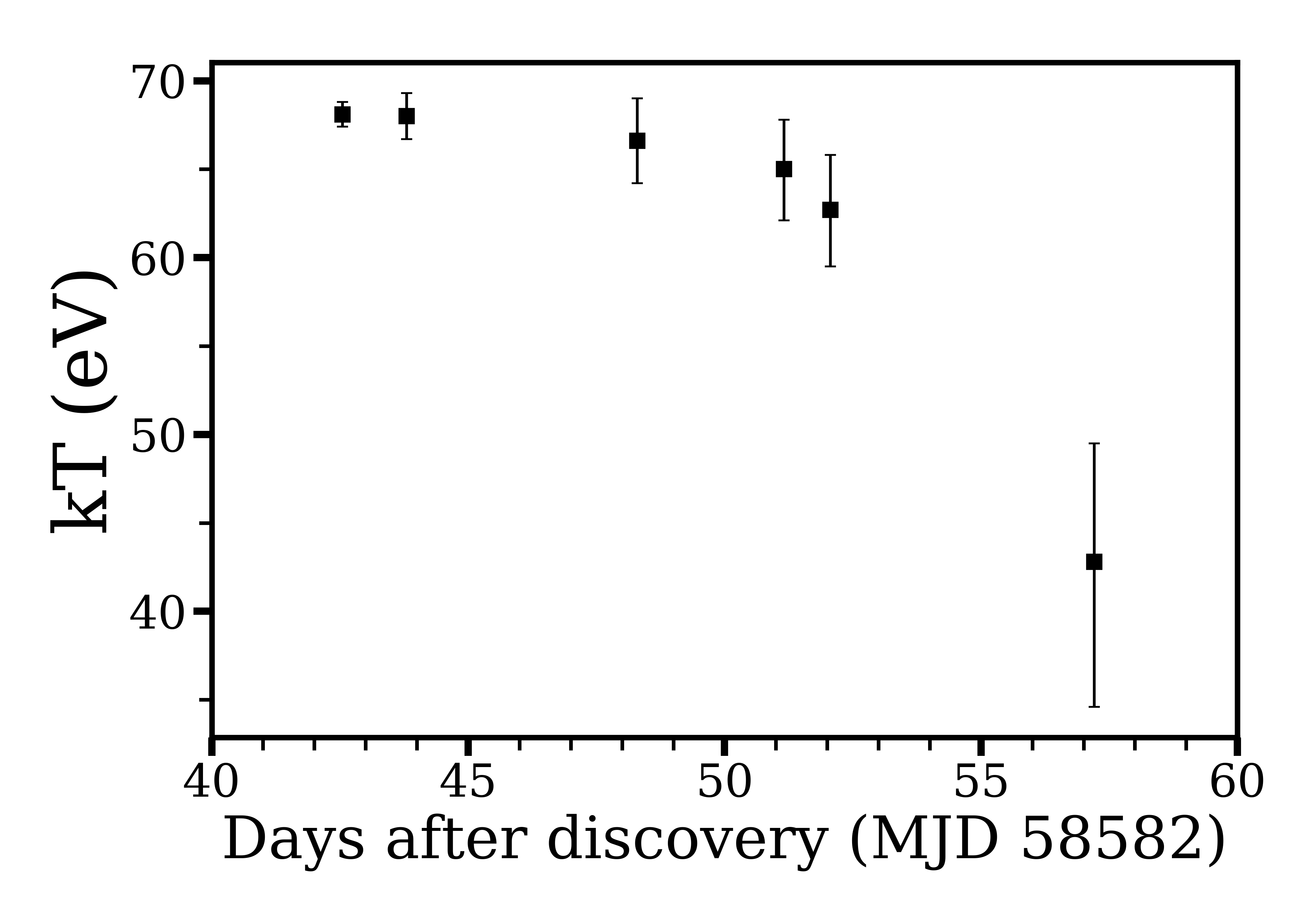}
        \caption{Evolution of the BB temperature from the fit to the NICER (0.3--1 keV) data with time.}
        \label{fig:xray_T}   
    \end{figure}

\begin{figure}
    \includegraphics[width=\linewidth]{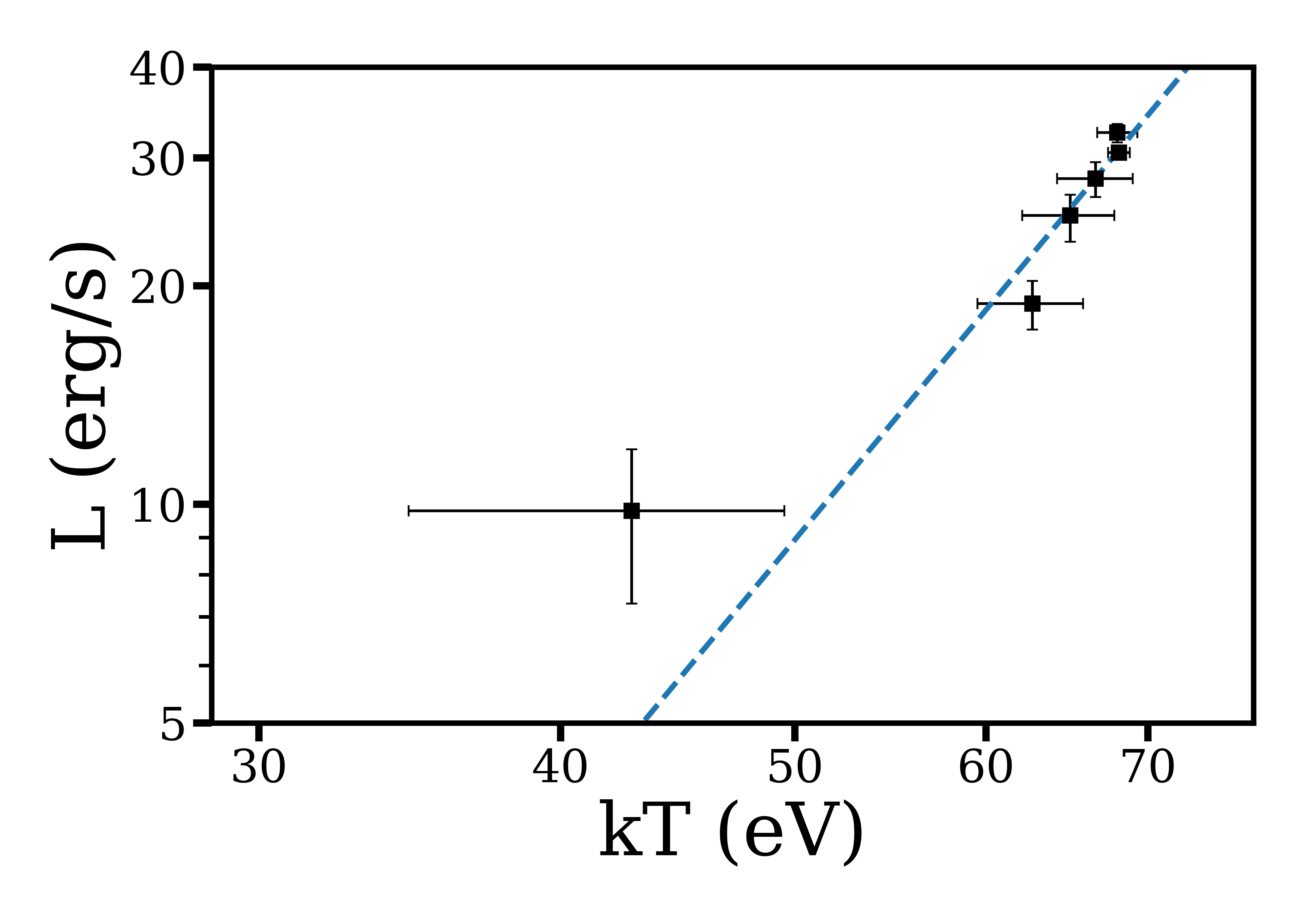}
        \caption{X-ray luminosity (0.3--1 keV) versus the BB temperature. The dashed line represents $\rm L\propto kT^4$ with arbitrary normalisation. The observed rate of cooling decreases over time.}
        \label{fig:xray_LvsT}   
    \end{figure}

The 0.3--10 keV X-ray lightcurve is shown in Fig.~\ref{fig:xray_lc}, where we have used the webPIMMS\footnote{https://heasarc.gsfc.nasa.gov/cgi-bin/Tools/w3pimms/w3pimms.pl} tool to convert the NICER 0.3--1 keV luminosities to the 0.3--10 keV energy range for consistency with the XRT data.
The X-ray luminosity shows a rapid decay from $\rm 3\times10^{43}$ \unitlum to $\rm 10^{42}$ \unitlum over 25 days. Then, the decay becomes more shallow and the X-ray luminosity reaches $\rm 3.6\times10^{41}$ \unitlum in the last epoch at which the source is detected (125 days after the discovery of the transient). The XRT light curve shows variability of factor 2-3 on a timescale of days during the initial rapid decrease, while this behaviour is not seen in the NICER data.
In Fig.~\ref{fig:xray_T}, we plot the evolution of the BB temperature with time. The temperature decreases from 70 eV to 40 eV over $\sim$15 days. In Fig.~\ref{fig:xray_LvsT} we plot the 0.3--1 keV X-ray luminosity versus the BB temperature from the fit to the NICER data. We overplot the data with a $L_X\propto T^4$ curve.

\begin{table*}
    \centering
    \begin{tabular}{lccccccc}
    \hline
ObsIDs & 0.3--1 keV count rate & Average Time & Exposure & kT & c-statistic & Degrees of freedom &  Luminosity \\
 & (cts s$^{-1}$) & (MJD) & (seconds) & (eV) & & & (10$^{42}$ erg s$^{-1}$) \\\hline
200680102 & 2.02 & 58\,625.8 & 2159 & 68$\pm$1 & 82.19 & 67 & 32.5$^{+1.0}_{-0.9}$ \\ 
200680103-4 & 1.71 & 58\,630.3 & 767 & 67$\pm$2 & 46.54 & 67 & 28.1$^{+1.6}_{-1.5}$ \\
200680105 & 1.48 & 58\,633.16 & 593 & 65$\pm$3 & 75.54 & 67 & 25.0$^{+2.0}_{-1.7}$ \\
200680106 & 1.09 & 58\,634.06 & 627 & 63$\pm$3 & 58.86 & 67 & 18.9$^{+1.5}_{-1.4}$ \\
200680107-8 & 0.4 & 58\,639.21 & 239 & 43$^{+8}_{-7}$ & 55.26 & 67 & 9.8$^{+2.5}_{-2.11}$ \\
200680110-12 & 0.02 & 58\,759.99 & 15493 & 30 & 108.7 & 68 & $<$0.63 \\
    \hline
    \end{tabular}
    \caption{Best-fit model parameters derived from time-resolved {\it NICER} X-ray (0.3--1 keV) spectra. The temperature in the last epoch is kept fixed at 30 eV and the luminosity value is a 3$\upsigma$ upper limit.}
    \label{tab:nicertab}
\end{table*}

\subsection{Optical spectroscopy}
\label{sec:spec_analysis}

The sequence of spectra from EFOSC2, DOLORES, UVES, X-shooter, ISIS and ACAM is shown in Fig.~\ref{fig:spectra_dsg}. The first spectra show a blue continuum that decays over time. The decay of the blue part of the spectrum pivots around 5000 \AA. The continuum is dominated by the host galaxy redwards of this wavelength at all stages of the TDE outburst. After roughly 60 days, the blue continuum light has decayed. Coincidentally, this is similar to the timescale over which the X-ray flux becomes undetectable. The spectra show strong Balmer emission lines (H$\upalpha$ through H$\updelta$) and a broad \heii\ emission line that becomes less visible with respect to the surrounding continuum over time. We also identify the \ion{O}{ii} doublet at 3726 and 3729 \AA\ and \ion{O}{iii} at 3760 \AA\ in emission.\\

\subsubsection{Subtraction of the stellar component}
\label{sec:host_sub}
The spectra clearly show absorption features due to the host galaxy, more prominently in the medium-high resolution spectra.
To remove the stellar contribution, we employ \ppxf: we build a synthetic host galaxy spectrum by fitting stellar spectra to our X-shooter spectrum. The method convolves a series of stellar template spectra to the observed spectrum (host galaxy+TDE), by combining the individual stellar templates with additive and/or multiplicative orthogonal polynomials and an initial guess of the line of sight velocity dispersion (LOSVD). The best fitting template (or combination of templates) is then found by \chisq\ minimization. The emission features and areas affected by telluric absorption are masked during the template convolution procedure. This method therefore removes from the observed spectra the stellar component from the host galaxy. We employ the \textsc{phoenix} high-resolution synthetic spectral library\footnote{http://phoenix.astro.physik.uni-goettingen.de/} \citep{husser13}; this library covers the whole wavelength range of the UVB and VIS arms of X-shooter (3000 -- 10000 \AA) with a resolution of R$\sim$50000. Of the whole library, which contains $\sim$30000 synthetic spectra covering a broad range of stellar properties, we select a subsample with effective temperature 2300 K $\leq$T$_{eff}\leq$ 12000 K, metallicity in range $-$2.0$\leq$[Fe/H]$\leq$1.0 and alpha elements abundance [$\alpha$/Fe]=0. This library does not contain emission line templates. We use the X-shooter spectrum as a basis for building our synthetic host galaxy spectrum as it is the highest resolution spectrum on which we can apply the \ppxf\ routine: the ISIS spectrum does not have enough continuum free from emission features and the continuum of the UVES spectrum has too low a SNR to make the routine converge. We then use again the \ppxf\ code to scale the synthetic host spectrum to the other spectra of our follow-up. Since this is not possible for spectra with a higher resolution, we did not perform the subtraction on the UVES spectrum. After the subtraction, the continuum of the spectra is scaled to 1, using the median of the flux.

\begin{figure}\includegraphics[width=1\columnwidth]{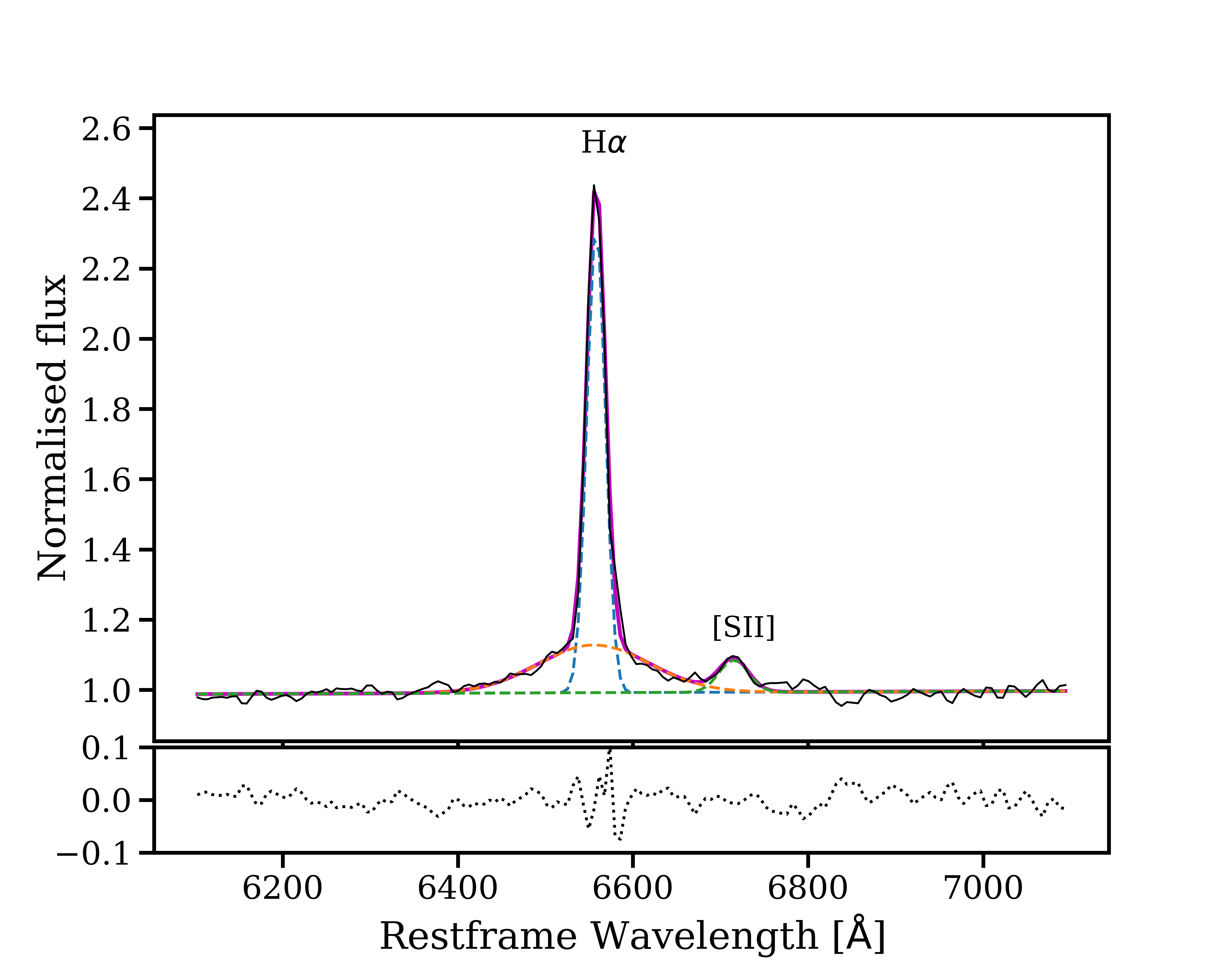} 
        \caption{Example of the fit to the \ha\ and [\ion{S}{ii}] emission lines in the EFOSC2 spectrum taken on 2019 June 10 (62 days after discovery), after performing the host galaxy subtraction. The dashed lines represent different Gaussian components, while the solid line represents the total fitting function. In the bottom panel, the residuals of the fit.}
        \label{fig:spec_fit_ha_sub}   
    \end{figure}

\begin{figure}\includegraphics[width=1\columnwidth]{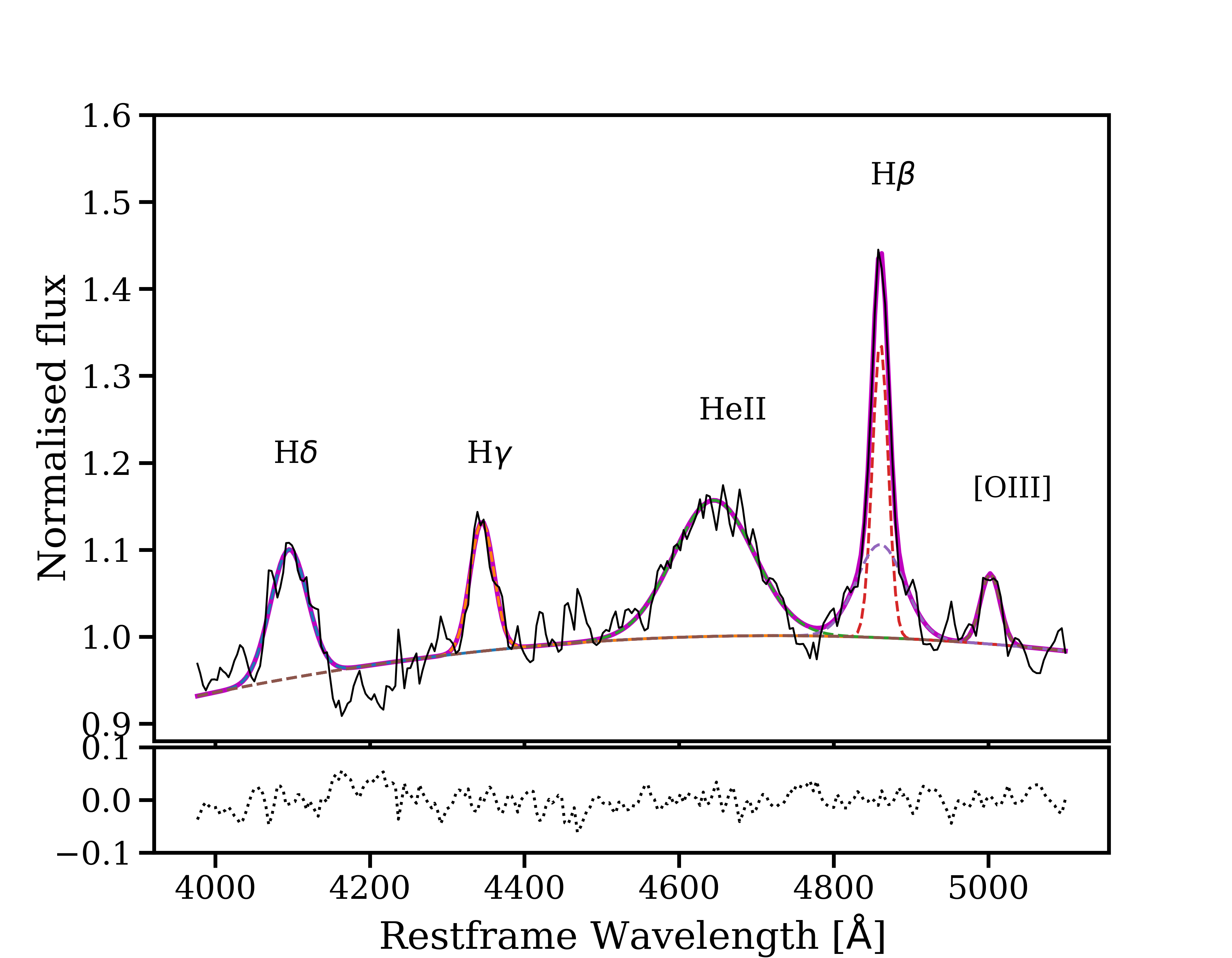} 
        \caption{Example of the fit to the emission lines in the \hb\ region of the EFOSC2 spectrum taken on 2019 June 08 (60 days after discovery), after the subtraction of the stellar component process. The dashed lines represent different Gaussian components, while the solid line represents the total fitting function. In the bottom panel, the residuals of the fit.}
        \label{fig:spec_fit_hb_sub}   
    \end{figure}

The host subtracted spectra (shown in Fig.~\ref{fig:spectra_dsg_sub}) still show strong \ha\ and \hb\ emission lines, both with a broad (FWHM$\sim$5000-10000 \kms) component at the base, and a narrower (FWHM$\sim$1000 \kms) peak. 
Due to the redshift of the host galaxy, the \ha\ line is contaminated by the telluric absorption band at $\sim$6900 \AA. This is most clearly visible in the ISIS, UVES and X-shooter spectra, due to the higher resolution of these spectra. Redwards of \ha, the two [\ion{S}{ii}] lines (6716 and 6731 \AA) are visible and resolved in some spectra. 
The \hg, \hd\ and the broad \heii\ lines are still detected and the \oiii\ doublet (4959, 5007 \AA) becomes clearly visible. Also the \ion{O}{ii} doublet (3726, 3729 \AA) and a weak \ion{O}{iii} at 3760 \AA\ are still detected. 
We fit the emission lines of the host subtracted spectra with multiple Gaussian components, combined with a polynomial to fit the local continuum, using \textsc{python} code employing the \textsc{lmfit}\footnote{https://lmfit.github.io/lmfit-py/} package \citep{lmfit}. During the fitting procedure, we tied the FWHM and the separation of the [\ion{S}{ii}] lines and of the \oiii\ lines, when the doublets were resolved. The central wavelengths of the broad base and the narrow peak of \hb\ are also tied together, to reduce the number of free parameters in this region. We were not able to fit the \nii\ doublet at 6548 and 6583 \AA, probably due to the presence of the telluric absorption.
Example \ha\ and \hb\ line fits are shown in Fig.~\ref{fig:spec_fit_ha_sub} and \ref{fig:spec_fit_hb_sub}, respectively.

\begin{figure}\includegraphics[width=1\columnwidth]{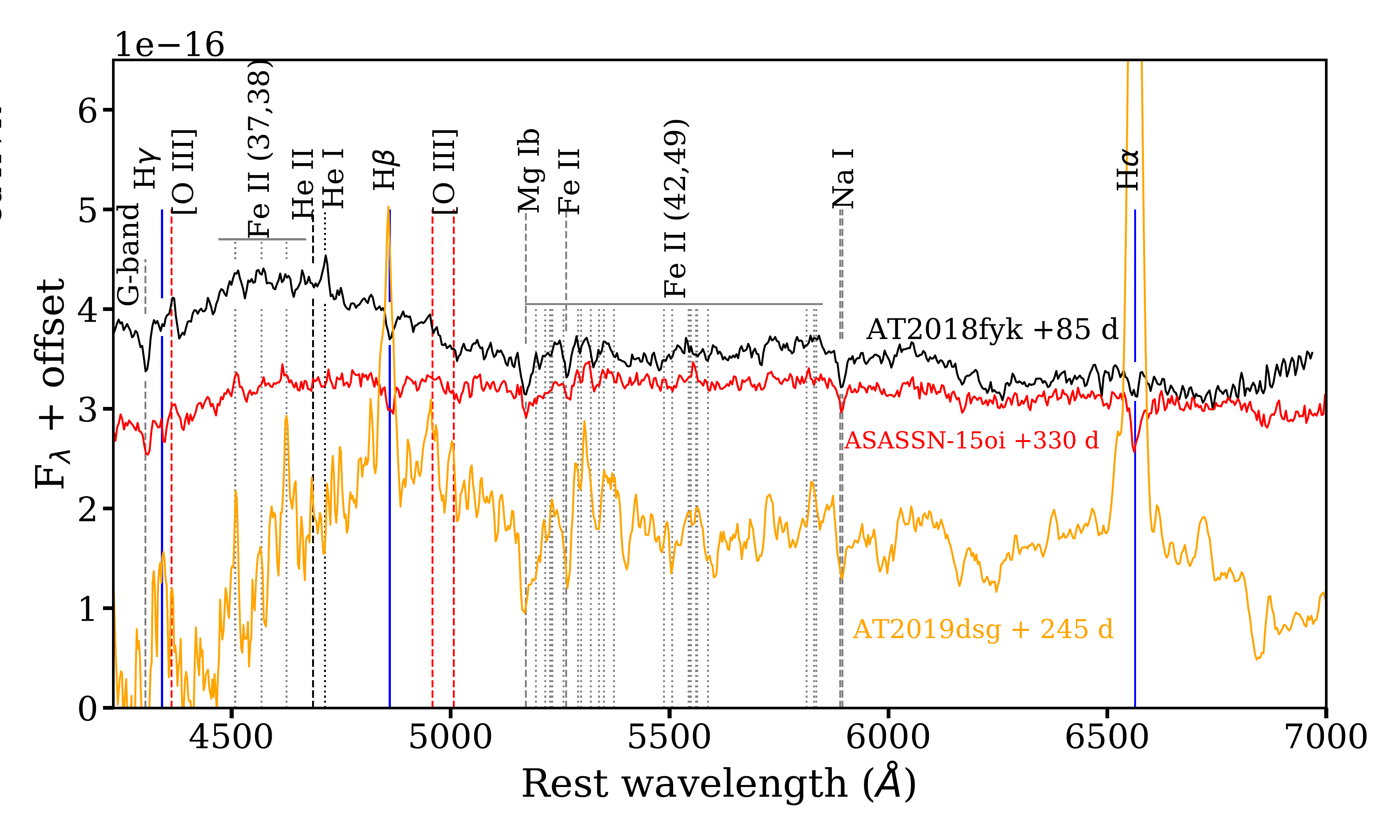}
        \caption{Comparison between the spectra of \dsg\ (orange) at +245 days and AT2018fyk (black) at +85 days and ASASSN 15oi (red) at +330 days, the other two TDES that have shown Fe lines in their optical spectra.}
        \label{fig:metal_lines}   
    \end{figure}

After the subtraction procedure, from the spectrum of 2019 September 09 (153 days after discovery) onward, the broad bases of \ha\ and \hb\ are not detected anymore. The \heii\ line is not detected after the spectrum of 2019 November 21 (226 days after discovery). We therefore associate these lines with the transient event, while we consider the remaining, more narrow, emission features as due to star formation in the host galaxy instead (see Sec. \ref{sec:activecomp}).
At late times (more than 200 days after discovery), in the continuum between \hb\ and \ha, metal lines appear. In Fig.~\ref{fig:metal_lines}, we show a comparison between \dsg\ and two other TDEs that have shown metal lines in the same wavelength range during their evolution: ASASSN-15oi and AT2018fyk \citep{wevers2019b}. In the spectrum of \dsg\ we can clearly identify absorption lines from \ion{Na}{i} and \ion{Mg}{i}b and many emission lines from \ion{Fe}{ii}, as well as the \ion{Fe}{ii} absorption feature at 5264 \AA.

We took a late time spectrum with TNG on 2020 May 24 (411 days after discovery, last spectrum of Fig.~\ref{fig:spectra_dsg} and \ref{fig:spectra_dsg_sub}), after the object came back from behind the Sun. In this spectrum we clearly see the narrow \ha\ and \hb, \oiii\ and [\ion{S}{ii}] lines, a still strong \hg\ and weak \hd\ emission lines. The aforementioned metal lines are not present in this spectrum.

\subsubsection{X-shooter, ISIS and absorption lines}

\begin{figure}\includegraphics[width=1\columnwidth]{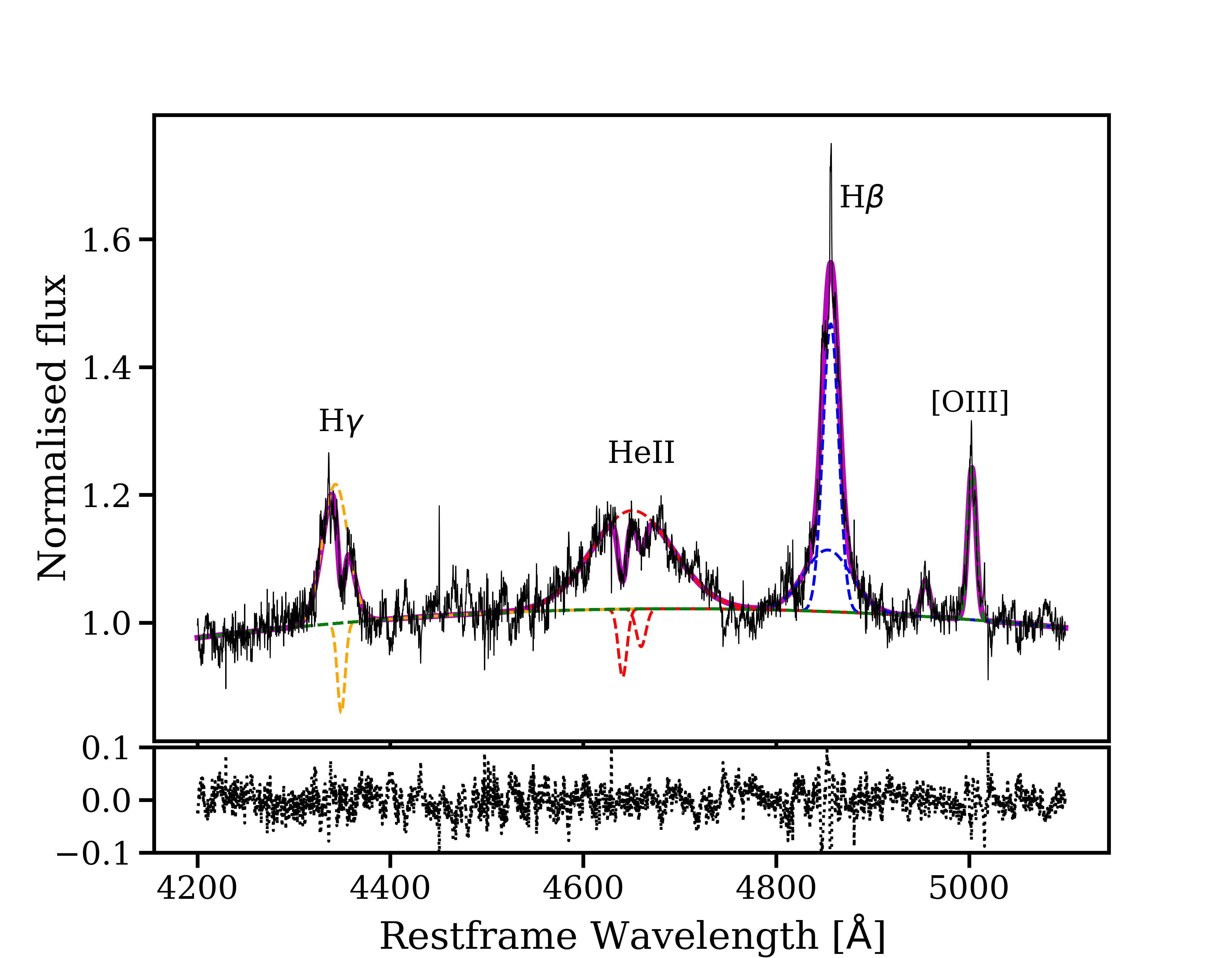} 
        \caption{Fit to the emission lines in the \hb\ region of the host galaxy subtracted X-shooter spectrum obtained 142 days after discovery of \dsg. The dashed lines represent different Gaussian components, while the solid line represents the total fitting function. In the bottom panel, the residuals of the fit.}
        \label{fig:spec_fit_xsh}   
    \end{figure}

The \hb\ region of the X-shooter spectrum is plotted in Fig.~\ref{fig:spec_fit_xsh}. The spectrum shows the \oiii\ doublet (4959 and 5007 \AA), the \hb\ line with the usual two components, the broad \heii, and the \hg\ line. The \hg\ emission line shows an absorption line superimposed and the \heii\ shows two absorption features. 

\begin{figure}\includegraphics[width=1\columnwidth]{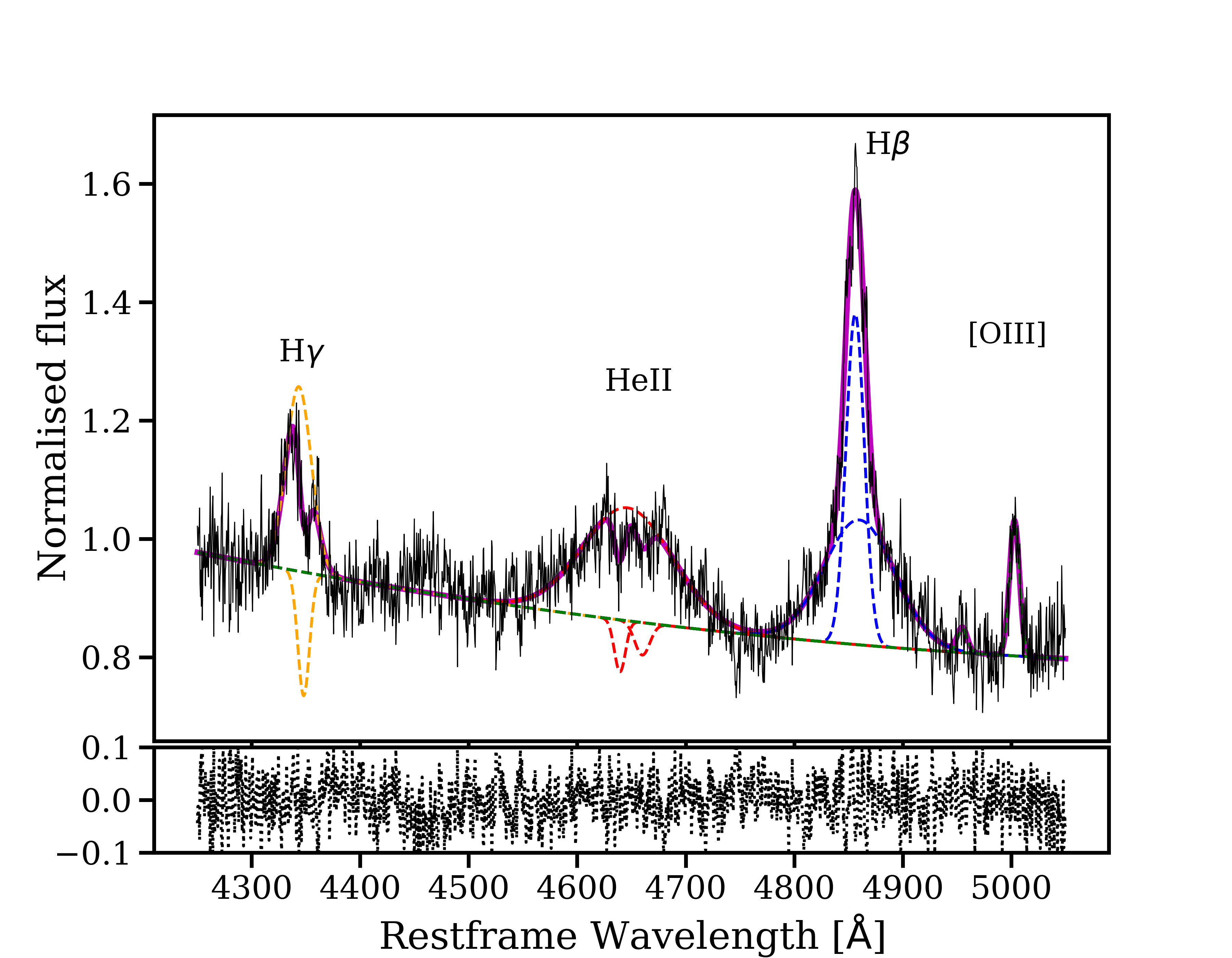}
        \caption{Fit to the emission lines in the \hb\ region of the host galaxy subtracted ISIS spectrum obtained 135 days after discovery of \dsg. The dashed lines represent different Gaussian components, while the solid line represents the total fitting function. In the bottom panel, the residuals of the fit.}
        \label{fig:spec_isis}   
    \end{figure}

The host subtracted ISIS spectrum shows a similar morphology, albeit the absorption features in the \heii\ line are less pronounced. The fit to the \hg\ and \heii\ emission lines of the ISIS spectrum is shown in Fig.~\ref{fig:spec_isis}. 
The absorption lines on top of the \hg\ and \heii\ emission lines are present in both the original spectrum, before the subtraction of the host galaxy light, and in the host-subtracted one. The FWHM, equivalent width (EW) and central wavelength of these absorption lines are reported in Table \ref{tab:abslines}. The absorption line parameters are within uncertainties between the two spectra.
\begin{table}
	\centering
	\small
 	\caption{\hg\ and \heii\ absorption lines}

 	\begin{center}
	\begin{tabular}{lccc}
		\hline
	    & $\rm H_\upgamma$ &  \ion{He}{ii} &    \ion{He}{ii} \\
	    \hline
	    $\rm FWHM_{I}$ & 829 $\pm$ 128 &  656 $\pm$ 140 & 983 $\pm$ 317 \\
	    $\rm FWHM_{X}$ & 649 $\pm$ 39  &  1065 $\pm$ 50  & 757 $\pm$ 42 \\
	    \hline
	    $\rm EW_I$  & -2.6 $\pm$ 0.6 & -1.1 $\pm$ 0.3 & -1.0 $\pm$ 0.4 \\
	    $\rm EW_X$  & -1.4 $\pm$ 0.1 & -2.5 $\pm$ 0.2 & -1.4 $\pm$ 0.1 \\
		\hline
        $\rm WL_I$ & 4348.3 $\pm$ 0.6 & 4639.2 $\pm$ 0.8 & 4658.8 $\pm$ 1.6 \\
		$\rm WL_X$ & 4349.1 $\pm$ 0.2 & 4641.9 $\pm$ 0.3 & 4662.1 $\pm$ 0.3\\
        \hline
\end{tabular}
\end{center}
\textit{Note.} FWHM (in \kms), EW (in \AA) and central wavelength (WL, in \AA) of the absorption lines superimposed the \hg\ and \heii\ emission lines in the X-shooter and ISIS spectra (Fig.~\ref{fig:spec_fit_xsh} and \ref{fig:spec_isis}, respectively). With I we indicate the results from the fit to the ISIS spectrum (135 days after discovery) and with X the ones from the X-shooter spectrum (142 days after discovery). The values have been corrected for the instrumental broadening.
\label{tab:abslines}
\end{table}

The resolution in the spectra obtained during the remainder of our follow-up campaign is not high enough to investigate these absorption features in detail (we will discuss the UVES spectrum separately in Sec. \ref{sec:uves}). 
In both the ISIS and X-shooter spectra, the \ha\ line is strongly affected by the atmospheric absorption that we were not able to correct completely with \textsc{molecfit} (see Sec. \ref{sec:xsh}), therefore a precise fit of the \ha\ line was not possible. Nonetheless, in the host-subtracted spectra, the two [\ion{S}{ii}] lines (6716 and 6731 \AA) are clearly detected and resolved. In the lower resolution spectra of our follow-up campaign, the two lines often blend together.

\subsubsection{Line fitting results}
\begin{figure}\includegraphics[width=1\columnwidth]{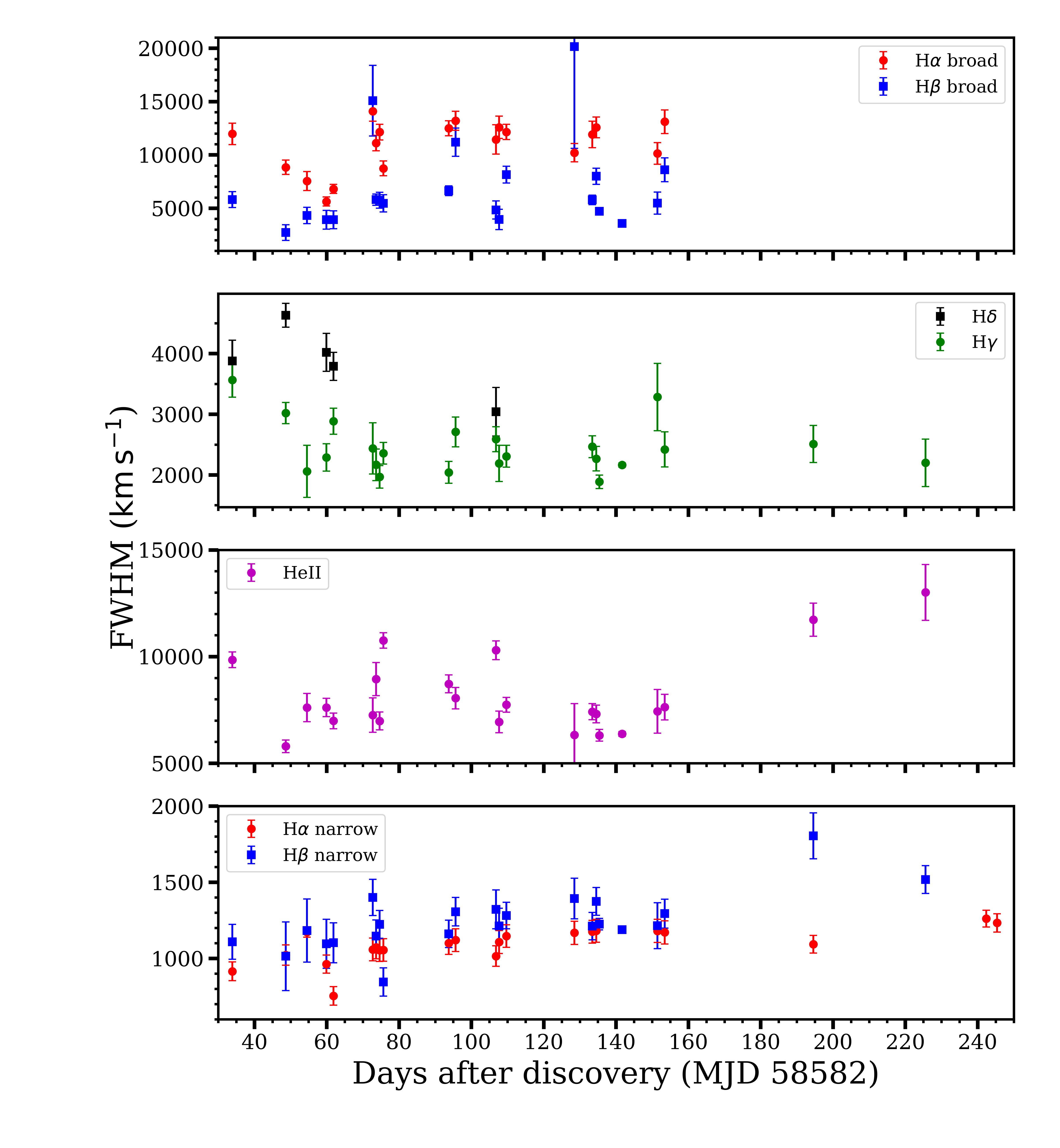}
        \caption{Evolution of the FWHM of the emission lines. In the first panel, the broad base of \ha\ (red circles) and \hb\ (blue squares); in the second panel, \hg\ (green circles) and \hd\ (black squares); third panel, \heii\ (magenta circles). These three lines are due to the TDE. In the bottom panel, the narrow peaks of \ha\ (red circles) and \hb\ (blue squares) are shown, these lines are due to host galaxy activity. On the x-axis, the number of days passed since the discovery of the transient.}
        \label{fig:line_fwhm}   
    \end{figure}

\begin{figure}\includegraphics[width=1\columnwidth]{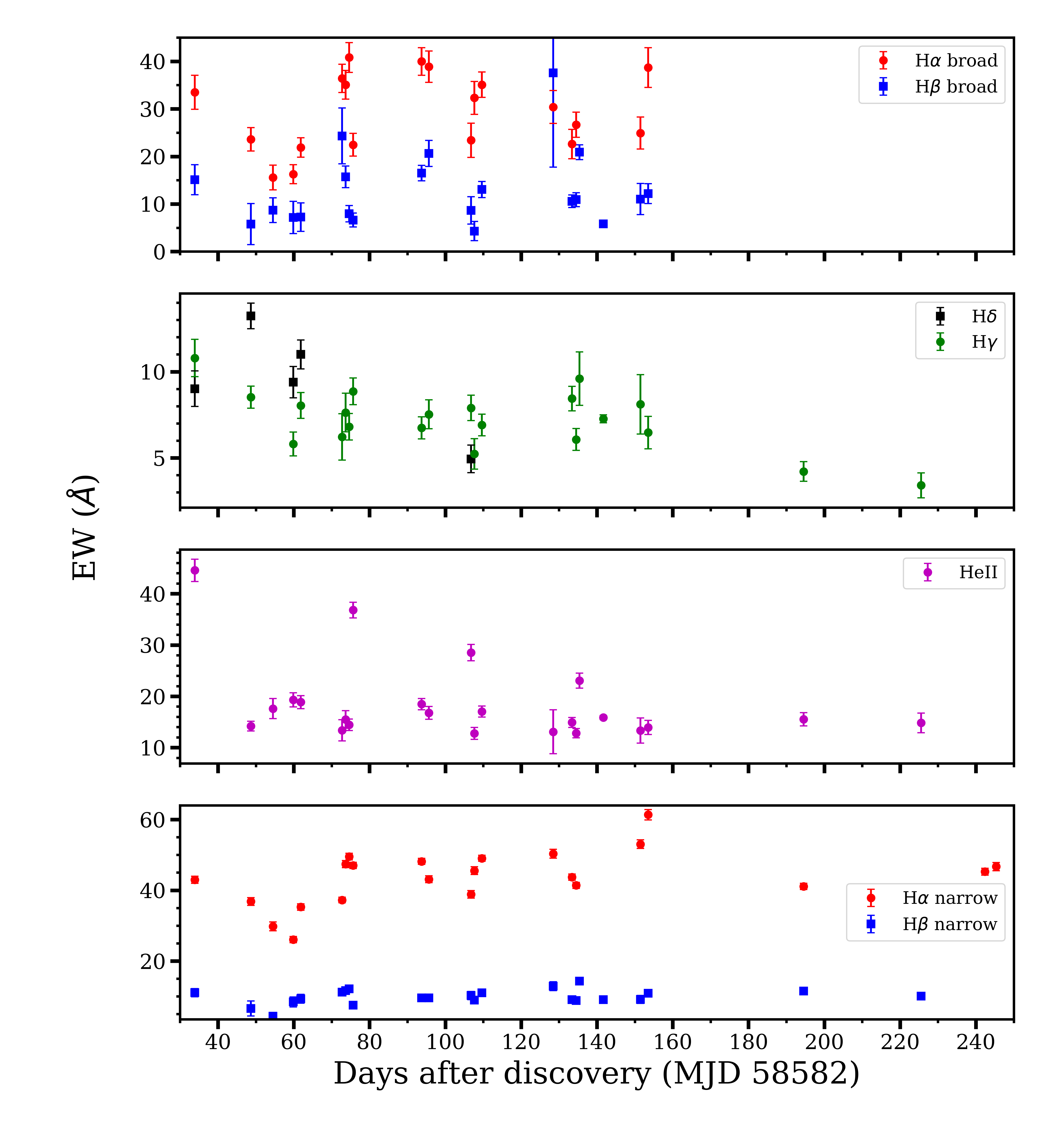}
        \caption{Evolution of the EW of the emission lines. From top to bottom: the broad \ha\ (red circles) and \hb\ (blue squares), the \hd\ (black squares) and the \hg\ (green circles), and the \heii~(magenta circles) emission lines caused by the TDE. Bottom panel: the narrow \ha\ (red circles) and \hb\ (blue squares) emission lines originating in the host galaxy. Note that the redshift of \dsg\ caused the \ha\ region to coincide in wavelength with the prominent telluric feature, rendering the derived EW of \ha\ uncertain. The number of days passed since the discovery of the transient is given on the X-axis.}
        \label{fig:line_ew}   
    \end{figure}

\begin{figure}\includegraphics[width=1\columnwidth]{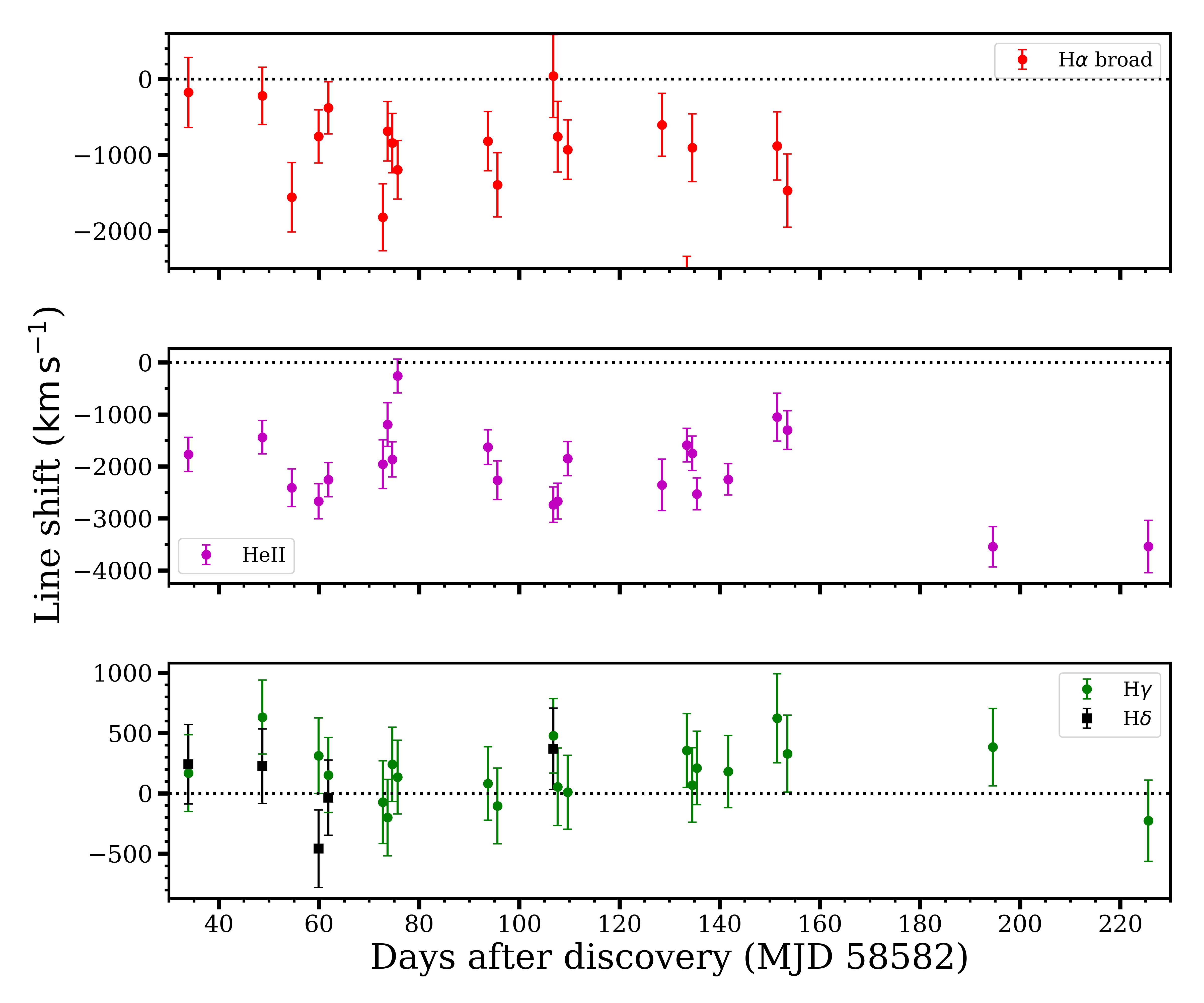}
        \caption{Evolution of the shift with respect to the laboratory central wavelength of the emission lines. From top to bottom: the broad \ha\ (red circles), the \heii~(magenta circles), the \hd\ (black squares), and \hg\ (green circles) emission lines caused by the TDE. The central wavelength of the broad H$\beta$ component was fixed during the fitting process and is not shown (see text). On the horizontal axis, the number of days passed since the discovery of the transient. Negative values indicate a blue-shift with respect to the rest frame wavelength.
        }
        \label{fig:line_shift}   
    \end{figure}

We report the results of the line fitting in Tables \ref{tab:lines_fwhm}, \ref{tab:lines_ew} and \ref{tab:lines_shift}.
We plot the evolution of the line parameters resulting from the fit: the FWHM, equivalent width (EW) and shift with respect to the rest frame wavelength are shown in Figures \ref{fig:line_fwhm}, \ref{fig:line_ew} and \ref{fig:line_shift}, respectively.

The values of the FWHM have been corrected for the instrumental broadening, measured from the sky lines of the spectra. The FWHM (Fig.~\ref{fig:line_fwhm}) of the broad components of \ha\ and \hb\ have different values but a similar evolution. The FWHM of the broad \ha\ starts at 12000 \kms, decreases to almost 5000 \kms over 30 days and then increases to the initial value of 12000 \kms in $\sim$20 days, to stay at this value until it is no longer detected, 150 days after the discovery of the transient. The FWHM of the broad \hb\ has an initial value around 5000 \kms, and for the whole duration of the follow-up campaign fluctuates between 4000 and 8000 \kms. There are some significant outliers that could be due to the corresponding spectra having a low signal to noise ratio (SNR), while others could be caused by variations due to uncertainties in the subtraction process. 
The initial evolution of the FWHM of \hg\ is similar to the one observed for the broad \ha\ (albeit with lower values): it starts at 3500 \kms, to then decreases to 2000 \kms over 30 days. It then remains around this value until the end of our follow-up campaign.
The \hd\ line in all the ACAM spectra is not well constrained, because it falls at the edge of the wavelength range covered. Therefore we do not fit it at these epochs. Its FWHM has an initial value similar to the one of \hg\, at 4000 \kms, it then increases to 4500 \kms at the next epoch (20 days later), to then gradually decrease to 3000 \kms between phase 60 and 110.
The FWHM of \heii\ goes from 10000 \kms to 5500 \kms in the first 30 days, then increases up to 10000 \kms in 30 days, to then decrease to 7000 \kms between 80 and 160 days. In the last two epochs, the FWHM of \heii\ increases up to 13000 \kms. 
The timescale of the initial decay of the FWHM of the broad \ha, \hb, and the \heii\ is similar to the timescale of the decay of the X-ray light and of the blue continuum.
Finally, the FWHMs of the narrow peaks of \ha\ and \hb\ do not show evolution over the time of observation and vary around 1000 \kms, except for two outliers at 195 and 225 days after discovery, when the narrow \hb\ has a FWHM above 1500 \kms. These points are due to the low SNR of the corresponding spectra. The values of the FWHM of the narrow \ha\ and \hb\ peaks obtained from the late time spectrum (411 days after discovery) are both around 1200 \kms. 

In Fig.~\ref{fig:line_ew} we plot the evolution of the EW of the emission lines with time. The EW of the broad base of \ha\ decreases from an initial value of 35 \AA\ to 20 \AA\ in 30 days, then increases to 40 \AA\ at phase +80, to then decrease again to 25 \AA\ over the rest of the follow-up campaign, except for the last epoch, when the EW is at 35 \AA. 
The EW of the broad base of \hb\ varies between 5 and 20 \AA, to then stabilise around 10 \AA\ in the last $\sim$40 days. The higher values are also in this case due to low SNR and uncertainties during the subtraction process, rather than being indicative of variability in the line. 
The EW of the \hg\ line starts at 10 \AA, to then decrease to 5 \AA\ after 30 days. It subsequently remains constant until the last two epochs in which the EW is 4 \AA. The \hd\ line follows a very similar evolution, except for the value at 50 days after discovery. The EW of the \heii\ has an initial value of 40 \AA\ and, already 20 days after this (50 days after discovery), it decreases to 10 \AA, where it remains almost constant for the next 100 days. The outliers at 80 and 110 days, with an EW above 30 \AA, are possibly due to uncertainties in the subtraction process. The initial evolution matches the time-scale of the decay of the blue continuum of the TDE. Finally, the EW of the narrow \hb\ remains constant around 10 \AA, while the EW of the narrow \ha\ varies around 40 \AA. The small scale variations of the \ha\ narrow peak are likely due to variations of the telluric absorption.

We plot the shift with respect to the restframe wavelength of the emission lines in Fig.~\ref{fig:line_shift} (we indicate blue shifts with negative values).
To reduce the number of free parameters, during the fitting procedure, the central wavelength of the broad \hb\ was tied to the one of the narrow peak, which, due to the high SNR in that line, dominates the fit result of this parameter (we do not show the shift of the broad \hb\ in the plot).
The broad \ha\ shows a blue shift that fluctuates between 0 and -2000 \kms, without a clear evolution with time. 
The \heii\ line is blueshifted by -2000 \kms at the first epoch of observation. The value of the shift fluctuates around this value for the first $\sim$ 150 days (except one outlier at 75 days), then decreases to -1000 \kms at 150 days. In the last two epochs at which the line is detected, the blue shift has increased to -4000 \kms.
The shift of the \hg\ and \hd\ line is always below $|500|$ \kms and is constant within uncertainties.
It is worth noting that the velocity offsets can be affected by O(10$^2$) \kms\ due to good observing conditions: when the seeing is better than the slit width, due to potential non perfect centering of the source in the slit, such changes can be induced. The narrow peaks of \ha\ and \hb\ are always shifted by less than $|500|$ \kms and are not plotted. The shift of \ha\ and \hb\ at each epoch is not of the same value. If we assume that the narrow \ha\ and \hb\ originate from the host galaxy, we can expect their central wavelength to be at the laboratory values. We can use the shift measured in their central wavelength as a proxy for wavelength calibration issues or for aforementioned uncertainties due to the observing conditions. We therefore use the values of the shift calculated for the narrow peak of \ha\ and \hb\ to estimate an additional uncertainty that is reflected in the error bars of Fig.~\ref{fig:line_shift}.

    \begin{figure}\includegraphics[width=1\columnwidth]{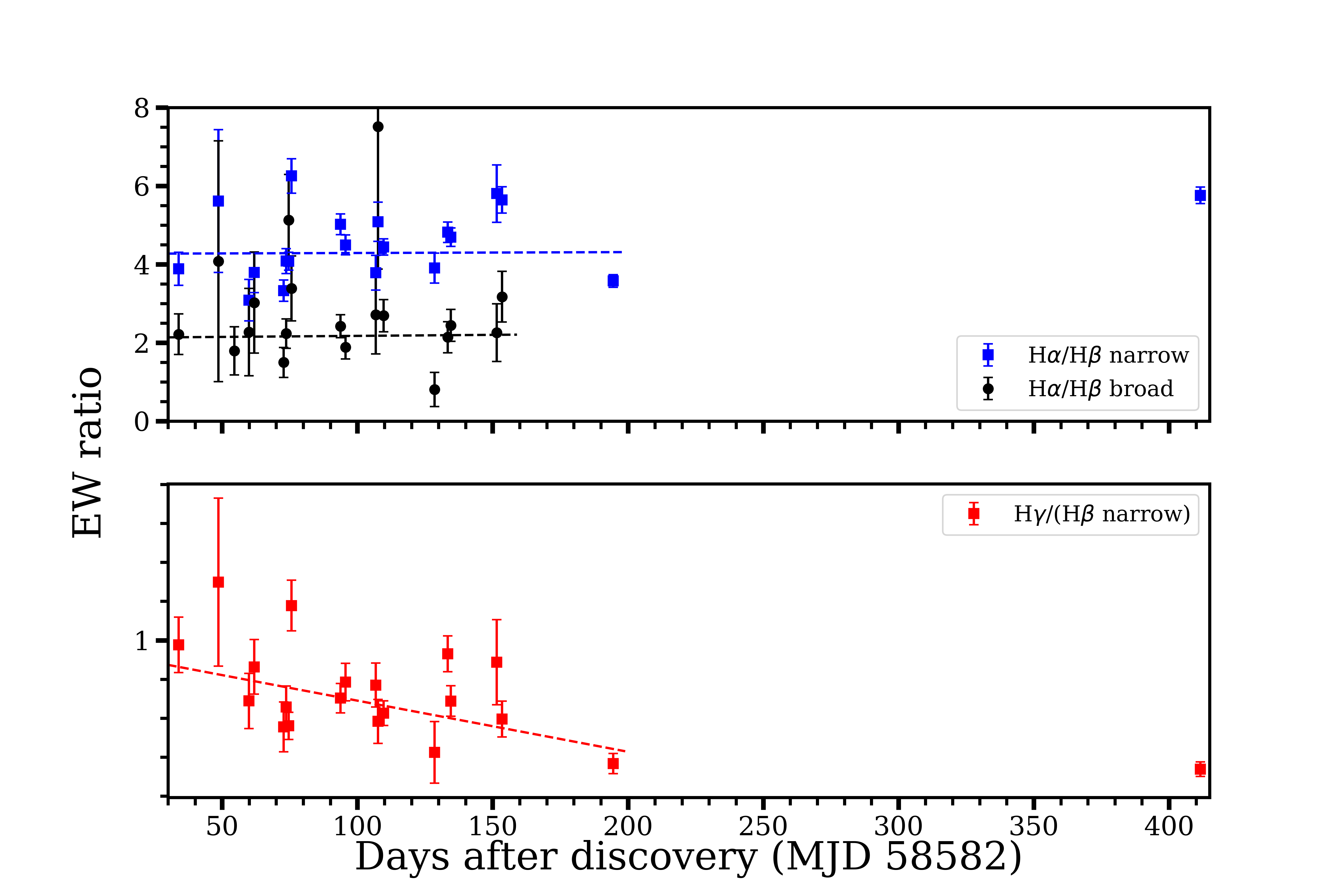}
        \caption{Ratio of the EW of Balmer lines. Top panel: narrow \ha/\hb\ in blue squares and broad \ha/\hb\  in black squares, both fitted with a straight line. The changes in the ratio between the narrow peaks are due to changes in the telluric absorption. Bottom panel: the ratio of the EW of the \hg and the narrow \hb\ emission line. On the X-axis, the number of days since the discovery of the transient.}
        \label{fig:line_ratio}   
    \end{figure}

We plot the flux ratio of the Balmer lines in Fig.~\ref{fig:line_ratio} and we fit the data with a straight line. The ratio of the narrow peaks of \ha\ and \hb\ shows a slight increase over time. During the first $\sim$200 days of observations, it remains almost constant at an average value of 4.5, to then increase to 5.8. The reduced \chisq\ of the fit is bad, with a value of 6.4. The ratio of the broad bases of \ha\ and \hb\ remains constant within uncertainties at an average value of 2.4 (reduced \chisq=1.9). The ratio between \hg\ and the narrow peak of \hb\ decreases from 1 to 0.3, which is also the value at the last epoch. The fit with a straight line to the data of the first $\sim$200 days of observations has a reduced \chisq\ of 1.8.

\subsubsection{UVES spectrum}
\label{sec:uves}
\begin{figure}\includegraphics[width=1\columnwidth]{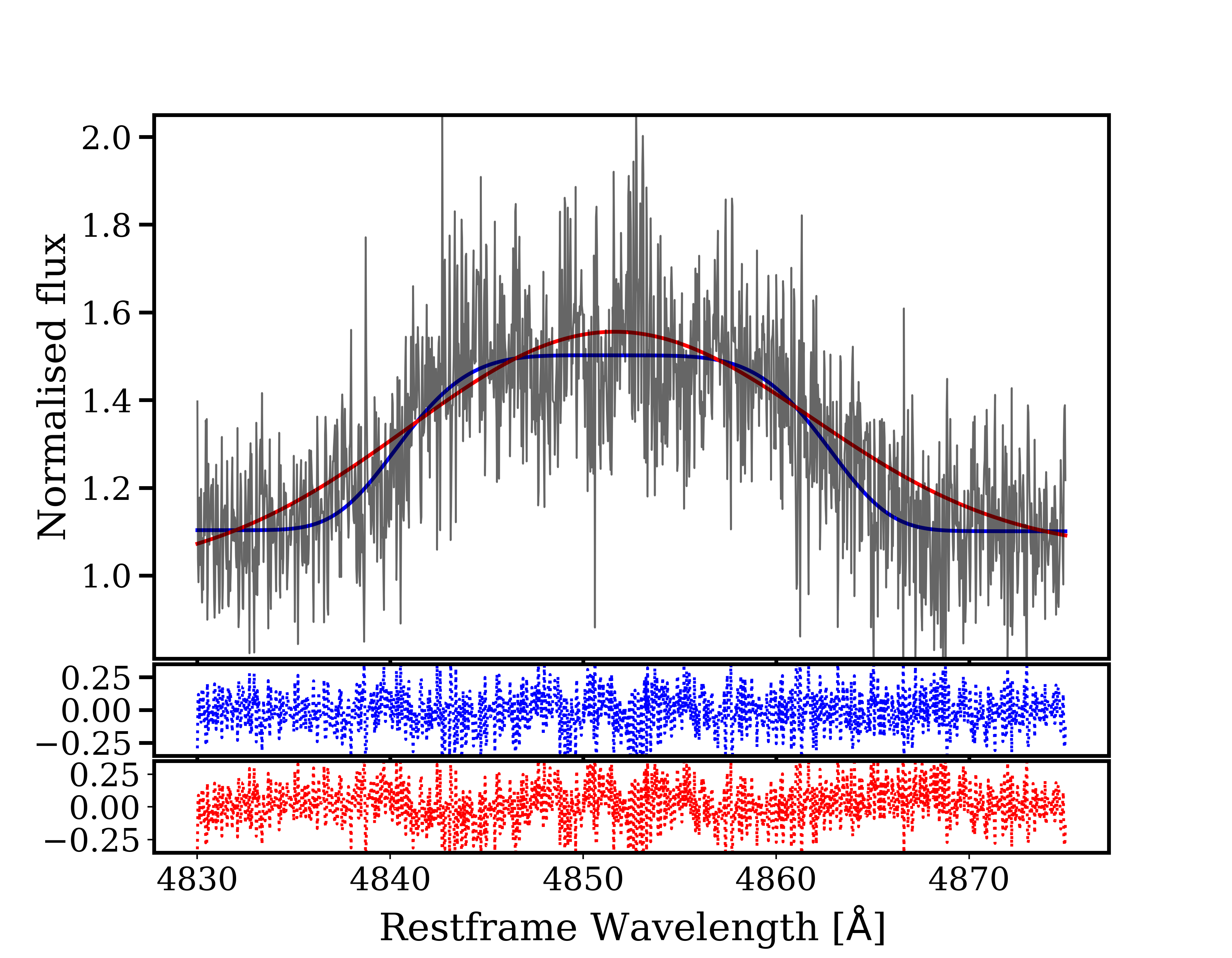}
        \caption{\hb\ region of the UVES spectrum with two different fits: in red, the fit with a Gaussian curve, similar to the function used in the other spectra and in blue, the fit using a flat-top Gaussian. The two bottom panels are the residuals of the two fit, the top one for the flat-top Gaussian fit and the bottom one for the Gaussian fit.}
        \label{fig:uves_hb}   
    \end{figure}

It is interesting to consider the UVES spectrum by itself, given its higher resolution. In Fig.~\ref{fig:uves_hb} we show the \hb\ region of the spectrum (\ha\ is strongly affected by the telluric absorption, while the other emission lines are not strong enough with respect to the continuum, which is very noisy).
The emission line in the UVES spectrum has a more box-like shape, instead of the Gaussian shape used to fit for lower resolution spectra. The line also shows evidence for structure on top. 
We fit this line both with a single Gaussian curve (in red in the plot), mimicking the fit of the other spectra, and with a "generalised" Gaussian (in blue):
$f(x)=a\cdot exp[-\left(\frac{x-c}{2\sigma}\right)^n]$
with n>2 to get a flat top (n$\simeq$5 from our fit). While the errors on the spectrum are too large to compare the reduced \chisq\ and assess the goodness of fit, visually the flat-top Gaussian describes the wings of the line better. On top of this, looking at the residuals, we can see that in the case of the "regular" Gaussian fit (bottom panel of Fig.~\ref{fig:uves_hb}, in red) there are small trends in the wings of the line (4840--45 and 4862--68 \AA), with respect to the flat-top Gaussian fit.

The difference between the shape of the \hb\ line in the UVES spectrum and in the other spectra can be accounted for by considering the different resolution. We rebin this spectrum, artificially lowering the resolution, to match the ACAM spectrum taken on 2019 July 11, which is the closest in time. The resulting spectrum is visually very similar to the ACAM one. We then fit the rebinned UVES spectrum with the line structure used for the other spectra (of which an example is shown in Fig.~\ref{fig:spec_fit_hb_sub}) and we obtain line parameters within uncertainties of the ones derived from the July 11 ACAM spectrum. Notably, the broad base of \hb\ and the \heii\ were difficult to constrain on the unbinned UVES spectrum, probably due to the low SNR in the continuum. The complex structure at the top of the emission line is consistent with that seen in the X-shooter (Fig.~\ref{fig:spec_fit_xsh}) and ISIS spectra.

\section{Discussion}

Before discussing the interpretation of the results of our observations and analysis, we summarise the main results.

The transient is bright in X-rays, UV, optical and radio. The X-ray light curve decays very quickly and the transient becomes undetected 125 days after discovery (2019 Apr 09, MJD 58\,582), while the low-frequency radio luminosity is seen to be increasing steadily. The optical spectra show a blue continuum that decays over the first 60 days of our observations. In addition, multiple broad emission lines are detected. After subtracting the stellar component of the host galaxy light, the spectra show Balmer (\ha\ through \hd ) and \heii\ 4686 \AA\ emission lines. The \oiii\ and [\ion{S}{ii}] doublets (4959, 5007 and 6716,6731 \AA, respectively) are also detected in emission. The \ha\ and \hb\ emission lines show two components: a narrow peak and a broad base. We consider the broad base of \ha\ and \hb, as well as the \heii\ line to originate from the TDE, while the narrow emission lines as due to emission from the host galaxy. However, we do consider the narrow \ion{Fe}{ii} emission lines appearing 200 days after discovery in the optical spectra to be caused by the TDE.
In a late time host galaxy spectrum, the broad bases of \ha\ and \hb, the \heii\ and the \ion{Fe}{ii} lines are no longer detected.
Our medium/high resolution spectra show absorption lines superimposed on some emission lines and some emission lines show a deviation from a Gaussian profile.

\subsection{Constraints on the presence of an AGN}
\label{sec:activecomp}
\begin{figure}\includegraphics[width=1\columnwidth]{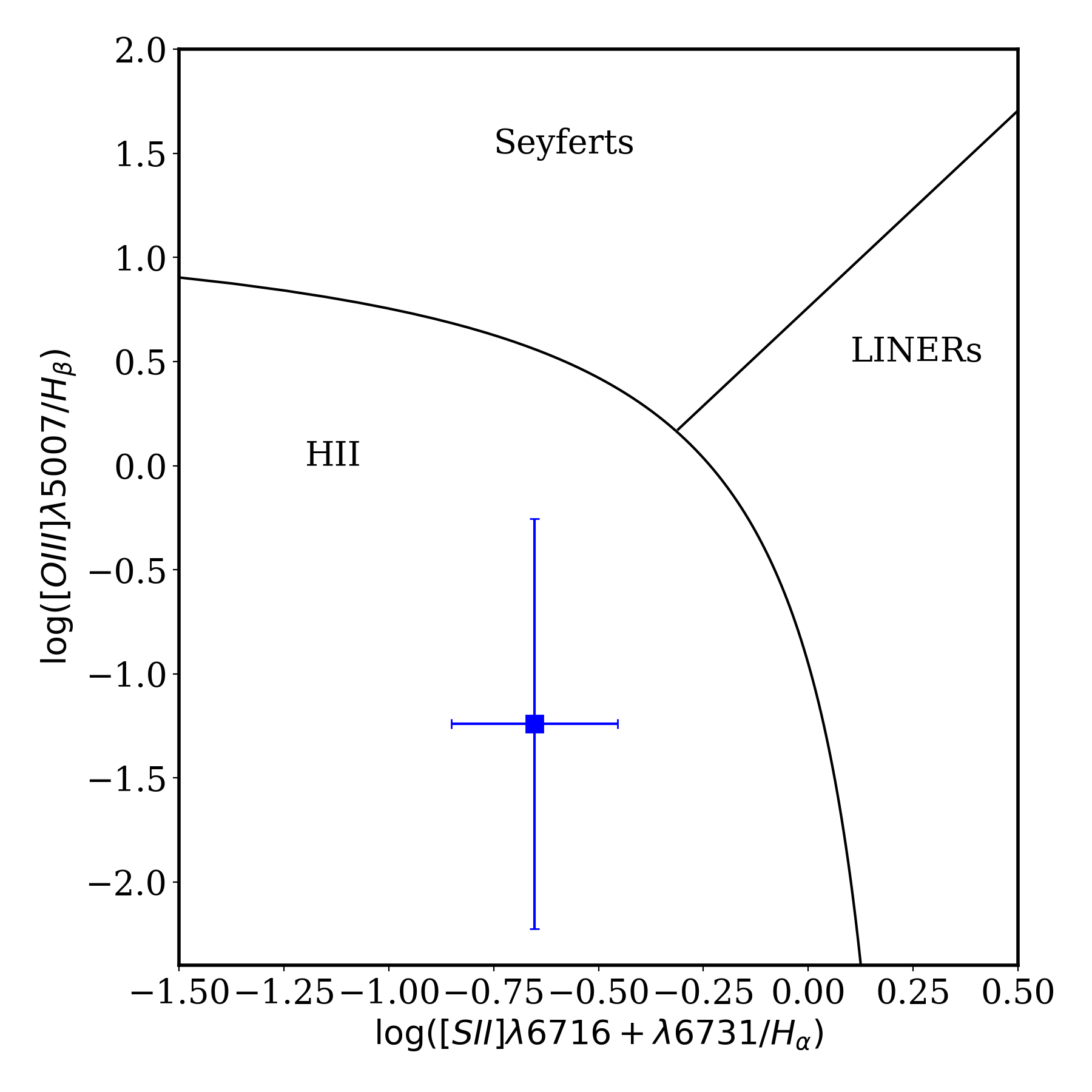}
        \caption{Position of the persistent narrow emission lines in the host galaxy spectrum of \dsg\ on a BPT diagram. The spectrum used was the late time TNG spectrum at 411 days, when no broad components to the Balmer lines could be detected. The lines separating the different regions come from \citet{kewley01}.}
        \label{fig:bpt}   
    \end{figure}

The persistence of the Balmer lines in our late time spectra, especially the narrow \ha\ and \hb\ peaks, the small offset and FWHM of these lines, coupled with the apparent absence of line evolution, indicate that those narrow emission lines do not originate in the transient event, but rather in a weak AGN or are caused by star-forming activity. 
In order to assess the underlying ionising mechanism for these narrow emission lines we plot the position of the source on a Baldwin, Phillips \& Terlevich (BPT) diagram \citep{baldwin81} in Fig.~\ref{fig:bpt}. Due to the presence of the telluric absorption on top of the \ha\ and \nii\ doublet lines we were not able to constrain the presence of the latter. Therefore, we are limited in the number of BPT diagrams we can use. For the BPT diagram, we use the values of the line parameters calculated from the last spectrum available, 411 days after discovery. The source falls in the \ion{H}{ii} region of the diagram, indicating that star formation is responsible for these narrow emission lines. 
In addition, the mid-infrared color from the Wide-field Infrared Survey Explorer (WISE), at W1$-$W2=$-$0.05, is lower than the threshold presented in \citet{Stern2012} for AGNs, also suggesting that the SMBH was not active prior to the TDE.

\citet{vanvelzen20} mention that the host galaxy falls in the green valley when plotting the intrinsic u$-$r color vs.~the host galaxy mass, a region that also hosts post starburst galaxies. The narrow \ha\ and \hb\ lines have a FWHM above 1000 \kms, which is higher than expected from an AGN narrow line region. Similarly, such a FWHM is high for most starburst galaxies, although such values have been observed in some (ultra) luminous infra-red galaxies \citep{arribas14}. Considering all the evidence, we exclude an AGN as the source of the narrow emission lines and favor an origin in star formation.

\subsection{Host galaxy star light subtraction and N Bowen fluorescence lines}

\begin{figure}\includegraphics[width=1\columnwidth]{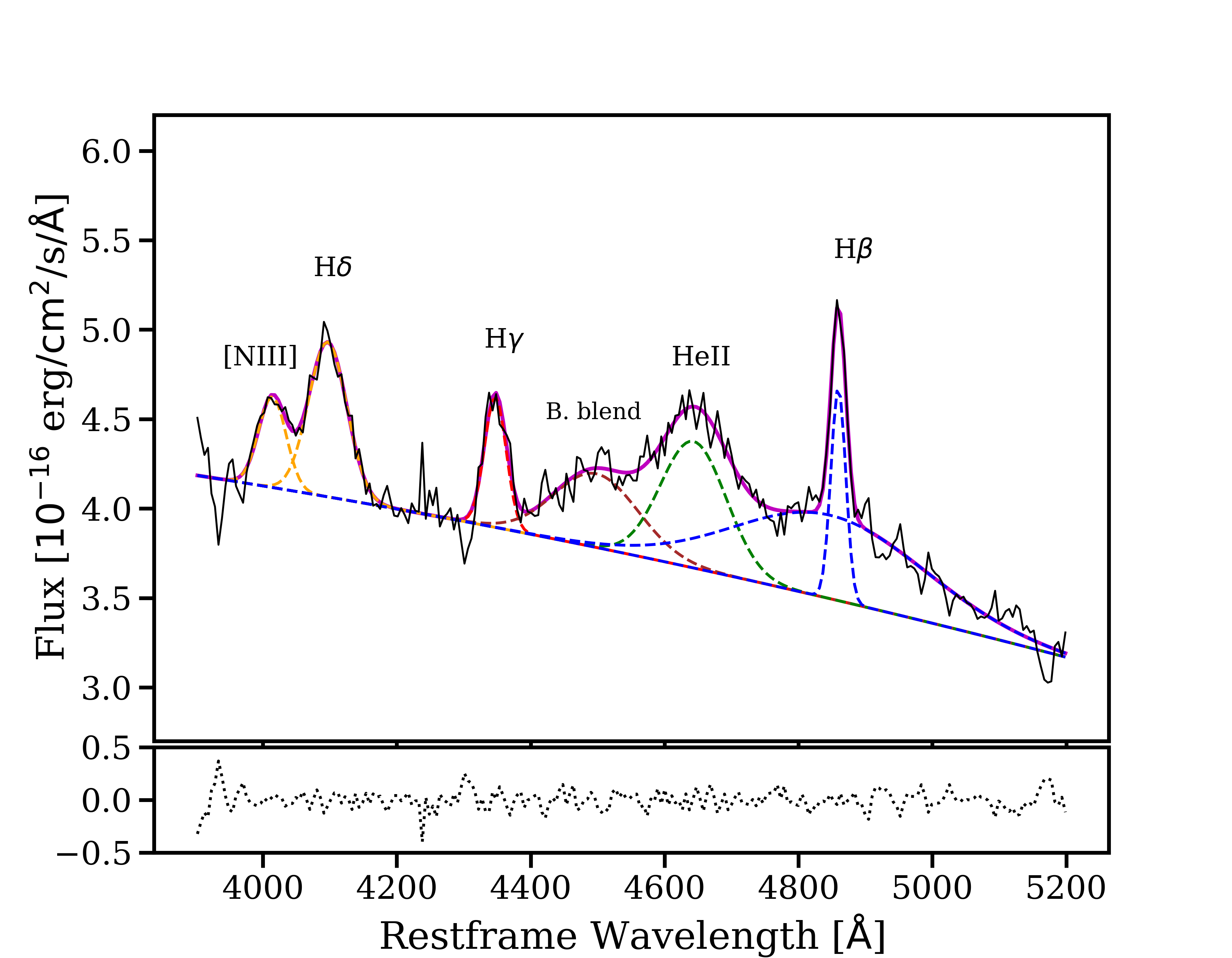} 
        \caption{Example of a fit to the emission lines in the \hb\ region of the EFOSC2 spectrum taken on 2019 June 08 (60 days after discovery), before performing the host galaxy subtraction. With B.~blend we indicate the blend of unresolved lines previously associated with Bowen fluorescence. In the bottom panel, the residuals of the fit are shown.}
        \label{fig:spec_fit_hb_unsub}   
    \end{figure}

After we corrected the observed spectra for the contribution of the host stellar light, we do not detect the N lines associated with the Bowen fluorescence lines in any of the spectra of our follow-up campaign. We carefully inspected the spectra of \dsg\, to investigate whether the Bowen blend is present or not. As an example we plot the line fit to the \hd\ to \hb\ region of the EFOSC2 spectrum of 2019 June 08 before the host galaxy star light is subtracted in Fig.~\ref{fig:spec_fit_hb_unsub}. The spectrum shows an emission line at $\sim$4050 \AA\ and an additional broad component between \hg\ and \heii. These lines have been associated with \niii\ lines due to the Bowen fluorescence mechanism \citep{bowen34,bowen35}: the line at 4050 \AA\ as the blend of two \niii\ lines (at 4097 and 4104 \AA, shifted by $\sim$50 \AA\ or 3700 \kms), while the broad component at 4500 \AA\ can be considered as a blend of many lines (see \citealt{steeghs02} for a list of Bowen fluorescence lines at these wavelengths). These emission lines were also identified in \citet{vanvelzen20}, leading to the classification of \dsg\ as a "TDE-Bowen". However, after the host galaxy subtraction process, both of these emission features are not detected in the spectra (see e.g., Fig.~\ref{fig:spec_fit_hb_sub}).

Emission lines due to the Bowen fluorescence have previously been identified in other TDEs: iPTF15af \citep{blagorodnova19}, AT2018dyb \citep{Leloudas2019}, ASASSN-14li \citep{Holoien201614li,prieto16}, iPTF16fnl \citep{onori19} and AT2019qiz \citep{nicholl20}.
iPTF15af and AT2018dyb were found in non-active galaxies, while the host galaxies of iPTF16fnl, ASASSN-14li, and AT2019qiz may harbour an AGN. For iPTF15af, iPTF16fnl and AT2019qiz host galaxy subtraction was performed on the spectra, while for AT2018dyb and ASASSN-14li this was not the case (the lines were very strong in these events). The subtraction for iPTF15af was performed by fitting a late time spectrum to the source spectrum. In the case of AT2019qiz, the host subtraction was performed by creating a synthetic host galaxy spectrum using stellar population synthesis models and subtracting it from the source spectrum (this, similarly to our case, removes the stellar contribution). Finally, for iPTF16fnl, a method similar to ours was employed. For the low resolution spectra, they used \ppxf\ to scale a late time host galaxy spectrum to the source spectra to subtract it, while for their X-shooter spectra, they employed the \textsc{phoenix} library to create a synthetic host spectrum.

    \begin{figure}\includegraphics[width=1\columnwidth]{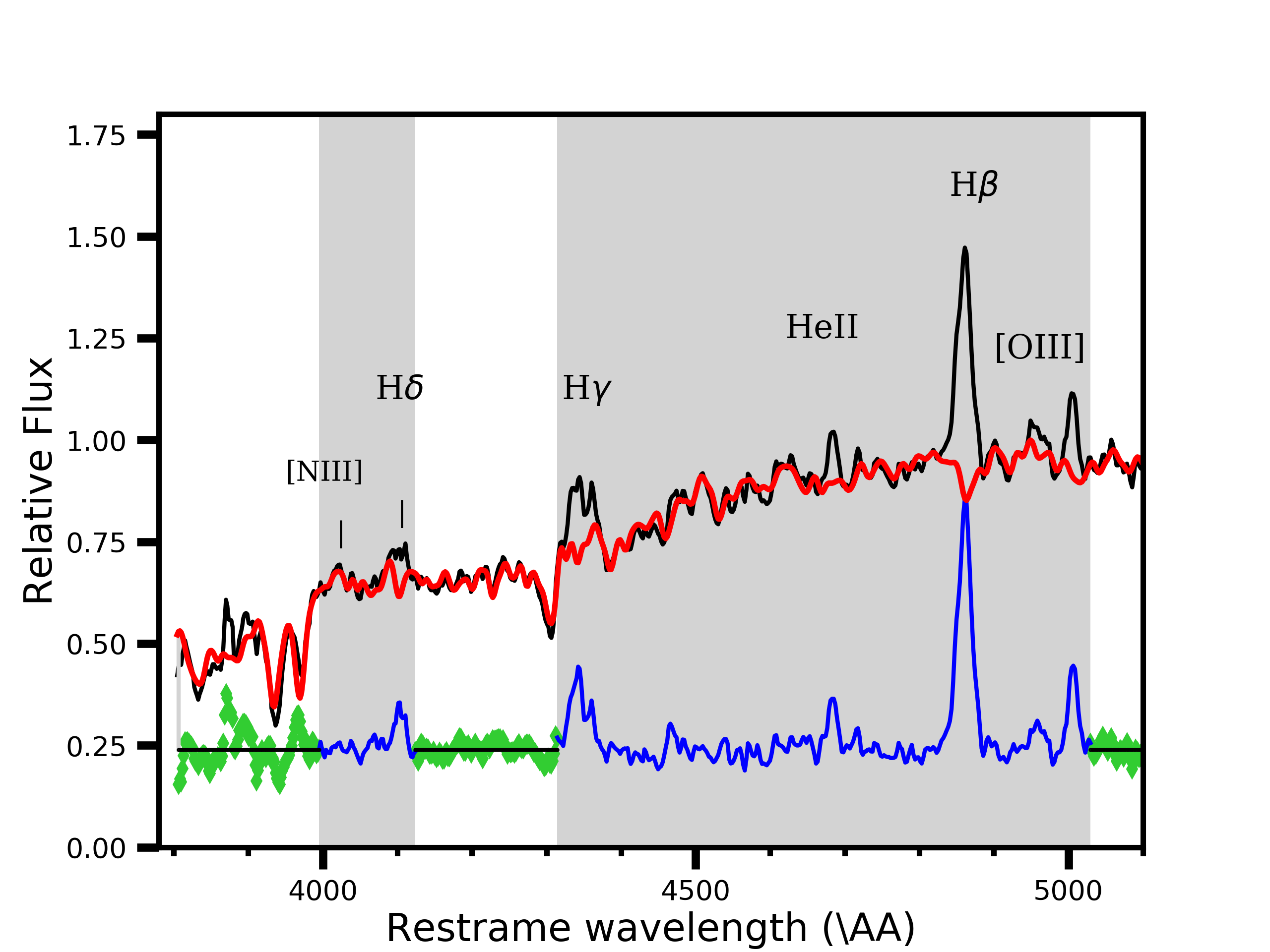}
        \caption{Zoom in of the  \hd\ - \hb\ region on the late time TNG spectrum (+411 days). In black, the source spectrum; in red, the convolved stellar library (the \ppxf\ best-fit host galaxy synthetic spectrum); in green and blue, the subtracted spectrum. The grey vertical bands indicate the regions of the spectrum excluded from the fit. The \hd\ - \hb\ region is almost completely masked during the fit, while the rest of the spectrum, except for the \ha\ line, is considered for the template fitting.}
        \label{fig:ppxf_tng}   
    \end{figure}

As a check on our host star light subtraction procedure, we also perform the subtraction of the host galaxy stellar component using the late-time TNG spectrum (411 days after the discovery of the transient) to build our synthetic host spectrum. For this spectrum, we use the MILES library \citep{sanchez06,falcon11}, which consists of $\sim$1000 stellar spectra covering the 3525 -- 7500 \AA\ wavelength range with a resolution of 2.5 \AA. Using \ppxf\ we first build the synthetic host spectrum and then we scale it to each spectrum of our follow-up campaign and subtract it. The scaling is necessary to account for variable slit losses in the data taking and for the different resolution of the detectors. In Fig.~\ref{fig:ppxf_tng}, we show the blue part of the observed late-time TNG spectrum, the best-fit results of the \ppxf\ procedure,  i.e., the synthetic host galaxy stellar light, and the residuals. The regions with relevant emission lines, including the wavelength range associated with potential \niii\ Bowen fluorescence features are excluded from the fit. The results of the spectral analysis on these newly subtracted spectra (including the disappearance of the \niii\ lines and the results of the line fit) were consistent within errors with the results presented in Sec.~\ref{sec:spec_analysis}. It is interesting to compare our \dsg\ results with those of iPTF16fnl obtained by \citet{onori19}: the spectrum used for the host subtraction in iPTF16fnl also contained a weak \niii\ 4100 \AA\ host galaxy line, but nevertheless the Bowen line was clearly present in the subtracted spectra (see fig.~6 of \citealt{onori19}). In all subtracted spectra, iPTF16fnl clearly shows N lines both at 4100 \AA\ and at 4640 \AA, also in the X-shooter spectra, on which the host-subtraction procedure was identical to ours. Clearly, the host stellar light subtraction procedure is not responsible for the non-detection of the N Bowen lines in \dsg.

We therefore associate the presence of these emission lines in our unsubtracted spectra with spectral features originating in the host galaxy. In our synthetic spectrum (in red in Fig.~\ref{fig:ppxf_tng}) the area between \hd\ and \hb\ is similar to the spectrum of a green valley galaxy (see fig.~9 of \citealt{pan10} for an example of a green valley spectrum). That green valley galaxy spectrum clearly shows a feature at $\sim$4020 \AA\ that matches the one observed in our unsubtracted spectra. In addition, the host galaxy spectrum shows a bump in the continuum around 4500 \AA\, which when left unsubtracted can explain the erroneous identification of this feature as a blend of Bowen lines caused by TDE emission.

\subsection{Radio emission from \dsg\ }

In discussing the radio properties of \dsg, we also consider the radio data first published in \citet{stein20} (also plotted in Fig.~\ref{fig:radio}).
The first radio observations of \dsg\ started 42 days after discovery, at which time \dsg\ was already detectable at frequencies $\nu \gtrsim 3$ GHz (Fig.~\ref{fig:radio}) and showed an inverted spectrum (the higher the observing frequency, the higher the observed flux density), indicating that the radio emission was in its optically thick phase at all observed frequencies. Over time the 
radio emission continued to grow at all frequencies, and around day 70 \dsg\ started to enter its optically thin phase at frequencies $\gtrsim$15 GHz. Correspondingly, the peak frequency of \dsg\ started to shift progressively towards smaller values, and by day $\sim$180 \dsg\ had reached its optically thin phase also at $\sim$6 GHz. Our e-MERLIN monitoring shows, however, that the radio emission at 5.1 GHz is still rising to its maximum at $\sim$ 180 days, indicating that the TDE had not yet entered the optically thin regime at frequencies $\lesssim$5.1 GHz at that time. This result is in agreement with the value of $\nu_{\rm peak} = 5.4 \pm 0.1$ GHz at $t = 178$ d reported by \citet{stein20}. The overall radio behaviour is consistent with the evolution of gas emitting synchrotron radiation, the emission at the lower frequencies is self-absorbed. This has also been seen in other TDEs with radio emission, including both TDEs that showed a relativistic jet (e.g., Arp~299B-AT1 \citealt{mattila18}), and those where the relativistic/sub-relativistic nature of the outflow is less clear (e.g., \citealt{vanvelzen2015,Alexander2016}). \citet{stein20} have modelled the radio emission of \dsg\ as a synchrotron self-absorbed sub-relativistic outflow that has a continuous injection of fresh relativistic electrons from a central engine. But we note that in the case of the TDE Arp~299-B~AT1, \citet{mattila18} modelled the radio emission with a relativistic jet whose initial population of electrons are continuously re-accelerated.
\citet{stein20} find that the expansion of the TDE proceeds with a constant speed for the first three epochs, until day 120, and then suffers an increase in expansion speed at the latest epoch. However, we note that this is explanation is not unique and probably model dependent, as the outflow radius is not measured directly. A relativistic jet with a fast core and a slow sheath seems to be able to account for an almost constant expansion (of the fast jet) as well \citep{Mimica2015}, although we defer detailed radio modelling to future work. Similarly, 
the increase of the expansion speed seen around day 180 for \dsg\ can also be explained by a relativistic jet that encounters a medium with a steep density profile, as found by \citet{mattila18} for Arp~299B-AT1.  
The 5.0 GHz radio emission of Arp~299-B~AT1 peaked at substantially later epochs, more than $\sim$1000 days after the event. In contrast, ASASSN-14li (e.g., \citet{bright18}) and AT2017gbl \citep{kool20}, showed their 5.0 GHz emission peaking around 150 days. At the end of our follow-up, the 5.0 GHz emission from \dsg\ has still not reached its peak. We stress that a comparison of phases across different TDEs is not straightforward, as we define time=0 as the discovery of the transient, and not as some moment related to the physics of the event.

\subsection{X-rays: disc cooling?}
The X-ray luminosity decays very quickly, decreasing by a factor $\sim$100 in roughly 60 days, and becomes undetected 125 days after discovery, although we lack deep late-time {\it Chandra} or XMM-{\it Newton} observations (cf.~\citealt{Jonker2020}). To explain such a dramatic decrease, we can invoke several possible mechanisms that can work alone or in conjunction, such as: (i) an increase in absorption in our line of sight to the X-ray emitting region or (ii) cooling of the disc that causes the emission to shift out of the X-ray band, caused by either a general decrease in mass accretion rate, reducing the energy output, or by an increase in the amount of ionized material causing the X-ray photosphere to move out with time. 

The observed cooling of the black body temperature in X-ray spectral fits to the NICER data (Fig.~\ref{fig:xray_T}) suggest that the first mechanism is not at play, as it would make the observed spectrum harder with time. Furthermore, following the TDE unification idea \citep{Dai2018} the cooling X-ray disc seems to suggest that we are observing the system relatively pole-on. The second mechanism mentioned above could be at play. We see a decrease in temperature over the first 60 days after discovery of the transient. If this cooling continued, the emission shifts out of the X-ray pass band. Indeed, in the last Swift detection (125 days after discovery), the photons were all detected below 0.5 keV. The X-ray luminosity seems to follow the Stefan-Boltzmann law (Fig.~\ref{fig:xray_LvsT}), meaning that the emission from the disc is consistent with a thermal nature.

What is causing this cooling is unclear: potentially the mass accretion rate is dropping, or alternatively, the photosphere is moving out, due to increased presence of ionized material. The black body radius obtained from the modeling of the X-ray spectrum using a black body model, $\rm R=3.4^{+1.5}_{-0.9}\times10^{11}$ cm is smaller than the Schwarzschild radius of a $\rm (5.4\pm3.2)\times10^6M_\odot$ BH, $\rm R_S = (1.6\pm0.9)\times10^{13}\, cm$. While a more edge-on view of the accretion disc could cause the observed emitting region to appear smaller another possible explanation involves the properties of the atmosphere above the accretion disc. Due to a rapid decrease of the opacity with X-ray photon energy due to light element free-free, bound-free, and bound-bound transitions, higher energy X-ray photons are coming from deeper, hotter layers (see for instance the explanation in \citealt{zavlin2002}). This effect is seen in quiescent neutron stars (NS), where the presence of an atmosphere causes the emitting radius to be underestimated if modeled with a simple black body (see \citealt{kaspi10}, for a review). It is worth noting that finding black body radii too small for the BH masses is not uncommon in TDEs (e.g.,~\citealt{Wevers2019a}).

To better understand the driver behind the decay in X-ray luminosity, we also checked the \heii\ 4686 \AA\ line EW. The \heii\ 4686 \AA\ line is known to be created by photo-ionisation due to (soft) X-ray photons, exhibiting photon-counting properties with a rough correspondence of 1 \heii\ photon emitted for each 0.3-10 keV X-ray photon \citep{pakull86,schaerer19}. If we compare the photon emission in the X-rays with the \heii\ line EW, we see that the rate of photons emitted in the X-rays is above the amount necessary to explain the detected \heii\ emission line at all epochs where we have data, by a factor of $\sim$100 at the beginning reducing to a factor $\sim$2 at the last X-ray detection (125 days after discovery). The \heii\ luminosity does not trace the initial dramatic decrease present in the X-ray light curve. Instead, its EW only decreases by a factor $\sim$4 between 30 and 50 days after the discovery date. This could suggest that the region responsible for the \heii\ emission is (partially) shielded from our line of sight, i.e., the X-ray emission is not "seen" by all material surrounding the black hole.

There is suggestive evidence for short term variability in the Swift X-ray data. Other TDEs have shown similar X-ray {\it flaring} activity (e.g., AT2019ehz, \citealt{vanvelzen20} and SDSS~J1201, \citealt{saxton12}). This behaviour can for instance be explained by a clumpy outflow. Another cause for short-term X-ray variability could be precession of the accretion disc. In such a scenario, when viewed under a large inclination, the disc can intermittently obscure the inner X-ray emitting region. The inclination angle changes due to precession creating the modulation of the light curve on the precession timescale \citep{franchini16,wen20}. In our case, we favour a more polar line of sight to the system and therefore, we deem the precession scenario as an explanation for the X-ray variability in \dsg\ less likely.
To explain the variability observed in the Swift lightcurve (Fig.~\ref{fig:xray_lc}), we could invoke the debris streams caused by the initial disruption of the star. As explored in more detail in Sec.~\ref{sec:abslines}, when the star is disrupted, the debris will form self gravitating streams \citep{Guillochon2014}. If some of these streams are deflected to high enough angles by nodal precession, they could (occasionally) intercept our line of sight to the inner accretion disc, where the X-rays are produced. The variability of the EW of \heii\ (Fig.~\ref{fig:line_ew}) could be reflecting the variability observed in the Swift X-ray light curve.

At late times (more than 200 days), we start seeing low ionisation \ion{Fe}{ii} emission lines in the spectra. These lines are produced in an optically thick, high density medium, ionised by X-ray radiation. Low ionisation \ion{Fe}{ii} lines have already been observed in TDEs, as reported in \citet{wevers2019b}, where the authors compare the emergence of these lines to the case of Cataclysmic Variables, where these lines are thought to originate from the surface of the accretion disc. \dsg\ is the third UV/optical detected TDE that show these low ionisation, optical \ion{Fe}{II} emission lines, the other two being AT~2018fyk and ASASSN--15oi \citep{wevers2019b}. We identify emission from multiplets 37, 38, 42 and 49, which are some of the strongest features also observed in some AGN. 
These three Fe-strong TDEs show several similarities in their observed properties. Similar to AT~2018fyk and ASASSN--15oi, \dsg\ produced observable X-ray emission (although, note that at the epochs at which we detect the \ion{Fe}{ii} lines we do not have X-ray data).
The Fe lines observed are much narrower than the other TDE lines. As explored below (see Section~\ref{sec:optlines}), following \citet{Roth2018}, we hypothesise that in \dsg\ the broad lines due to the TDE (\heii, \ha\ and \hb) are created in an expanding photosphere that causes the shift of the line centroid and the large FWHM. Therefore, we assume that the Fe lines do not originate in this expanding photosphere, but rather in optically thick clumps of gas. This gas is either pre-existing but only irradiated by X-rays at later times for instance due to disc slimming (\citealt{wen20}) or low-velocity condensations of earlier outflowing material. While the late-time presence of the Fe lines reinforces the idea that there is still X-ray emission, their ionisation potential is only a few eV, so that even a "cool" disc could be sufficient to ionise them.

\subsection{N Bowen fluorescence lines}
\label{sec:n-bowen}
Besides showing late-time Fe emission lines, AT~2018fyk, ASASSN--15oi, and \dsg\ have another common characteristic: their spectra show no evidence for \ion{N}{III} Bowen fluorescence emission. On the other hand, emission features near 3760 \AA\ do appear in all three sources (see e.g., the UVES and ISIS spectra in Fig.~\ref{fig:spectra_dsg}).

The wavelengths of the 3760 \AA\ features coincide with a group of transitions that can be directly linked to both excited states pumped by the primary Bowen fluorescence process of Oxygen \citep{bowen34}. An alternative explanation for these features could be \ion{Fe}{II} emission lines from multiplets 120 and 29 \citep{Netzer1983}, which can be particularly strong due to \ion{Fe}{II} self-fluorescence (i.e., wavelengths coincidences between \ion{Fe}{II} transitions). However, due to the presence of several transient forbidden transitions of \ion{O}{III} (including $\lambda 4363, 4960$) in the Fe-strong TDEs, we prefer the former explanation. 

Why do we observe Bowen \ion{O}{III} lines but not the typical \ion{N}{III} lines in Fe-strong TDEs?
The Bowen fluorescence mechanism is based on a chain of photo-(de)excitations that arise from a strong X-ray ionising source exciting \ion{He}{II} Ly-$\alpha$ photons and a dense medium. The primary fluorescence occurs through Oxygen, whose transitions in turn have resonances that can pump N atoms to excited states. In particular, the \ion{O}{III} cascade produces the 2p3s $^{3}$P$^0$ state through de-excitations including (among others) the 3760 \AA\ transitions. This unstable state subsequently decays through UV resonance lines ($\lambda$ 374.432, 374.162 \AA) with \ion{N}{III}. This results in \ion{N}{III} emission near the 4640 \AA\ complex that is the hallmark of Bowen fluorescence, observed in X-ray binaries \citep{McClintock1975}, novae \citep{Selvelli2007}, planetary nebulae, QSOs \citep{Weymann69} and TDEs \citep{blagorodnova19, Leloudas2019,onori19}. 

If resonant line fluorescence is efficient, one would then expect to see \ion{N}{III} 4640 \AA\ emission whenever the \ion{O}{III} 3760 \AA\ feature is present. However, this does not appear to be the case for Fe-strong TDEs.
One straightforward explanation is that the \ion{N}{III} 4640 \AA\ emission feature is {\it hidden} in the \ion{Fe}{II} complex. However, the majority of TDEs with Bowen lines have \ion{N}{III} 4640 \AA\ emission line fluxes that exceed the \ion{O}{III} 3760 \AA\ fluxes \citep{Leloudas2019}. In addition, the 4640 \AA\ feature is clearly visible even next to the dominant \ion{He}{II} 4686 \AA\ line. This suggests that the line strength of \ion{N}{III} 4640 \AA\ is much lower in Fe-strong TDEs. In other words, the resonant line fluorescence of \ion{N}{III} is much less efficient.

Probably the absence of significant \ion{N}{III} emission reflects different physical conditions along our line of sight in N- and Fe-strong TDEs. In particular, \citet{Selvelli2007} argued that the key to pumping \ion{N}{III} resonance lines is multiple scatterings in an optically thick medium. In optically thin conditions, the probability that an \ion{O}{III} resonance photon will directly excite the parent level that results in \ion{N}{III} 4640 \AA\ emission is very low. However, if the medium is optically thick, multiple scatterings of the \ion{O}{III} photon results in a very high probability of ultimately triggering the \ion{N}{III} emission. A more detailed, quantitative analysis to infer the physical properties of the gas will require high SNR spectroscopy of the Bowen lines, including those in the UV part of the spectrum which we are lacking for \dsg, combined with radiative transfer modelling.

\subsection{Broad optical emission lines}
\label{sec:optlines}
It is interesting to consider the evolution of the parameters of the (broad) emission lines detected in \dsg. The EW and the FWHM of the \hg\ line both show a decrease (by a factor $\sim$2) between 30 and 70 days after discovery, a timescale that is similar to the one over which the blue continuum of the optical spectra decays. More interestingly, the EW of the \hg\ line shows a gradual decrease over the whole period of our follow-up campaign. Potentially, there might be two components to that emission line (similarly to \ha\ and \hb\ case): one from the host galaxy and another one from the TDE. However, in the case of the \hg, we are not able to separate these two components in a narrow and a broad line. The decrease of the EW can be explained by the decay of the broad component arising from the TDE. From 190 days onwards, we are probably detecting only the contribution from the host galaxy narrow emission line: the value of the EW is almost a factor 2 lower than the previous data point and does not change any further (see also Tab.~\ref{tab:lines_ew}). The broad component of \ha\ and \hb\ also is no longer detected at these epochs. If the central wavelength is dominated by the narrow line component originating in the host galaxy, this would also explain the low shift of the central wavelength detected in this line (and in \hd\ in the few epochs in which it is constrained). Finally, also the decrease of the ratio between the EW of the \hg\ and that of the narrow \hb\ (Fig.~\ref{fig:line_ratio}) could be explained by the \hg\ having these two components.

The evolution of the \heii\ is also interesting: it is the only line to show a consistent blueshift with respect to the restframe wavelength, and this blueshift is increasing in the last two epochs. The amplitude of this blueshift is in line with the velocity of the radio outflow derived in \citet{stein20}, which is also increasing around 170 days after discovery. The FWHM of the \heii\ also increases at the same time. This could mean that the \heii\ line originates in expanding gas that traces the outflow that could be responsible for the radio emission as well. Its FWHM is then partly determined by electron scattering \citep{roth18} and partly by differences in the outflow velocities projected onto our line of sight of the \heii\ emitting gas.
The broad components of \ha\ and \hb\ have FWHMs with of similar or greater values compared to \heii. They could also originate from an expanding shell of gas. Also the velocity shift of the broad base of \ha\ has a value similar to the one of \heii. Unfortunately, the broad \ha\ and \hb\ were not detected at the last two epochs when the blueshift of the \heii\ line increases. The scenario of expanding material, where X-ray radiation is reprocessed and the broad optical lines are emitted, is similar to the one discussed in \citet{nicholl20} for the TDE AT\,2019qiz.

\subsection{High resolution spectroscopy and absorption lines}
\label{sec:abslines}
In our follow-up campaign of \dsg\ we obtained two medium resolution spectra (ISIS and X-shooter) and one high resolution spectrum (UVES). In these spectra, we see evidence for absorption lines superimposed on the broad emission lines: in both the X-shooter (Fig.~\ref{fig:spec_fit_xsh}) and ISIS (Fig.~\ref{fig:spec_isis}) spectra, the \hg\ and \heii\ lines show absorption lines. The UVES spectrum (Fig.~\ref{fig:uves_hb}) also provides suggestive evidence for the presence of several absorption lines superimposed on the \hb\ line. Unfortunately, the SNR of the UVES spectrum is too low to establish the presence of these potential absorption lines beyond doubt. The spectrum does show evidence that the overall line shape may deviate from a simple Gaussian profile. The \hb\ line shows a flat top. Deviations from a Gaussian profile of the broad emission lines of a TDE have been seen before in the literature, with lines showing double peaks and box-shaped profiles \citep{short20}. 

We thoroughly checked our synthetic host spectrum for features that could artificially create the absorption lines detected in the \hg\ and \heii\ lines, during the host-subtraction procedure, but found none.
The absorption lines could be due to the host galaxy. On one hand, if this were the case, we would expect the line parameters to not change between the two spectra, especially since they are separated by just 7 days (the line parameters are reported in Tab.~\ref{tab:abslines}). Indeed, the line parameters are consistent to within uncertainties. On the other hand, these lines are present in the spectra after we removed the stellar component, disfavouring such an origin. 
We check for absorption from diffuse atmospheric bands (DIBs), finding that only the absorption feature in \heii\ at 4660 \AA\ could potentially be explained by the DIB at 4659.8 \AA\ \citep{Hobbs08}.

When a star is disrupted, the debris may form strongly self gravitating streams \citep{Guillochon2014}. What we could be seeing is absorption lines caused by such streams, where the different orbital motions and projected velocities of these different streams cause the variation of the width of the lines. To have this, we would need some of the self-gravitating streams to be deflected by large angles, while the bulk of the disrupted material circularises into an accretion disc, as we deem it likely that the disc is observed under a low inclination angle.
Detecting these streams could be an indicator of the BH spin: in fact, to produce the detected absorption lines, the streams have to be somewhat long-lived (in our case, for at least 140--150 days). This happens if the streams do not self intersect during their orbital motion. As suggested in \citet{Guillochon2015}, the relativistic precession induced by the BH spin can deflect the streams from the original orbital plane, thus avoiding their intersection. This effect is more pronounced for retrograde orbits \citep{hayasaki16} than for prograde ones \citep{liptai19}.
It is important to note that in \citet{Leloudas2019}, a search for such absorption features in their UVES spectra is presented, without finding any absorption line due to the TDE. Overall, it is unclear if such streams can be present while at the same time active accretion is ongoing.

TDEs are geometrically complicated phenomena, with streams of debris self-intersecting, outflows, an accretion disc which is not expected to have a simple and symmetric geometry - especially at the beginning of the outburst - and it is only natural for this complex environment to be reflected in the spectral properties.
In the future, employing high resolution spectrographs and developing physical models that account for the fine structure observed could be paramount to obtain a more thorough picture of the dynamics of the debris stream, the eventual formation of the accretion disc and to infer the spin of the BH.

\section{Summary}
We present results of a spectroscopic monitoring campaign of the tidal disruption event AT2019dsg. We perform a detailed analysis of the emission line content and evolution at optical wavelengths, using 26 spectra covering 34 to 411 days after discovery. Combining these results with other multi-wavelength information, including radio interferometric observations and X-ray and UV measurements, we attempt to explain the multi-wavelength observed properties and their evolution in a coherent manner.

\begin{itemize}
\item The TDE is X-ray bright, but the X-ray luminosity rapidly decays and becomes undetectable in less then 200 days. This rapid decline can be explained with a cooling disc. 
\item 
In line with \citet{stein20} we suggest that an outflow is powered by the accretion flow and the outflow is responsible for the optical (line) emission. The emission lines, particularly \heii, trace the outflow velocity (which, as modeled in \citealt{stein20}, increases between 120 and 180 days after discovery). We therefore propose that the emission lines are formed in this expanding medium. 
\item Our  e-MERLIN radio observations show that the emission is still increasing at 5.1 GHz, in contrast with radio observations at higher frequencies, which show a clear decreasing trend. This indicates that the TDE is still optically thick at $t \sim 200$ d for $\nu \lesssim 5.1$ GHz. While \citet{stein20} interpreted their radio observations as being due to a sub-relativistic outflow, we warn that a relativistic jet with a fast core and a slow sheath could possible also explain the radio observations (see, e.g., \citealt{Mimica2015}) and, advocating a change in the density profile, the apparent increase in the expansion speed around day 180.
\item At later times, more than 200 days after discovery, low ionisation Fe emission lines appear. We do not have X-ray data at these epochs, but the presence of the Fe lines indicates the continued presence of an ionising continuum source.
\item We show that after carefully subtracting the host galaxy light, the features previously identified as \niii\ Bowen fluorescence lines by \citet{vanvelzen20} disappear. This underlines the importance of performing host galaxy subtraction for the spectral study of TDEs.
\item We discuss the similarities between the Fe-strong TDEs in terms of the presence/absence of Fe and N Bowen fluorescence lines.
\item In our medium/high resolution spectra, we see absorption lines that could be due to the self-gravitating debris streams, caused by the disruption of the star, intercepting our line of sight. High resolution spectroscopy may play a big role in the future for understanding the dynamics of the debris stream in TDEs.
\end{itemize}

\section*{Acknowledgements}
GC and PGJ acknowledge support from European Research Council (ERC) Consolidator Grant 647208.
TW is funded by ERC grant 320360 and by European Commission grant 730980.
This research was made possible through the use of the AAVSO Photometric All-Sky Survey (APASS), funded by the Robert Martin Ayers Sciences Fund and NSF AST-1412587.
We acknowledge the use of public data from the {\it Swift} data archive. We thank P. M. Vreeswijk for the data reduction of the UVES spectrum.
This paper includes data obtained with the William Herschel Telescope (proposal IDs: W19B/P7, W19A/N3) as well as observations made with the Italian Telescopio Nazionale Galileo (TNG) operated on the island of La Palma by the Isaac Newton Group of Telescopes and the Fundación Galileo Galilei of the INAF (Istituto Nazionale di Astrofisica) at the Spanish Observatorio del Roque de los Muchachos of the Instituto de Astrofisica de Canarias.
Based on observations collected at the European Organisation for Astronomical Research in the Southern Hemisphere under ESO programmes 1103.D-0328 "advanced Public ESO Spectroscopic Survey for Transient Objects" (ePESSTO+). GL was supported by a research grant (19054) from VILLUM FONDEN. D.M-S. acknowledges support from the ERC under the European Union’s Horizon 2020 research and innovation programme (grant agreement No. 715051; Spiders). JM acknowledges financial support from the State Agency for Research of the Spanish MCIU through the ``Center of Excellence Severo Ochoa'' award to the Instituto de Astrof\'isica de Andaluc\'ia (SEV-2017-0709) and from the grant RTI2018-096228-B-C31 (MICIU/FEDER, EU). IA is a CIFAR Azrieli Global Scholar in the Gravity and the Extreme Universe Program and acknowledges support from that program, from the European Research Council (ERC) under the European Union’s Horizon 2020 research and innovation program (grant agreement number 852097), from the Israel Science Foundation (grant numbers 2108/18 and 2752/19), from the United States - Israel Binational Science Foundation (BSF), and from the Israeli Council for Higher Education Alon Fellowship. TWC acknowledges the EU Funding under Marie Sk\l{}odowska-Curie grant H2020-MSCA-IF-2018-842471. TMB was funded by the CONICYT PFCHA / DOCTORADOBECAS CHILE/2017-72180113. MN is supported by a Royal Astronomical Society Research Fellowship. 
\newline

\subsection*{Data Availability}
All data will be made available in a reproduction package uploaded to Zenodo. 

\bibliographystyle{mnras.bst}
\bibliography{bibliography.bib}
\newpage

\appendix
\section{FLOYDS spectra}
\label{app:floyds}

    \begin{figure*}
        \includegraphics[scale=0.55, angle = 0]{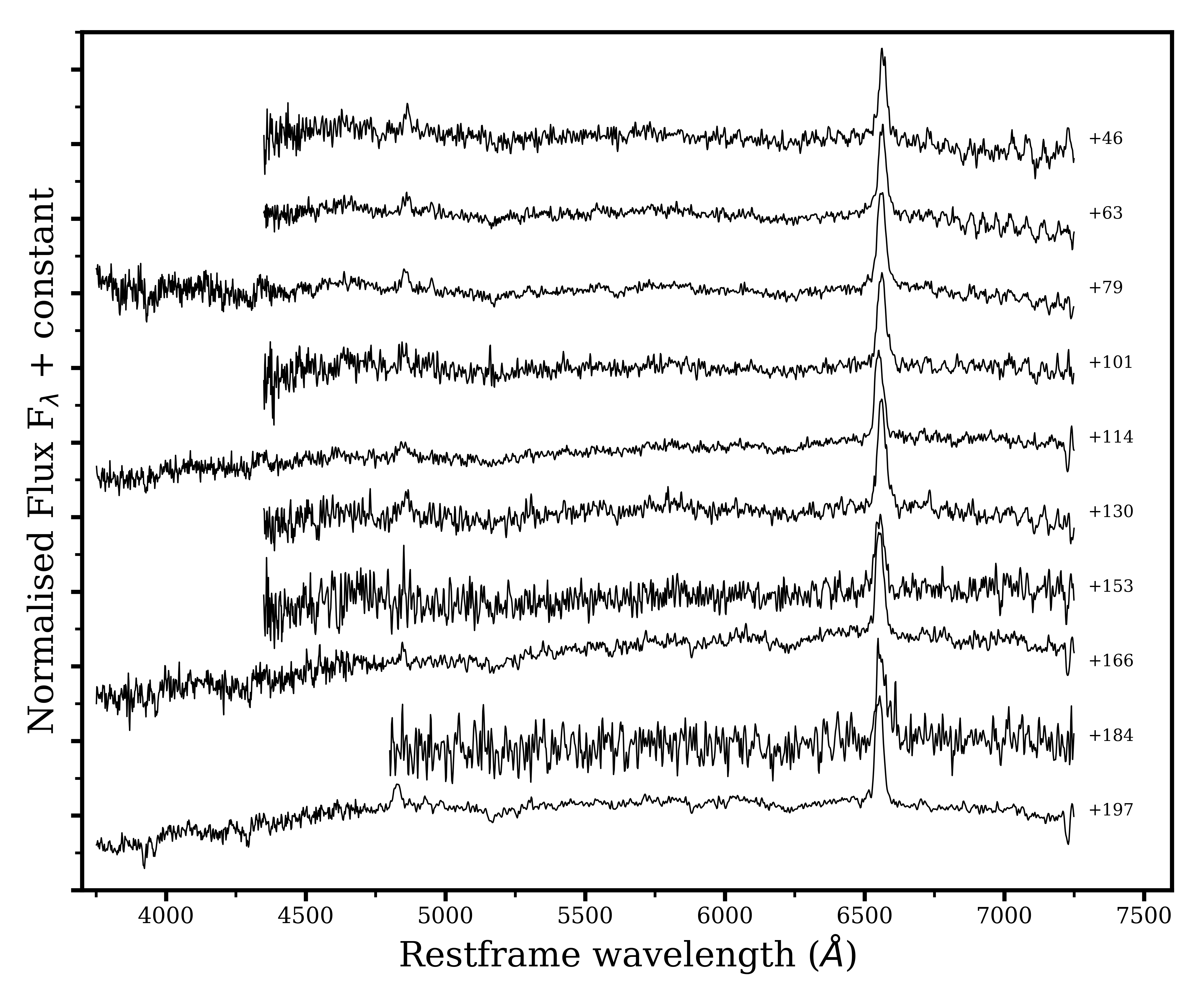} 
        \caption{Sequence of spectra taken with the FLOYDS spectrograph. For each spectrum the phase with respect to the transient discovery date (2019 Apr 09, MJD 58582) is reported on the right. For plotting purposes, all spectra have been divided by their median value.}
        \label{fig:spectra_dsg_floyds}   
    \end{figure*}

\begin{table*}
	\centering
	\small
 	\caption{FLOYDS spectroscopic observations}

 	\begin{center}
	\begin{tabular}{lcccl}
		\hline
		MJD$^{(1)}$ & phase$^{(2)}$ &UTC Date   & exposure time   & slit   \\
        	{[days]} & {[days]}	 &          & [s]             &[$''$]     \\
        \hline
        58\,628.73 &  +46 &  2019 May 25 & 3600 & 2.0 \\
        58\,645.65 &  +63 &  2019 Jun 11 & 3600 & 2.0 \\
        58\,661.72 &  +79 &  2019 Jun 27 & 3600 & 2.0 \\
        58\,683.66 &  +101 &  2019 Jul 19 & 3600 & 2.0 \\
        58\,696.65 &  +114 &  2019 Aug 01 & 3600 & 2.0 \\
        58\,712.49 &  +130 &  2019 Aug 17 & 3600 & 2.0 \\
        58\,735.55 &  +153 &  2019 Sep 09 & 3600 & 2.0 \\
        58\,748.41 &  +166 &  2019 Sep 22 & 3600 & 2.0 \\
        58\,766.46 &  +184 &  2019 Oct 10 & 3600 & 2.0 \\
        58\,779.35 &  +197 &  2019 Oct 23 & 3600 & 2.0 \\
		\hline
\end{tabular}
\end{center}
\textit{Note.}(1) Modified Julian Day of observations; (2) calculated with respect to the discovery date MJD 58\,582. 
\end{table*} 

\newpage
\section{Line fitting results}

\begin{table*}
\begin{center}
\caption{FWHM of the emission lines, in \kms as measured from our spectra at epoch given by the Modified Julian Date in the first column, corrected for instrumental broadening. With $\cdots$ we indicate an epoch in which the line in question could not be fit.}
\label{tab:lines_fwhm}

\begin{tabular}{lcccccccc}\hline
MJD$^{(1)}$ & phase$^{(2)}$ & \hd\   & \hg\   & \heii\ & \hb\ broad & \ha\ broad  & \hb narrow & \ha\ narrow \\
\hline               
58\,616.36 & +34  & 3880 $\pm$ 340     &  3560 $\pm$ 280   &  9850  $\pm$ 370  &    5800  $\pm$ 740   &   11970 $\pm$ 1010 &  1110 $\pm$ 110  &  920  $\pm$ 60 \\
58\,631.13 & +49  & 4630 $\pm$ 200     &  3020 $\pm$ 170   &  5790  $\pm$ 290  &    2720  $\pm$ 740   &   8840  $\pm$ 670  &  1010 $\pm$ 220  &  1020 $\pm$ 70 \\  
58\,638.28 & +55  &	 $\cdots$	       &	 $\cdots$	   &  7610  $\pm$ 660  &    4310  $\pm$ 760   &   7540  $\pm$ 880  &  1180 $\pm$ 210  &  1170 $\pm$ 30 \\
58\,642.35 & +60  & 4020 $\pm$ 310     &  2290 $\pm$ 230   &  7610  $\pm$ 430  &    3910  $\pm$ 870   &   5630  $\pm$ 420  &  1100 $\pm$ 160  &  960  $\pm$ 60 \\
58\,644.32 & +62  & 3790 $\pm$ 230     &  2880 $\pm$ 210   &  6980  $\pm$ 370  &    3920  $\pm$ 840   &   6810  $\pm$ 430  &  1100 $\pm$ 130  &  750  $\pm$ 60 \\
58\,655.19 & +73  &	 $\cdots$	       &  2440 $\pm$ 420   &  7260  $\pm$ 810  &    15080 $\pm$ 3310  &   14100 $\pm$ 930  &  1400 $\pm$ 120  &  1060 $\pm$ 70   \\
58\,656.12 & +74  &	 $\cdots$	       &  2170 $\pm$ 260   &  8940  $\pm$ 770  &    5800  $\pm$ 530   &   11110 $\pm$ 730  &  1150 $\pm$ 110  &  1070 $\pm$ 70  \\ 
58\,657.09 & +75  &	 $\cdots$      	   &  1970 $\pm$ 190   &  6980  $\pm$ 420  &    5750  $\pm$ 750   &   12140 $\pm$ 740  &  1220 $\pm$ 90   &  1050 $\pm$ 70  \\ 
58\,658.14 & +76  &	 $\cdots$          &  2360 $\pm$ 180   &  10760 $\pm$ 360  &    5450  $\pm$ 810   &   8740  $\pm$ 690  &  850  $\pm$ 90   &  1060 $\pm$ 70   \\
58\,676.19 & +94  &	 $\cdots$          &  2040 $\pm$ 180   &  8720  $\pm$ 420  &    6640  $\pm$ 450   &   12490 $\pm$ 710  &  1160 $\pm$ 90   &  1100 $\pm$ 70 \\  
58\,678.10 & +96  &	 $\cdots$          &  2710 $\pm$ 240   &  8050  $\pm$ 500  &    11200 $\pm$ 1330  &   13200 $\pm$ 890  &  1310 $\pm$ 90   &  1120 $\pm$ 80  \\ 
58\,689.25 & +107 & 3040 $\pm$ 400     &  2590 $\pm$ 210   &  10300 $\pm$ 440  &    4830  $\pm$ 850   &   11450 $\pm$ 1370 &  1320 $\pm$ 130  &  1020 $\pm$ 70  \\ 
58\,690.11 & +108 &	 $\cdots$          &  2190 $\pm$ 300   &  6930  $\pm$ 510  &    3950  $\pm$ 950   &   12580 $\pm$ 1040 &  1210 $\pm$ 110  &  1110 $\pm$ 80  \\ 
58\,692.12 & +110 &	 $\cdots$    	   &  2310 $\pm$ 180   &  7740  $\pm$ 340  &    8150  $\pm$ 780   &   12150 $\pm$ 710  &  1280 $\pm$ 90   &  1150 $\pm$ 70  \\ 
58\,710.94 & +128 &	 $\cdots$          &         $\cdots$  &  6320  $\pm$ 470  &    20100 $\pm$ 9500  &   10210 $\pm$ 850  &  1390 $\pm$ 130  &  1170 $\pm$ 80  \\ 
58\,715.88 & +134 &	 $\cdots$          &  2460 $\pm$ 180   &  7410  $\pm$ 380  &    5780  $\pm$ 450   &   11910 $\pm$ 1250 &  1210 $\pm$ 90   &  1170 $\pm$ 70  \\ 
58\,717.00 & +135 &  $\cdots$          &  2270 $\pm$ 200   &  7300  $\pm$ 410  &    8000  $\pm$ 760   &   12570 $\pm$ 970  &  1370 $\pm$ 90   &  1180 $\pm$ 70  \\
58\,717.89 & +135 &  $\cdots$          &  1890 $\pm$ 110   &  6310  $\pm$ 270  &    4710  $\pm$ 220   &      $\cdots$      &  1220 $\pm$ 40   &        $\cdots$     \\  
58\,724.14 & +142 &  $\cdots$          &  2160 $\pm$ 36    &  6370  $\pm$ 110  &    3570  $\pm$ 150   &      $\cdots$      &  1190 $\pm$ 20   &        $\cdots$     \\
58\,733.93 & +151 &  $\cdots$          &  3280 $\pm$ 550   &  7440  $\pm$ 1020 &    5470  $\pm$ 1030  &   10140 $\pm$ 1010 &  1210 $\pm$ 150  &  1180 $\pm$ 80   \\
58\,735.96 & +153 &  $\cdots$          &  2420 $\pm$ 290   &  7630  $\pm$ 600  &    8610  $\pm$ 1110  &   13100 $\pm$ 1110 &  1290 $\pm$ 100  &  1170 $\pm$ 80   \\
58\,777.02 & +195 &  $\cdots$          &  2510 $\pm$ 300   &  11730 $\pm$ 780  &   	   $\cdots$           &      $\cdots$  &  1800 $\pm$ 150  &  1100 $\pm$ 60   \\
58\,808.04 & +226 &  $\cdots$     	   &  2200 $\pm$ 390   &  13000 $\pm$ 1310 &  	   $\cdots$           &      $\cdots$  &  1520 $\pm$ 90   &    $\cdots$         \\
58\,824.83 & +242 &  $\cdots$     	   &         $\cdots$  &	   $\cdots$    &       $\cdots$ 	      &      $\cdots$  &	   $\cdots$   &  1260 $\pm$ 50 \\
58\,827.82 & +245 &  $\cdots$   	   &         $\cdots$  &    $\cdots$       &       $\cdots$           &      $\cdots$  &       $\cdots$   &  1230 $\pm$ 60 \\
58\,994.00 & +411 &  $\cdots$          &  1860 $\pm$ 160   &	   $\cdots$    &   	  $\cdots$	          &	     $\cdots$  &  1190 $\pm$ 60   &  1170 $\pm$ 50 \\
 
\hline
\end{tabular}

\textit{Notes:} (1) Modified Julian Date of the observations, (2) days passed from hte disovery of the transient, MJD 58\,582.46. 
\end{center}
\end{table*}

\begin{table*}
\begin{center}
\caption{Equivalent Width of the emission lines, in \AA. With $\cdots$ we indicate an epoch in which the line in question could not be fit.}
\label{tab:lines_ew}

\begin{tabular}{lccccccccc}\hline
MJD$^{(1)}$ & phase$^{(2)}$ & \hd  & \hg\   & \heii\   & \hb\ broad & \ha\ broad  & \hb\ narrow  & \ha\ narrow \\

\hline     
58\,616.36  & +34  & 9.0 $\pm$ 1.0  & 10.8 $\pm$ 1.1 &  44.6 $\pm$ 2.2  &  15.1 $\pm$ 3.1  &  33.5 $\pm$ 3.6  &  11.1 $\pm$ 1.2  &   43.0 $\pm$ 1.0   \\
58\,631.13  & +49  & 13.2 $\pm$ 0.7 & 8.5 $\pm$ 0.6  &  14.2 $\pm$ 1.0  &  5.8  $\pm$ 4.3  &  23.6 $\pm$ 2.5  &  6.6 $\pm$ 2.1   &   36.8 $\pm$ 1.1   \\
58\,638.28  & +55  &    $\cdots$    &    $\cdots$    &  17.6 $\pm$ 1.9  &  8.7  $\pm$ 2.6  &  15.6 $\pm$ 2.6  &  4.4 $\pm$ 1.0   &   29.8  $\pm$ 1.2 \\
58\,642.35  & +60  & 9.4 $\pm$ 0.9  & 5.8 $\pm$ 0.7  &  19.3 $\pm$ 1.4  &  7.1  $\pm$ 3.4  &  16.3 $\pm$ 2.0  &  8.4 $\pm$ 1.4   &   26.0 $\pm$ 0.9   \\
58\,644.32  & +62  & 11 $\pm$ 0.8   & 8.0 $\pm$ 0.7  &  18.9 $\pm$ 1.3  &  7.2  $\pm$ 3.0  &  21.9 $\pm$ 2.0  &  9.3 $\pm$ 1.2   &   35.3 $\pm$ 0.9   \\
58\,655.19  & +73  &    $\cdots$    &  6.2 $\pm$ 1.3 &  13.4 $\pm$ 2.1  &  24.3 $\pm$ 5.9  &  36.4 $\pm$ 3.0  &  11.2 $\pm$ 0.9  &  37.3 $\pm$ 0.8   \\
58\,656.12  & +74  &    $\cdots$    & 7.6 $\pm$ 1.1  &  15.5 $\pm$ 1.7  &  15.7 $\pm$ 2.3  &  35.1 $\pm$ 3.0  &  11.6 $\pm$ 0.9  &  47.4 $\pm$ 1.0   \\
58\,657.09  &  +75 &    $\cdots$    & 6.8 $\pm$ 0.8  &  14.4 $\pm$ 1.1  &  8.0  $\pm$ 1.7  &  40.8 $\pm$ 3.1  &  12.2 $\pm$ 0.6  &  49.5 $\pm$ 1.0   \\
58\,658.14  & +76  &    $\cdots$    & 8.9 $\pm$ 0.8  &  36.8 $\pm$ 1.5  &  6.6  $\pm$ 1.5  &  22.5 $\pm$ 2.4  &  7.5 $\pm$ 0.5   &   47.0 $\pm$ 0.9   \\
58\,676.19  & +94  &    $\cdots$    & 6.7 $\pm$ 0.6  &  18.5 $\pm$ 1.1  &  16.5 $\pm$ 1.6  &  39.9 $\pm$ 2.9  &  9.6 $\pm$ 0.5   &   48.2 $\pm$ 0.8   \\
58\,678.10  & +96  &    $\cdots$    & 7.5 $\pm$ 0.8  &  16.8 $\pm$ 1.2  &  20.6 $\pm$ 2.7  &  38.9 $\pm$ 3.3  &  9.6 $\pm$ 0.5   &   43.1 $\pm$ 1.0   \\
58\,689.25  & +107 & 4.9 $\pm$ 0.8  & 7.9 $\pm$ 0.7  &  28.5 $\pm$ 1.6  &  8.6  $\pm$ 2.9  &  23.4 $\pm$ 3.6  &  10.3 $\pm$ 1.2  &  38.8 $\pm$ 1.1   \\
58\,690.11  & +108 &    $\cdots$    & 5.2 $\pm$ 0.9  &  12.8 $\pm$ 1.2  &  4.3  $\pm$ 2.0  &  32.3 $\pm$ 3.5  &  8.9 $\pm$ 0.9   &   45.6 $\pm$ 1.1   \\
58\,692.12  & +110 &    $\cdots$    & 6.9 $\pm$ 0.6  &  17.0 $\pm$ 1.1  &  13.0 $\pm$ 1.7  &  35.1 $\pm$ 2.7  &  11.0 $\pm$ 0.5  &  49.0 $\pm$ 0.8   \\
58\,710.94  & +128 &    $\cdots$    & 5.5 $\pm$ 2.0  &  13.1 $\pm$ 4.3  &  37.6 $\pm$ 20.0 &  30.4 $\pm$ 3.5  &  12.9 $\pm$ 1.2  &  50.3 $\pm$ 1.2   \\
58\,715.88  & +134 &    $\cdots$    & 8.4 $\pm$ 0.7  &  14.9 $\pm$ 1.0  &  10.6 $\pm$ 1.3  &  22.6 $\pm$ 3.1  &  9.1 $\pm$ 0.5   &   43.7 $\pm$ 0.9   \\
58\,717.00  & +135 & $\cdots$       & 6.1 $\pm$ 0.6  &  12.8 $\pm$ 0.9  &  10.9 $\pm$ 1.5  &  26.7 $\pm$ 2.6  &  8.8 $\pm$ 0.4   &   41.4 $\pm$ 0.8   \\
58\,717.89  & +135 & $\cdots$       & 9.6 $\pm$ 1.6  &  23.0 $\pm$ 1.5  &  20.9 $\pm$ 1.6  &       $\cdots$   &  14.3 $\pm$ 0.6  &       $\cdots$   \\
58\,724.14  & +142 & $\cdots$       & 7.3 $\pm$ 0.2  &  15.9 $\pm$ 0.4  &  5.8 $\pm$ 0.5   &       $\cdots$   & 9.1 $\pm$ 0.2    &       $\cdots$   \\
58\,733.93  & +151 & $\cdots$       & 8.1 $\pm$ 1.7  &  13.3 $\pm$ 2.4  &  11.0 $\pm$ 3.3  &  24.9 $\pm$ 3.4  &  9.1 $\pm$ 1.1   &   53.0 $\pm$ 1.2      \\
58\,735.96  & +153 & $\cdots$       & 6.5 $\pm$ 0.9  &  13.9 $\pm$ 1.4  &  12.2 $\pm$ 2.1  &  38.7 $\pm$ 4.2  &  10.9 $\pm$ 0.6  &  61.4 $\pm$ 1.5      \\
58\,777.02  & +195 & $\cdots$       & 4.2 $\pm$ 0.6  &  15.5 $\pm$ 1.3  &      $\cdots$    &       $\cdots$   &	11.5 $\pm$ 0.5   &  41.1 $\pm$ 0.9       \\
58\,808.04  & +226 & $\cdots$       & 3.4 $\pm$ 0.7  &  14.8 $\pm$ 1.9  &      $\cdots$    &       $\cdots$   &	10.1 $\pm$ 0.6   &      $\cdots$         \\
58\,824.83  & +242 & $\cdots$       &   $\cdots$     &       $\cdots$   &      $\cdots$    &       $\cdots$   &       $\cdots$   &    45.2 $\pm$ 1.0   \\
58\,827.82  & +245 & $\cdots$       &   $\cdots$     &       $\cdots$   &      $\cdots$    &       $\cdots$   &       $\cdots$   &    46.7 $\pm$ 1.2   \\
58\,994.00  & +411 & $\cdots$       & 4.6 $\pm$ 0.5  &       $\cdots$   &      $\cdots$    &       $\cdots$   &   13.6 $\pm$ 0.5 &  65.9 $\pm$ 2.6      \\
 \hline\end{tabular}

\textit{Notes:} (1) Modified Julian Date of the observations, (2) days passed from hte disovery of the transient, MJD 58\,582.46.
\end{center}
\end{table*}

\begin{table*}
\begin{center}
\caption{Shift with respect to the restframe wavelength of the emission lines, in \kms. With $\cdots$ we indicate an epoch in which the line in question could not be fit. The narrow peaks of \ha\ and \hb\ are not listed, as we assumed they were at the restframe and we used their offset to probe uncertainties in the wavelength calibration.}
\label{tab:lines_shift}

\begin{tabular}{lcccccc}\hline
MJD$^{(1)}$ & phase$^{(2)}$ & \hd\ & \hg\   & \heii\   & \hb\ broad$^{(3)}$ & \ha\ broad   \\

\hline     
58\,616.36  & +34  &  240  $\pm$ 330 &  170  $\pm$ 310   &  -1770 $\pm$ 320  &  20   $\pm$ 300  &  -170  $\pm$ 460  \\                      
58\,631.13  & +49  &  230  $\pm$ 310 &  630  $\pm$ 300   &  -1440 $\pm$ 320  &  420  $\pm$ 300  &  -220  $\pm$ 370  \\                       
58\,638.28  & +55  &      $\cdots$   &       $\cdots$    &  -2410 $\pm$ 360  &  -200 $\pm$ 300  &  -1560 $\pm$ 460  \\
58\,642.35  & +60  & -460 $\pm$ 320  &  310  $\pm$ 310   &  -2670 $\pm$ 330  &  -60  $\pm$ 300  &  -760  $\pm$ 350  \\                        
58\,644.32  & +62  & -36  $\pm$ 310  &  150  $\pm$ 310   &  -2250 $\pm$ 320  &  -100 $\pm$ 300  &  -380  $\pm$ 340  \\                     
58\,655.19  & +73  &      $\cdots$   &  -72  $\pm$ 340   &  -1960 $\pm$ 470  &  -400 $\pm$ 300  &  -1820 $\pm$ 440  \\                     
58\,656.12  & +74  &      $\cdots$   &  -200 $\pm$  310  &  -1190 $\pm$ 420  &  -320 $\pm$ 290  &  -690  $\pm$ 390  \\                     
58\,657.09  & +75  &      $\cdots$   &  240  $\pm$ 300   &  -1860 $\pm$ 330  &  -180 $\pm$ 290  &  -840  $\pm$ 390  \\                     
58\,658.14  & +76  &      $\cdots$   &  130  $\pm$ 300   &  -260  $\pm$ 320  &  -140 $\pm$ 290  &  -1190 $\pm$ 380  \\                     
58\,676.19  & +94  &      $\cdots$   &  80   $\pm$ 300   &  -1630 $\pm$ 330  &  -280 $\pm$ 290  &  -820  $\pm$ 390  \\                     
58\,678.10  & +96  &      $\cdots$   &  -100 $\pm$ 310   &  -2270 $\pm$ 370  &  -420 $\pm$ 290  &  -1390 $\pm$ 420  \\                     
58\,689.25  & +107 &  370  $\pm$ 330 &  480  $\pm$ 300   &  -2740 $\pm$ 340  &  -90  $\pm$ 300  &  40    $\pm$ 540  \\                       
58\,690.11  & +108 &      $\cdots$   &  55   $\pm$ 320   &  -2670 $\pm$ 340  &  -380 $\pm$ 290  &  -760  $\pm$ 460  \\                 
58\,692.12  & +110 &      $\cdots$   &  9    $\pm$ 300   &  -1850 $\pm$ 330  &  -370 $\pm$ 290  &  -930  $\pm$ 390  \\                     
58\,710.94  & +128 &      $\cdots$   &        $\cdots$   &  -2350 $\pm$ 490  &  -450 $\pm$ 300  &  -600  $\pm$ 410  \\                     
58\,715.88  & +134 &      $\cdots$   &  350  $\pm$ 300   &  -1590 $\pm$ 320  &  -150 $\pm$ 290  &  -2890 $\pm$ 550  \\                     
58\,717.00  & +135 &      $\cdots$   &  69   $\pm$ 300   &  -1750 $\pm$ 330  &  -230 $\pm$ 290  &  -900  $\pm$ 440  \\                     
58\,717.89  & +135 &      $\cdots$   &  240  $\pm$ 60    &  -2530 $\pm$ 70   &  -60  $\pm$ 60   &         $\cdots$	  \\ 
58\,724.14  & +142 &      $\cdots$   &  210  $\pm$ 17    &  -2250 $\pm$ 30   &  -460.1 $\pm$ 40 &         $\cdots$	  \\   
58\,733.93  & +151 &      $\cdots$   &  620  $\pm$ 360   &  -1050 $\pm$ 460  &  -250 $\pm$ 300  &  -880  $\pm$ 450  \\                     
58\,735.96  & +153 &      $\cdots$   &  330  $\pm$ 320   &  -1300 $\pm$ 370  &  -90  $\pm$ 290  &  -1470 $\pm$ 480  \\                       
58\,777.02  & +195 &      $\cdots$   &  380  $\pm$ 320   &  -3550 $\pm$ 380  &        $\cdots$      &         $\cdots$	  \\                     
58\,808.04  & +226 &      $\cdots$   &  -230 $\pm$ 330   &  -3540 $\pm$ 500  &        $\cdots$      &         $\cdots$	  \\                           
58\,824.83  & +242 &      $\cdots$   &         $\cdots$	 &         $\cdots$      &        $\cdots$      &         $\cdots$	  \\                 
58\,827.82  & +245 &      $\cdots$   &         $\cdots$	 &         $\cdots$      &        $\cdots$      &         $\cdots$	  \\                 
58\,994.00  & +411 &      $\cdots$   &  150  $\pm$ 300   &         $\cdots$      &        $\cdots$      &         $\cdots$	  \\       
 \hline

\end{tabular}

\textit{Notes:} (1) Modified Julian Date of the observations, (2) days passed from the disovery of the transient, MJD 58\,582.46. (3) The central wavelength of the broad \hb\ has been tied to the one of the narrow peak during the fitting procedure.
\end{center}
\end{table*}

\section{Photometry data}

\begin{table*}
\begin{center}
\caption{Extinction corrected, AB magnitudes}
\label{tab:mag}
\begin{tabular}{lccccccc}\hline
MJD$^{(1)}$ & seeing $['']$  &          w2 	  &	         m2 	 &		     w1 	&	    	U      &	   B	      &	      g         \\
\hline
58\,619.12*   &    1.9   &       $\cdots$   &       $\cdots$   &       $\cdots$   &       $\cdots$   &       $\cdots$   & 16.57 $\pm$ 0.06 \\
58\,620.71*   &    2.4   &       $\cdots$   &       $\cdots$   &       $\cdots$   &       $\cdots$   & 16.74 $\pm$ 0.08 & 16.56 $\pm$ 0.06 \\	
58\,620.18    & $\cdots$ & 16.46 $\pm$ 0.05 & 16.82 $\pm$ 0.06 & 16.83 $\pm$ 0.07 & 16.96 $\pm$ 0.09 & 16.50 $\pm$ 0.09 &       $\cdots$   \\
58\,621.07*   &    2.4   &       $\cdots$   &       $\cdots$   &       $\cdots$   &       $\cdots$   & 16.69 $\pm$ 0.10 & 16.48 $\pm$ 0.06 \\	
58\,624.08    & $\cdots$ & 16.56 $\pm$ 0.05 & 16.82 $\pm$ 0.06 & 16.92 $\pm$ 0.08 & 17.22 $\pm$ 0.11 & 16.66 $\pm$ 0.10 &       $\cdots$   \\       
58\,627.21    & $\cdots$ & 16.66 $\pm$ 0.05 & 16.82 $\pm$ 0.05 & 17.00 $\pm$ 0.07 & 17.11 $\pm$ 0.09 & 16.74 $\pm$ 0.10 &       $\cdots$   \\       
58\,627.80*   &    1.9   &       $\cdots$   &       $\cdots$   &       $\cdots$   &       $\cdots$   & 16.93 $\pm$ 0.08 & 16.64 $\pm$ 0.06 \\
58\,630.32    & $\cdots$ & 16.71 $\pm$ 0.05 & 16.84 $\pm$ 0.05 & 17.03 $\pm$ 0.07 & 17.09 $\pm$ 0.09 & 16.74 $\pm$ 0.09 &       $\cdots$   \\       
58\,633.52    & $\cdots$ & 16.61 $\pm$ 0.06 & 16.86 $\pm$ 0.07 & 17.09 $\pm$ 0.10 & 17.24 $\pm$ 0.13 & 16.75 $\pm$ 0.13 &       $\cdots$   \\       
58\,636.10    & $\cdots$ & 16.76 $\pm$ 0.05 & 17.14 $\pm$ 0.05 & 17.13 $\pm$ 0.07 & 17.33 $\pm$ 0.09 & 16.86 $\pm$ 0.09 &       $\cdots$   \\       
58\,639.27*   &    2.2   &       $\cdots$   &       $\cdots$   &       $\cdots$   &       $\cdots$   &       $\cdots$   & 16.60 $\pm$ 0.06 \\
58\,639.83    & $\cdots$ & 16.81 $\pm$ 0.04 & 17.09 $\pm$ 0.05 & 17.09 $\pm$ 0.06 & 17.30 $\pm$ 0.07 & 16.91 $\pm$ 0.08 &       $\cdots$   \\       
58\,642.40    & $\cdots$ & 16.84 $\pm$ 0.04 & 17.08 $\pm$ 0.05 & 17.19 $\pm$ 0.06 & 17.35 $\pm$ 0.08 & 16.96 $\pm$ 0.08 &       $\cdots$   \\       
58\,645.32    & $\cdots$ & 16.88 $\pm$ 0.05 & 17.16 $\pm$ 0.06 & 17.19 $\pm$ 0.07 & 17.42 $\pm$ 0.10 & 16.87 $\pm$ 0.09 &       $\cdots$   \\       
58\,647.69*   &    2.2   &       $\cdots$   &       $\cdots$   &       $\cdots$   &       $\cdots$   & 16.95 $\pm$ 0.08 & 16.66 $\pm$ 0.05 \\
58\,648.79    & $\cdots$ & 16.98 $\pm$ 0.05 & 17.19 $\pm$ 0.06 & 17.17 $\pm$ 0.07 & 17.30 $\pm$ 0.09 & 16.85 $\pm$ 0.09 &       $\cdots$   \\       
58\,653.58    & $\cdots$ & 17.07 $\pm$ 0.07 & 17.17 $\pm$ 0.07 & 17.29 $\pm$ 0.09 & 17.46 $\pm$ 0.12 & 16.90 $\pm$ 0.12 &       $\cdots$   \\       
58\,655.09*   &    2.8   &       $\cdots$   &       $\cdots$   &       $\cdots$   &       $\cdots$   & 16.78 $\pm$ 0.11 & 16.54 $\pm$ 0.10 \\
58\,656.04+   &    1.1   &       $\cdots$   &       $\cdots$   &       $\cdots$   &       $\cdots$   & 16.97 $\pm$ 0.07 & 16.64 $\pm$ 0.09 \\
58\,661.07+   &    1.0   &       $\cdots$   &       $\cdots$   &       $\cdots$   &       $\cdots$   & 16.98 $\pm$ 0.07 & 16.68 $\pm$ 0.07 \\
58\,661.75*   &    2.4   &       $\cdots$   &       $\cdots$   &       $\cdots$   &       $\cdots$   & 16.98 $\pm$ 0.07 & 16.59 $\pm$ 0.05 \\
58\,665.42*   &    2.4   &       $\cdots$   &       $\cdots$   &       $\cdots$   &       $\cdots$   & 17.01 $\pm$ 0.07 & 16.62 $\pm$ 0.05 \\
58\,667.10+   &    1.1   &       $\cdots$   &       $\cdots$   &       $\cdots$   &       $\cdots$   & 16.96 $\pm$ 0.07 & 16.69 $\pm$ 0.08 \\
58\,672.13+   &    1.3   &       $\cdots$   &       $\cdots$   &       $\cdots$   &       $\cdots$   & 16.98 $\pm$ 0.08 & 16.74 $\pm$ 0.06 \\
58\,675.16    & $\cdots$ & 17.38 $\pm$ 0.07 & 17.71 $\pm$ 0.08 & 17.67 $\pm$ 0.10 & 17.68 $\pm$ 0.13 & 16.90 $\pm$ 0.11 &       $\cdots$   \\       
58\,675.95*   &    2.1   &       $\cdots$   &       $\cdots$   &       $\cdots$   &       $\cdots$   & 17.10 $\pm$ 0.08 & 16.74 $\pm$ 0.06 \\
58\,680.19    & $\cdots$ & 17.48 $\pm$ 0.06 & 17.87 $\pm$ 0.07 & 17.75 $\pm$ 0.08 & 17.90 $\pm$ 0.11 & 17.22 $\pm$ 0.09 &       $\cdots$   \\       
58\,682.42*   &    2.9   &       $\cdots$   &       $\cdots$   &       $\cdots$   &      $\cdots$    & 16.93 $\pm$ 0.18 & 16.66 $\pm$ 0.10 \\
58\,693.60    & $\cdots$ & 17.64 $\pm$ 0.06 & 17.99 $\pm$ 0.08 & 18.00 $\pm$ 0.07 &      $\cdots$    &    $\cdots$      &       $\cdots$   \\       
58\,695.39    & $\cdots$ & 17.94 $\pm$ 0.09 & 18.06 $\pm$ 0.12 & 18.20 $\pm$ 0.10 &      $\cdots$    &    $\cdots$      &       $\cdots$   \\       
58\,699.31    & $\cdots$ & 17.98 $\pm$ 0.07 & 18.18 $\pm$ 0.10 & 18.23 $\pm$ 0.08 &      $\cdots$    &    $\cdots$      &       $\cdots$   \\       
58\,703.44    & $\cdots$ & 17.75 $\pm$ 0.09 & 18.28 $\pm$ 0.14 & 18.06 $\pm$ 0.10 &      $\cdots$    &    $\cdots$      &       $\cdots$   \\       
58\,707.42    & $\cdots$ & 17.98 $\pm$ 0.08 & 18.23 $\pm$ 0.10 & 18.02 $\pm$ 0.08 &      $\cdots$    &    $\cdots$      &       $\cdots$   \\       
58\,712.80    & $\cdots$ & 18.22 $\pm$ 0.08 & 18.52 $\pm$ 0.11 & 18.36 $\pm$ 0.08 &      $\cdots$    &    $\cdots$      &       $\cdots$   \\       
58\,729.28    & $\cdots$ & 18.40 $\pm$ 0.10 &       $\cdots$   &       $\cdots$   &      $\cdots$    &    $\cdots$      &       $\cdots$   \\       
58\,731.93    & $\cdots$ & 18.12 $\pm$ 0.11 & 18.60 $\pm$ 0.14 & 18.45 $\pm$ 0.17 & 18.07 $\pm$ 0.18 & 17.12 $\pm$ 0.14 &       $\cdots$   \\       
58\,733.44    & $\cdots$ & 18.40 $\pm$ 0.14 & 18.49 $\pm$ 0.16 & 18.30 $\pm$ 0.18 & 18.45 $\pm$ 0.25 & 17.26 $\pm$ 0.17 &       $\cdots$   \\       
58\,761.06    & $\cdots$ & 18.64 $\pm$ 0.12 & 18.84 $\pm$ 0.13 & 18.55 $\pm$ 0.16 & 18.41 $\pm$ 0.20 & 17.46 $\pm$ 0.15 &       $\cdots$   \\       
58\,766.05    & $\cdots$ & 18.73 $\pm$ 0.17 & 19.15 $\pm$ 0.21 & 18.75 $\pm$ 0.23 & 18.20 $\pm$ 0.21 & 17.48 $\pm$ 0.19 &       $\cdots$   \\       
58\,771.43    & $\cdots$ & 18.67 $\pm$ 0.09 & 18.95 $\pm$ 0.10 & 18.62 $\pm$ 0.12 & 18.24 $\pm$ 0.12 & 17.13 $\pm$ 0.09 &       $\cdots$   \\

\hline
\end{tabular}

\textit{Notes:} (1) Modified Julian Date of the observations. Epochs marked with * indicate data from Las Cumbres, epochs marked with + indicate data from LT, the rest of the data is taken with \textit{Swift}. For all filters the extinction correction was done using the values in \citet{schlafly11} who assume a reddening law with R$_v$=3.1
\end{center}
\end{table*}

\label{lastpage}
\end{document}